\documentclass[pra,aps,amsmath,amssymb,superscriptaddress,letter,nopacs,nofootinbib,longbibliography, twocolumn,10pt]{revtex4-2}
\usepackage[english]{babel}
\usepackage[utf8]{inputenc}
\usepackage[T1]{fontenc}

\usepackage{amsthm}
\usepackage{amsmath,amssymb}
\theoremstyle{plain} 
\usepackage{graphicx}

\usepackage{caption}
\usepackage{subcaption}

\usepackage{float} 
\usepackage[colorlinks=true, allcolors=blue]{hyperref}
\usepackage{braket}
\usepackage{float}
\usepackage{bbm,amsfonts}
\usepackage[shortlabels]{enumitem}
\usepackage{mathtools}
\usepackage[title]{appendix}
\usepackage{dsfont} 
\usepackage{tikz}
\usepackage{rotating} 
\usepackage{mathtools} 

\hypersetup{
	colorlinks=true,  
	linkcolor=blue,   
	citecolor=blue,   
	urlcolor=blue     
}

\newtheorem*{example*}{Example}

\newtheorem*{remark*}{Remark}

\usetikzlibrary{quantikz2}
\usepackage{circuitikz}

\usepackage{pifont}

\def\diams{\color{red}\ding{169}}
\def\hearts{\color{red}\ding{170}}
\def\spades{\ding{171}}

\usepackage{silence}
\DeclareMathOperator{\tr}{tr}
\newcommand{\one}[0]{\mathbb{I}}
\newcommand{\C}{\mathds{C}}

\newcommand{\albert}[1]{{\color{blue} #1}}

\DeclarePairedDelimiter\ceil{\lceil}{\rceil}
\DeclarePairedDelimiter\floor{\lfloor}{\rfloor}

\usepackage{soul} 
\usepackage{amsmath}
\usepackage{amssymb}
\usepackage{physics}
\usepackage{tikz}
\usepackage{quantikz}
\usepackage{graphicx}
\usepackage{nicematrix}
\usepackage{geometry}
\usepackage{graphicx}
\usepackage{url} 
\usepackage{hyperref} 
\usepackage{orcidlink}

\usepackage[justification=justified]{caption}
\DeclareCaptionJustification{justified}{\justified}

\newcommand{\com}[1]{\textcolor{blue}{#1}}

\newcommand{\rl}[1]{\left(#1\right)}

\newcommand{\mcC}{\mathcal{C}}
\newcommand{\mclC}{\overline{\mathcal{C}}}

\newcommand\Circle[1]{%
  \tikz[baseline=(char.base)]\node[circle,draw,inner sep=2pt] (char) {#1};}

\newcommand\Circlee[1]{%
  \tikz[baseline=(char.base)]\node[circle,inner sep=2pt] (char) {#1};}

\usepackage{pifont}

\usepackage{ragged2e}

\def\redCom{black}

\begin{document}

\makeatletter
\long\def\@makecaption#1#2{%
  \vskip\abovecaptionskip
  \justifying\noindent\small #1: #2\par
  \vskip\belowcaptionskip}
\makeatother

\title{\texorpdfstring{Absolutely maximally entangled pure states \\
of multipartite quantum systems}{}} 
\date{June 25, 2026} 


\author{Grzegorz Rajchel-Mieldzio{\'c}${}^{\orcidlink{0000-0003-2652-9482}}$}
\thanks{grzegorz@beit.tech}
\affiliation{BEIT sp.\ z o.o., ul.\ Mogilska 43, 31-545 Krak{\'o}w, Poland}

\author{Rafa{\l} Bistro{\'n}${}^{\orcidlink{0000-0002-0837-8644}}$}
\affiliation{Faculty of Physics, Astronomy and Applied Computer Science, Jagiellonian University, 30-348 Kraków, Poland}
\affiliation{Doctoral School of Exact and Natural Sciences, Jagiellonian University, ul. Łojasiewicza 11, 30-348 Kraków, Poland}

\author{Albert~Rico${}^{\orcidlink{0000-0001-8211-499X}}$}
\thanks{albert.ricoandres@uni-siegen.de}
\affiliation{Naturwissenschaftlich-Technische Fakult\"{a}t, Universit\"{a}t Siegen, Walter-Flex-Stra\ss e 3, 57068 Siegen, Germany}

\author{Arul Lakshminarayan${}^{\orcidlink{0000-0002-5891-6017}}$}
\affiliation{Department of Physics, \& Center for Quantum Information, Computation and Communication,
Indian Institute of Technology Madras, Chennai, India 600036}

\author{Karol {\.Z}yczkowski${}^{\orcidlink{0000-0002-0653-3639}}$}
\affiliation{Faculty of Physics, Astronomy and Applied Computer Science, Jagiellonian University, 30-348 Krak{\'o}w, Poland}
\affiliation{Center for Theoretical Physics, Polish Academy of Sciences, Warsaw
}

\begin{abstract}
Absolutely maximally entangled (AME) pure states
of a system composed of $N$ parties
are distinguished by the property that for any splitting at least one partial trace is maximally mixed. Due to maximal possible correlations between any two selected subsystems these states have numerous applications in various fields of quantum information processing
including multi-user teleportation, quantum error correction and secret sharing. We present an updated survey of various techniques to generate such strongly entangled states, including those going beyond  the standard construction of graph and stabilizer states.
Our contribution includes, in particular,
analysis of the degree of entanglement of reduced states obtained by partial trace of AME projectors, states obtained by a symmetric superposition of GHZ states, an orthogonal frequency square representation of the ``golden'' AME state and an updated summary of the number of local unitary equivalence classes.
\end{abstract}

\maketitle

\section{Introduction}


{\em 
\noindent
Dedicated to Ryszard Horodecki

\noindent
for his eightieth birthday.
}

\medskip
Quantum entanglement - the key feature of quantum theory - plays a pivotal role in
the theory of quantum information and various emerging quantum technologies.
Pure state entanglement of a bipartite quantum system is relatively well understood:
The entire information concerning entanglement is encoded in the vector of Schmidt coefficients --  singular values of the matrix
representing analyzed state in a product basis \cite{horodecki2009entanglement,Karol-GeoQstates2006}. 
Any state of an
$d \times d$ system, with all Schmidt coefficients squared equal to $1/d$
 is maximally entangled and is
 called {\em generalized Bell state}.

 The multipartite case with a system composed of $N\ge 3$ subsystems
is more complicated  \cite{CLPS99,BPRST00},
and also much more involved~\cite{AFOV08,WGE16,Srivastava2024,horodecki2024multipartiteentanglement}. For instance, for multipartite systems,  the notion of maximally entangled state
depends on the measure of multipartite entanglement used \cite{CKW00,Dur_2000}.
Several natural entanglement measures, including various
 distances to the set of fully separable states \cite{VP98,ZB02,WG03},
 are not easy to evaluate \cite{TPT08,MGBBB10}.
%
%

For a variety of purposes, one analyzes various splittings of the entire system
into two parties, evaluates bipartite entanglement and performs averaging
over various splittings \cite{MW02}. 
From an algebraic perspective, 
any $N$-partite state $|\psi\rangle$ is represented by a complex tensor $T$ with $N$ indices,
and therefore one can study its various flattenings and analyze
singular values of matrices generated in this way. 
This procedure is straightforward, in contrast to attempts
to evaluate the rank of a tensor \cite{BFZ23}
or to obtain a generalized singular value decomposition of a tensor
 \cite {AACJLT00,CHS00}.

One option is to look for pure quantum states that display maximal
entanglement for any possible cut of the system into two parts \cite{Scott2003QECCentPow}.
This is equivalent to the condition that for any symmetric splitting of the system
the partial traces are maximally mixed
\cite{CGSS05,AC13}.
Such states are called  {\em maximally multipartite entangled states} \cite{FFPP08,FFMPP10}
or {\em absolutely maximally entangled} (AME) states \cite{Helwig2012}.

It is known that such AME states do not exist for $N=4$ qubits 
\cite{Higuchi_2000_twoCouples}, 
nor for $N \ge 7$ qubits \cite{Huber_2017}, but they do exist for $N=5$ and $N=6$ qubits~\cite{Borras07}
and four subsystems with local dimension $d\ge 3$. Information
concerning the existence of AME$(N,d)$ states for low values of
the parameters $N$ and $d$ is kept updated in an online repository \cite{TableOfAME}, 
but in several cases their existence is still open.    

The identification of AME states for $N$ subsystems with $d$ levels each,
distinguished by their particular  properties,
is important from the point of view of foundations of quantum theory. 
The AME states correspond to 
multi-unitary matrices \cite{Goyeneche2015}
and {\em perfect tensors} \cite{Pastawski2015}
which form an indispensable tool in the field of
tensor networks \cite{Jahn2019}
and studies of the bulk--boundary correspondence 
\cite{bhattacharyya2016,Harris2018,Jahn2022}.
Furthermore, AME states are directly useful for performing various
tasks of quantum information processing. An AME state of $N=2k$
parties allows us to teleport an $k$-partite quantum state from any
given group of $k$ users to the remaining ones, and is crucial for a quantum secret sharing
protocol \cite{Helwig2012}.
It provides a quantum error correction code
\cite{laflamme1996AperfectQECC(5-2),Rains_1999,Raissi2018,Mazurek2020}
and a construction of a unitary matrix of order $d^k$
with the maximal entangling power,
introduced first for bipartite systems \cite{ZZF00}
and later generalized for multipartite  case \cite{LGMZ20}.

Last but not least, AME states are interesting
from the point of view of combinatorics and 
other branches of pure mathematics,
as they are related to quantum generalizations
of orthogonal Latin squares, cubes and hypercubes and 
to quantum orthogonal arrays \cite{Goyeneche2015}.

Although the topic of strongly entangled quantum
 states was intensively studied for more than two decades,
the last five years have brought entire families of brand-new solutions~\cite{Rather_2022,Zyczkowski_2023,Rather_2023,bistron_2023,Gross_2025},
not related to stabilizers and graph states.
The main goal of this contribution is twofold. First, we provide a brief survey of recent developments in the study of absolutely maximally entangled (AME) states. Second, we present new results concerning the degree of entanglement of subsystems obtained by tracing out certain parties (Section~\ref{Ent_AME}), and we update the current understanding of the non-equivalence problem for specific AME$(N,d)$ states (Section~\ref{sec:lu_equivalence}).

In addition, we exhibit a particular AME(4,5) state that can be expressed as a superposition of five GHZ-equivalent states (Appendix~\ref{app:AMEtoolbox}). We also present an orthogonal frequency square design associated with the ``golden'' AME(4,6) state (Section~\ref{sec:unconventional}), which may offer deeper insight into the quantum design underlying it and potentially inspire new, non-standard constructions.

This work is organized as follows.
In Section~\ref{sec:def} we set the scene 
providing the necessary notions and  definitions.
Section~\ref{sec:MinSupMDS} presents 
AME states with minimal support,
and Section~\ref{sec:Graphs} discusses
stabilizer and graph AME states.
Novel constructions of AME states not belonging 
to this standard class are analyzed in Section~\ref{sec:unconventional}. 
Section \ref{Ent_AME}
concerns entanglement in reduced AME states.
The problem of local equivalence and identification
of non-equivalent AME solutions 
is studied in Section~\ref{sec:lu_equivalence}. 
The relation between AME states and
quantum error correction codes is described in Section~\ref{sec:QEC}, and their applications in
tensor networks are discussed in Section~\ref{sec:holo}.  Appendices
include a survey of the properties of AME states in low dimensions
with a comprehensive list of known cases from various equivalence families.

\section{Setting the stage}
\label{sec:def}


The aim of this section is to recall relevant notions, quantities and measures.
Let $\ket{\psi}$ be a pure state of  $N$
particles, each of local dimension $d$.  It is specified by the complex amplitudes $T_{i_1, \cdots i_N}$ such that 
\begin{equation}
    \label{eq:General_Nd_state}
\ket{\psi}=\sum_{i_1 \cdots i_N=0}^{d-1} T_{i_1 \cdots i_N}\ket{i_1 \cdots i_N}.
\end{equation}
Let $[N]$ denote the set $\{1, \cdots, N\}$. Consider a bipartition of this set of  $N$ particles into two complementary 
subsets, say $\mathcal{C} \subset [N]$ and $\overline{\mathcal{C}}=[N] \setminus \mathcal{C}$. Without loss of generality we will assume that the size of $\mathcal{C}$ is not larger than $\overline{\mathcal{C}}$: $|\mathcal{C}|\leq |\overline{\mathcal{C}}|$. Let  $i_{\mathcal{C}}$ denote the $|\mathcal{C}|$ indices $\{i_k | k\in \mathcal{C}\}$ and $i_{\overline{\mathcal{C}}}$ be similarly defined, and let $T_{i_\mathcal{C} i_{\overline{\mathcal{C}}}}=T_{i_1, \cdots, i_N}$
be the component corresponding to the combined set of indices.
The $d^{|\mathcal{C}|} \times d^{|\overline{\mathcal{C}|}}$ shaped array $T_{\mathcal{C}}$ with elements
\begin{equation}
\label{eq:reshaping}
    \bra{i_\mathcal{C}}T_{\mathcal{C}}\ket{i_{\overline{\mathcal{C}}}}=T_{i_\mathcal{C} i_{\overline{\mathcal{C}}}}
\end{equation}
determines the entanglement between the set of particles in $\mathcal{C}$ with those in $\overline{\mathcal{C}}$. 
\medskip

\subsection{Absolutely Maximally Entangled states}
Quantum correlations in a pure state $\ket{\psi}$ shared 
between subsystems $\mcC$ and $\mclC$
can be characterized by the degree of mixing
of the reduced density matrix 
$\rho_\mcC={\rm Tr}_{\mclC} |\psi\rangle \langle \psi|
=T_{\mcC} T^{\dagger}_{\mcC}$.
Entanglement is maximal if the matrix $T_{\mcC}$
is proportional to a unitary, so that 
the partial trace is maximally mixed,
$\rho_{\mcC}=\mathbb{I}/d^{|\mcC|}$,
where $d^{|\mcC|}$ denotes the dimension
of the reduced state.

The degree of mixing of a density matrix
can be measured by the von Neumann entropy, $S(\rho)=-\tr (\rho \log \rho)$.
Thus the {\em entropy of entanglement}
of the pure state $|\psi\rangle$
with respect to the partition $\mcC|\mclC$
reads, $E(\psi)= S(\rho_\mcC)=S(\rho_{\mclC})$.
The maximum, $S(\rho_\mcC)=|\mcC| \log d$,
is achieved for a state maximally 
entangled with respect to this partition \cite{Karol-GeoQstates2006}. 

As already mentioned in the introduction, in the search for highly entangled multipartite states, {\em absolutely maximally entangled} state,
shortened to AME, (pronounced {\em Aa-may}),
have been singled out as those
such that for any bipartition $\mcC|\mclC$ as defined above, one has uniformly $S(\rho_{\mcC})=|\mcC|\log d$.
 This is the case if, and only if, for any choice of $\mcC$ the matrix $T_{\mcC}$ obtained by 
flattening of the tensor 
$T_{i_1 \cdots i_N}$, as in Eq.~\ref{eq:reshaping},
is unitary up to a constant,
so that all the density matrices $\rho_\mcC$ are maximally mixed   \cite{Helwig2012,helwig2013existApplic}.
From an information theoretic perspective, these states are such that any bipartition of it leads to maximal ignorance of the whole state if only local operations are performed.

 AME states shared among $N$ parties of local dimension $d$ each are denoted AME$(N,d)$. It remains an open problem (item 35 in a list of Open Problems at \cite{IQOQI}) to determine the pairs $N$ and $d$, for which states AME$(N,d)$ exist.
 To begin with AME$(2,d)$ exists for all $d>1$, and are simply generalizations of the qubit Bell states: \begin{equation}\ket{\Phi^+} = \frac{1}{\sqrt{d}}\sum_{j=0}^{d-1} \ket{jj}.
 \label{eq:GenBell}
 \end{equation}
{\color{\redCom} Note that a locally rotated state,
$\ket{\Phi_U} = \frac{1}{\sqrt{d}}
\sum_{j=0}^{d-1} \ket{j} \otimes U|j\rangle$
is also maximally entangled for any local unitary matrix $U$ of order $d$.
}
 
Similar AME$(3,d)$ exists for all $d>1$ as a generalization of the GHZ state:
\begin{equation}\ket{\Phi_{GHZ}} = \frac{1}{\sqrt{d}}\sum_{j=0}^{d-1} \ket{jjj}.
\label{eq:GenGHZ}
\end{equation}

\vspace{0.3cm}
\emph{k}-\textbf{uniformity} -- 
In the case of the GHZ state, the reduced states of  
the sets $\mcC$ contain only one particle are maximally mixed, $\rho_\mcC=\mathbb{I}/d$. 
States satisfying this property are called $1$-uniform.

More generally, a multipartite state $\ket{\psi}$ is {\em $k$-uniform} if the reduced state of {\em all} subsets $\mcC$, such that $|\mcC|=k\leq N/2$, is maximally mixed \cite{Rains_1999,Scott2003QECCentPow}. Exact~\cite{Goyeneche2014,Grassl:codetables,LW19,RTGA20} 
and approximated~\cite{Guo2025ApproxK-uni} $k$-uniform states are a valuable resource in quantum information and computation tasks, due to their high entanglement content. Tighter approximations to AME states have been recently achieved in~\cite{rico2026AAMEs}. We note that a $k$-uniform state is also $k'$-uniform if $k'<k$. Thus an AME$(N,d)$ state is a 
$k$-uniform state for all $k\leq \lfloor N/2 \rfloor$. Both $k$-uniform and AME states have been implemented experimentally to test their entanglement properties~\cite{miller2024weightEnumExperiment}.
Beyond pure states, the notion of $k$-uniformity can be also
generalized for mixed states \cite{KBK+19}.

Due to their symmetric structure and wide applicability, in this 
contribution we shall focus on AME states in finite-dimensional homogeneous systems, in which all $N$ subsystems
have the same local dimension $d$.
However, existence and constructions for heterogeneous systems~\cite{Goyeneche2014,GBZ16,Huber_2018Shadow,shen2021heterogeneous,ball2025AMEhetero} and continuous-variable systems~\cite{Facchi2009Gaussian} have
also been considered.

 

 
\subsection{Four party AME states and 2-unitary operators}

\begin{figure}[h!]
\begin{tikzpicture}
    \node[] at (0,0) {\includegraphics[scale=.23]{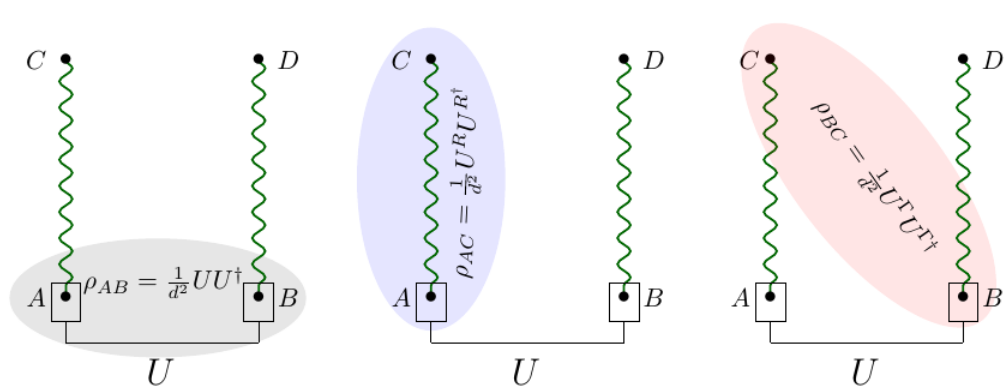}};
    \node[] at (-2.8,1.5) {($a$)};
    \node[] at (0.15,1.5) {($b$)};
    \node[] at (2.9,1.5) {($c$)};
\end{tikzpicture}
\caption{Four party state (\ref{eq:psi_as_UAB}). Three symmetric bi-partitions:  ($a$) $AB|CD$, ($b$) $AC|BD$, and ($c$) $AD|BC$;
and their associated marginal states in terms of the tensor $U\equiv U_{AB}$ and its rearrangements, see Eq.~(\ref{eq:3_rho_matrices}). Wavy lines represent
maximally entangled states between the respective particles. Figure borrowed from \cite{Suhail_Thesis}.}
\label{fig:4party}
\end{figure}
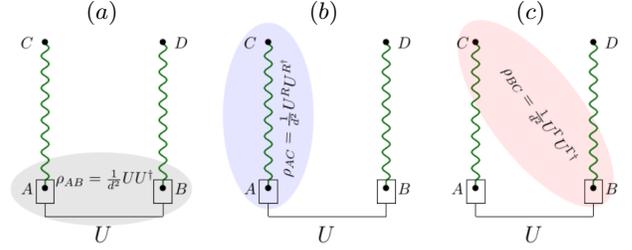

As far as existence itself is concerned, the first nontrivial case is $N=4$. The requirements  are spelled out more concretely in this case for clarity.
The general state of 4 particles (denoted here by $A,B,C,D$) can be written as:
\begin{equation}\label{eq:psi_as_UAB}
\begin{split}
    \ket{\psi}&=\sum_{i,j,k,l=0}^{d-1} T_{ijkl}\ket{ijkl}\\
    &=(\mathbb{I}_{CD}\otimes d\, T_{AB})\ket{\Phi^+}_{AC}\ket{\Phi^+}_{BD} \\ &=(\mathbb{I}_{CD}\otimes U_{AB})\ket{\Phi^+}_{AC}\ket{\Phi^+}_{BD},
\end{split}
\end{equation}




where $\ket{\Phi^+}$ are the maximally entangled two-particle states in Eq.~\eqref{eq:GenBell}, and $\bra{kl}T_{AB}\ket{ij}=T_{ijkl}$. 
\textcolor{\redCom}{In the notation of Eq.~(\ref{eq:reshaping}), $\mcC=\{A,B\}$, $\mclC=\{C,D\}$, $T_{\mcC}\equiv T_{AB}$.}
The symbol  $U_{AB}\equiv d\, T_{AB}$, is used suggestively as this 4 party state is pictured in Fig.~\ref{fig:4party}. Here the state is formed as the result of an operator $U$ acting on two particles ($A$ and $B$) while $A$ is maximally entangled with an ancilla $C$ and similarly $B$ with $D$. There are 3 symmetric bipartitions: $AB|CD$, $AC|BD$ and $AD|BC$, and if these are all maximally entangled, then $\ket{\psi}$ is an AME$(4,d)$ state. 

The reduced density matrices $\rho_{AB}$, $\rho_{AC}$ and $\rho_{BC}$ can be expressed as follows:
\begin{equation}
\begin{split}
    \rho_{AB}=\frac{1}{d^2}U_{AB}& U^{\dagger}_{AB}, \;
    \rho_{AC}=\frac{1}{d^2}U_{AB}^R U^{R\,\dagger}_{AB}, \quad \text{and} \\
    &\rho_{BC}=\frac{1}{d^2}U^{\Gamma}_{AB} U^{\Gamma\, \dagger}_{AB}.
\end{split}
\label{eq:3_rho_matrices}
\end{equation}
Here $U_{AB}$, $U_{AB}^R$ and $U_{AB}^{\Gamma}$ are special cases of the reshaping in Eq.~(\ref{eq:reshaping}). For bipartite operators $U$, the corresponding operators $U^R$ and $U^{\Gamma}$ have appeared widely in separability criteria of density matrices as the reshuffling
(also called realignment~\cite{Chen2002Realignment}) and partial transpose~\cite{Peres_1996PT}.
\begin{itemize}
\item[(i)] Reshuffling, $R$ :
$\bra{ij}U^R\ket{\alpha \beta}=\bra{i \alpha}U \ket{j \beta}.$
\item[(ii)] Partial (or blockwise) transpose, $\Gamma$:
$\bra{i\beta}U^{\Gamma}\ket{j\alpha}=\bra{i \alpha}U \ket{j \beta}.$
\end{itemize} 
This motivates the following definitions:
\begin{enumerate}
    \item A bi-partite matrix $U$ of dimension $d^2$ is {\em dual unitary} if $U$ and $U^R$ are unitary \cite{BKP19},
\item A matrix $U$ is {\em T-dual unitary} if $U$ and $U^\Gamma$ are unitary \cite{DNP17},
\item A matrix $U$ is called {\em 2-unitary} if
$U, U^R$ and $U^\Gamma$ are unitary \cite{Goyeneche2015}, so 
$U$ is both dual and T-dual unitary. 
\end{enumerate}
Thus the state $\ket{\psi}$
 will form a $2$-uniform AME$(4,d)$ state 
 if and only if the square matrix $U_{AB}$ of order $d^2$
 obtained by flattening of the tensor  $T_{ijkl}$ as 
 in Eq.~(\ref{eq:psi_as_UAB}) is 2-unitary.
 

\subsection{Classical and Quantum Orthogonal Latin squares}
Before continuing the discussion on strongly entangled
multi-partite states let us make a detour into the theory of classical combinatorial designs. Take $d$ copies of $d$ symbols, say $1,\dots, d$ and arrange them into a square of order $d$, such that all symbols in each row and each column are different. Such designs
 exist for any $d\ge 2$, and are called
{\em Latin squares}, as they were usually 
constructed out of Latin letters.

Consider another such a square, this time written with Greek letters, and place both of them one after another,
so that each 
cell of the square contains now
two letters from different alphabets. If all $d^2$ pairs of two letters are different, the Latin squares are called
{\em orthogonal} and the entire pattern has the 
natural name
{\em Graeco-Latin square}  -- 
see Fig.~\ref{tab:ame_4_4}.

\medskip 

\begin{table}[H]
\centering
\begin{tabular}{|l|l|l|}
\hline
{\color{red}A\hearts} & {K\spades} & {\color{red}Q\diams}\\ \hline
{\color{red}K\diams} & {\color{red}Q\hearts} & {A\spades}\\ \hline
{Q\spades} & {\color{red}A\diams} & {\color{red}K\hearts}\\ \hline
\end{tabular}
\;
$\equiv$
\;
\begin{tabular}{|l|l|l|}
\hline
A$\alpha$ & B$\beta$ & C$\gamma$ \\ \hline
B$\gamma$ & C$\alpha$ & A$\beta$ \\ \hline
C$\beta$ & A$\gamma$ & B$\alpha$ \\ \hline
\end{tabular}
\;
$\equiv$
\;
\begin{tabular}{|l|l|l|}
\hline
0,0 & 1,1 & 2,2 \\ \hline
1,2 & 2,0 & 0,1 \\ \hline
2,1 & 0,2 & 1,0 \\ \hline
\end{tabular}
\captionof{figure}{Exemplary Graeco-Latin square of order three: there are no repetitions of any digit in any row nor column of the square, while all nine cards are different. Instead of rank and suit of cards one can use Greek and Roman letters
as Euler did studying the problem more than 200 years ago.
This pattern determines AME(4,3) state
(\ref{eq:p9_state}).
}
\label{tab:ame_4_4}
\end{table}

More formally, a Graeco-Latin square of order $d$, also 
called {\em orthogonal Latin squares}
 and written as OLS$(d)$,  
 consists of $d^2$ pairs,  
 consists of $d^2$ pairs of symbols,
formed out of $d$ Greek letters and $d$ Latin letters, 
such that 
\begin{enumerate}[label=(\Alph*)]
    \item all $d^2$ pairs of symbols in the square are different,
    \item each row of the square contains all $d$ Greek and $d$ Latin
letters, which do not repeat, and
\item the same condition holds for  each column of the square.
\end{enumerate}
{\color{\redCom}To introduce a notation we will use later on: let $ K $ and $ L $ be two Latin squares of order $ d $, with entries $ K_{ij} $ and $ L_{ij} $, $0 \leq i,j \leq d-1$. They are orthogonal Latin squares if their cell-wise superposition  $ (K_{ij},L_{ij})$ has no repetition for all $0\leq i,j\leq d-1$. Hence they are a  permutation of all possible $d^2$ pairs $(i,j)$. }

\textcolor{\redCom}{These designs were studied by Euler, 
who conjectured that they exist for all $d\neq 2 \mod{4}$.
The nonexistence of OLS(2) is easy to see, and the first notable exception is $d=2+4=6$, and a formal proof that OLS(6) does not exist
was provided only in 1900 by Tarry \cite{Ta01}.}
In 1959 OLS were found for $d=22$ and
later for $d=14$ and $d=10$  \cite{BS59,BS60},
and since then it is known  \cite{CD01,Pa59}
that OLS exist in any dimension $d>2$, except $d=6$.

Going back to quantum states, 
the generalized GHZ state $\sum_{j=0}^{d-1} \ket{jjjj}/\sqrt{d}$ is 1-uniform, but it fails to be 2-uniform and hence is not an AME state. The first surprise is that AME$(4,2)$ states do not exist, namely no four qubit state is 2-uniform \cite{Higuchi_2000_twoCouples}, equivalently in the group $U(4)$ there are no 2-unitary matrices of order four 
-- see Fig. \ref{fig:P9epgt}a. This fact can 
be compared to {\em frustration} in spin systems \cite{FFMPP10}, \textcolor{\redCom}{as in the four-qubit case, the conditions that the three two-party reduced density matrices  $\rho_{AB},\, \rho_{AC}$ and $\rho_{BC}$, illustrated in Fig.~\ref{fig:4party}, each be maximally mixed ({\it i.e.}, equal to $\one/d^2$) cannot all be fulfilled at once.}


However, AME$(4,d)$ exists for all $d>2$. 
This is due to a connection 
(\ref{eq:ame_from_ols})
between AME states and orthogonal Latin squares (OLS): if an OLS of size $d$ exists, it is possible to find an AME$(4,d)$  state.  
An AME$(4,d)$ state can be constructed from an OLS $ (K_{ij},L_{ij})$ as \cite{CGSS05},
\begin{align}\label{eq:ame_from_ols}
\begin{aligned}
\ket{\text{AME}(4,d)} = \frac{1}{d} \sum_{i,j=0}^{d-1} \ket{ij} \ket{K_{ij}L_{ij}},
\end{aligned}
\end{align} 
see Appendix~\ref{app:AMEtoolbox}.

Therefore, since orthogonal Latin squares of order $d$ exist for all $d>2$ except
$d=6$, the existence of AME$(4,d)$ is guaranteed in all these dimensions, with the sole potential exception of $d=6$. Consequently, the case AME$(4,6)$ could not be settled using OLS methods and remained unresolved for some time, but was only recently proven to exist \cite{Rather_2022}.
This particular case will be treated 
separately in Section \ref{sec:unconventional}.

 The OLS in dimension 6, was associated with a famous puzzle of Officers of Euler: 36 officers, from 6 different ranks and 6 different regiments, are to be placed in a $6 \times 6$ square array such that no regiment or rank repeats along any row or column. The existence of AME$(4,6)$ may be interpreted as a quantum solution to this classically impossible problem, 
 provided superposition states are allowed. Thus quantum orthogonal Latin squares are investigated
 as generalizations of these classical designs.

As demonstrated in 1999 by Zauner, 
for any  classical combinatorial notion 
one can look for its quantum analogue \cite{Za99}. 
The notion of a Latin square can be generalized by replacing discrete symbols with vectors or pure quantum states  \cite{MV16,BN17}. 
A quantum Latin square of size $d$ consists of a square array of vectors such that each row and column of the array forms an orthonormal basis of $\mathcal{H}_d$. All classical Latin squares become quantum when we simply identify $\left\lbrace i \mapsto \ket{i}, 0\leq i\leq d-1\right \rbrace$ as the computational basis. It has been shown that for $d=2$ and $3$ all quantum Latin squares are equivalent to such classical ones for some appropriate choice of bases \cite{Paczos_2021}. However, for $d\geq 4$ there exist {\em genuine quantum} Latin squares \cite{Paczos_2021,Zang_2021,Zhang_2025}, which contain more than $d$
different states and cannot be transformed by unitary rotations into classical designs.


%

However, a more general notion of quantum orthogonal Latin squares (QOLS) is necessary to allow for entangled states in the bases of $\mathcal{H}_d \otimes \mathcal{H}_d$.  
Several alternative definitions of QOLS were introduced~\cite{Goyeneche_2018,Musto_2019,Han_2025}, and we shall follow the one
\cite{rico2020absolutely,Rather_2022},
directly related to multi-unitary 
matrices~\cite{Goyeneche2015},
most suitable for studying AME states.
A $d \times d$ array of bipartite pure normalized states $\ket{\Psi_{ij}} \in \mathcal{H}_d^{A} \otimes \mathcal{H}_d^{B}$ is said to form a {\em quantum orthogonal Latin square}, if they satisfy the following conditions,
\begin{subequations}
\begin{align}
\text{(A')}& \quad
\braket{\Psi_{ij}}{\Psi_{kl}} = \delta_{ij} \delta_{kl}  \, , \label{eq:Unitarity}\\
\text{(B')}& \quad 
\text{Tr}_A 
\sum_{k=0}^{d-1} \ket{\Psi_{ik}} \bra{ \Psi_{jk}}
= \delta_{ij} \mathbb{I}_d\label{eq:Duality}\quad\text{and}\\
\text{(C')}& \quad \text{Tr}_A 
\sum_{k=0}^{d-1}  \ket{\Psi_{ki}} \bra{ \Psi_{kj}}
= \delta_{ij} \mathbb{I}_d 
  \label{eq:T-duality}
\end{align}
\end{subequations}
where the partial trace can be taken over the party $A$ or $B$ equivalently. 
Note these conditions are
analogous to their classical counterparts:
 $d^2$ different pairs of classical symbols (A)
corresponds to orthogonality (A') of
bipartite states  (\ref{eq:Unitarity}),
while the no-repetition along rows (B) and columns (C) corresponds to (B') and (C').
The fact that the mean value of the first or second entries 
averaged along any row/column is a constant is reflected in 
 (\ref{eq:Duality}) and  (\ref{eq:T-duality}), as
both partial traces are proportional to identity.
%
For further remarks on the  interpretation
of these conditions see \cite{Zyczkowski_2023}. 
Recently, existence of quantum orthogonal Latin squares formed by separable states was discussed~\cite{Han_2025}.
These states form a proper subset of QOLS defined above.

\medskip

The following three statements are equivalent: 
\begin{enumerate}
    \item $\{\ket{\Psi_{ij}}, 0 \leq i,j \leq d-1 \}$ is a QOLS.
    \item $\ket{\psi}=\frac{1}{d}\sum_{i,j=0}^{d-1} \ket{ij}\ket{\Psi_{ij}}$ is an AME$(4,d)$ state.
    \item $U=\sum_{i,j=0}^{d-1} \ket{\Psi_{ij}}\bra{ij}$  is 2-unitary.
\end{enumerate}
It is quite straightforward to verify that the conditions $U$ to be unitary,
dual unitary and T-dual unitary are equivalent to the conditions in Eqs.~(\ref{eq:Unitarity}), ~(\ref{eq:Duality}), and ~(\ref{eq:T-duality}), respectively. 



The $2$-unitary matrix corresponding to the state in Eq. (\ref{eq:ame_from_ols}) constructed from an OLS gives a 2-unitary permutation of order $d^2$,
\begin{align}
\begin{aligned}
P_{d^2} = \sum_{i,j=0}^{d-1}  \ket{K_{ij}L_{ij}}\bra{ij}.
\end{aligned}
\end{align}
From the existence of OLS$(d)$ it follows that $2$-unitary permutations exist in all local dimensions $d$ except $d=2$ and $d=6$. If a $2$-unitary permutation is multiplied on either side with a diagonal unitary matrix, it remains 
$2$-unitary. In fact, permutations that are dual/T-dual unitary remain dual/T-dual unitary under such generally non-local unitary operations. 

Notions of $2$-unitarity and its generalization to multi-unitary matrices for constructing AME states with more than $4$ parties was formulated in \cite{Goyeneche2015}.
Circuits constructed using dual unitary matrices have been recently  studied as models of nonintegrable many-body quantum systems
\cite{BKP19},
in which correlation functions can be solved for exactly, for a recent review see \cite{BCP_review2025}.
This is facilitated by a ``space-time'' duality that is operational for dual unitary gates and explains their name.
The circuits constructed of $2$-unitary matrices possess extreme ergodic properties \cite{ASA_2021}, similar to that of Bernoulli systems at the apex of the classical ergodic hierarchy.


\subsection{Operator entanglement and entangling power}
For any bipartite unitary $U \in \mathbb{U}(d^2)$, it is useful to define the following entanglement entropies:
 \begin{equation}
     \label{eq:EUandEUS}
     \begin{split}
     E(U)&=1-\frac{1}{d^4}\Tr\left( U^R U^{R \dagger} \right)^2 \quad \text{and}\\
          E(US)&=1-\frac{1}{d^4}\Tr\left( U^\Gamma U^{\Gamma \dagger} \right)^2.
     \end{split}
     \end{equation}
Here $S$ is the SWAP gate: $S\ket{\phi_A}\ket{\phi_B}=\ket{\phi_B}\ket{\phi_A}$. The quantity $E(U)$ is a measure of the operator entanglement of $U$ \cite{Musz_2013}, at the same time it can be interpreted as the linear entropy of the state $\rho_{AC}$ of the 4-party state it defines via Eq.~\ref{eq:psi_as_UAB}, and hence the entanglement in the $AC|BD$ partition.
It ranges from $0$ when $U$ is a product operator to $E(S)=1-1/d^2$. $E(U)$ is maximized ($=E(S)$) iff $U$ is dual unitary, which provides another characterization of this class, which includes the SWAP gate $S$. Similarly $E(US)$ is maximized on the set of T-dual unitaries which include all product unitaries (and hence the Identity). Equivalently it may be interpreted as the linear entropy of $\rho_{BC}$ and hence the entanglement between the $AD|BC$ partition. Therefore $U_{AB}$ will be 2-unitary iff $E(U_{AB})=E(U_{AB} S)=E(S)$, which is equivalent to the single condition $E(U_{AB})+E(U_{AB} S)=2E(S)$. 

This combination has a deep significance of being essentially the entangling power of the bipartite unitary $e_p(U_{AB})$.
We recall the expression for $e_p$
along with that of a complementary quantity: 
 \begin{subequations}
     \begin{align}
         e_p(U)&=\frac{1}{E(S)}\left[ E(U)+E(US)-E(S)\right],
         \label{eq:EP}\\
         g_t(U)& = \frac{1}{2E(S)} \left[ E(U)-E(US)+E(S)\right].
         \label{eq:GT}
     \end{align}
 \end{subequations}

\begin{figure}[ht!]
 \includegraphics[width=.44\textwidth]{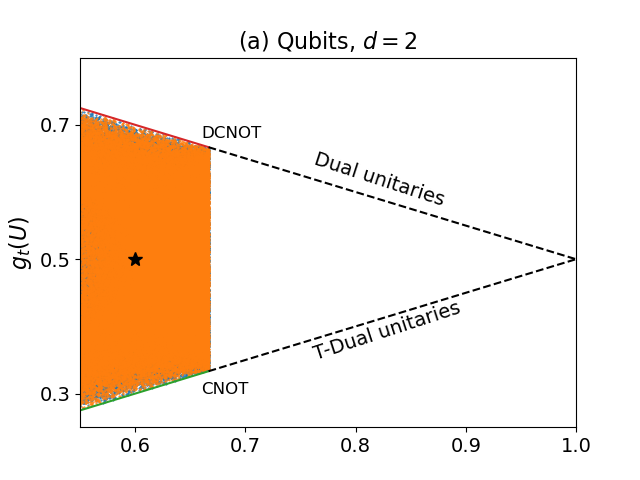}
 \includegraphics[width=.43\textwidth]{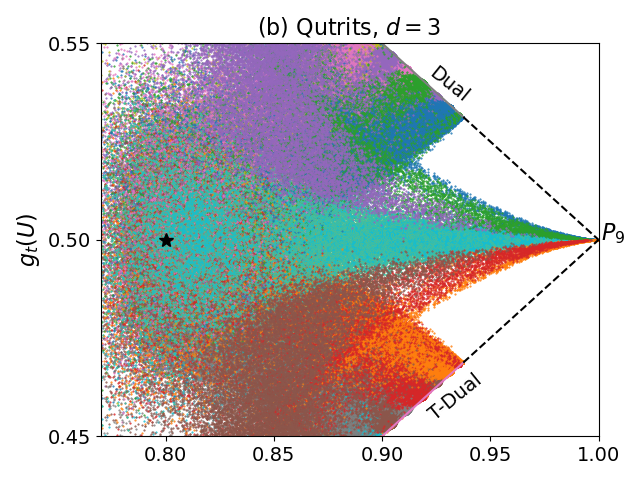}
 \includegraphics[width=.44\textwidth]{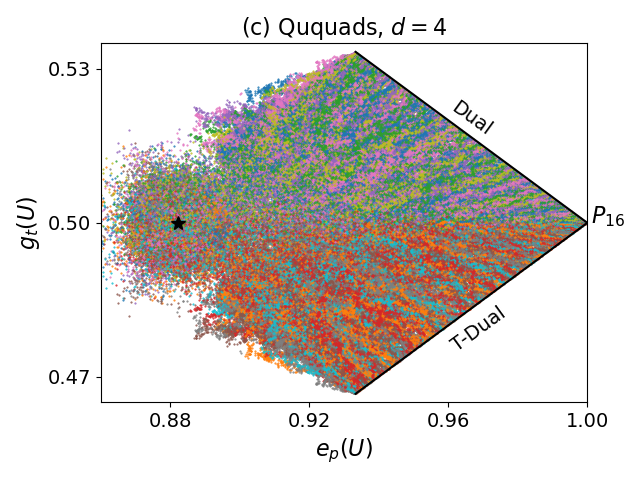}
 \caption{Entangling power $e_p(U)$ and gate-typicality
 $g_t(U)$ defined in (\ref{eq:EP}) and  (\ref{eq:GT})
 for two subsystems of local dimension:
 a) $d=2$, b) $d=3$ and c) $d=4$.
  Shown unitaries of size $d^2$ enjoy atypically large entangling power. Black stars mark the average values
  over the Haar measure:    $\bar{e_p}=(d^2-1)/(d^2+1)$ 
  and $\bar {g_t}=1/2$. Panel a) demonstrates that 
  2-unitary matrices of order $d^2=4$ do not exist
  as there are no matrices for which $e_p(U)=1$,
  while panel b) shows that for $d^2=9$ there are no dual unitaries (located along the upper dashed boundary line) in the neighborhood of the $2-$unitary permutation matrix $P_9$, in contrast to the case $P_{16}$ shown in panel c).
    %
 }
 \label{fig:P9epgt}
 \end{figure}

 The entangling power is the average entanglement created when $U$ acts on the ensemble of pure product states \cite{ZZF00}, wherein the state of each subsystem is drawn according
 to the Haar measure. It vanishes for the identity and the SWAP gate, and is maximized to $1$, iff $U$ is 2-unitary. The gate-typicality $g_t$, introduced in \cite{Bhargavi2017ImactGate},
 is a complementary quantity that vanishes for local operators and is maximized for the swap, as $g_t(S)=1$. 
 Under this normalization the mean value,
 equal to the average value with respect to the Haar measure,
$\langle g_t\rangle =1/2$,
  corresponds to  a typical random unitary  matrix
 hence  the state $\ket{\psi}$ in Eq.~(\ref{eq:psi_as_UAB}) is AME$(4,d)$, iff $e_p(U_{AB})=1$, which implies that $g_t(U_{AB})=1/2$.
Both quantities (\ref{eq:EP}) and  (\ref{eq:GT})
form the  plane $(e_p,g_t)$
useful to analyze the set of bi-partite unitary 
matrices of order $d^2$ --
see Fig. \ref{fig:P9epgt} obtained for
$d=2,3,4$.

Since every unitary matrix
$U$ of size $d^2$ with maximal entangling 
power $e_p=1$ is two-unitary and 
by Eq.~\eqref{eq:psi_as_UAB}
defines a state AME$(4,d)$,
\begin{equation}
\label{AME_U}
|\Psi\rangle = {\color{\redCom} \frac{1}{d}} \sum_{i,j=0}^{d-1}|ij\rangle \otimes U|ij\rangle~,
\end{equation}
a search for a new solution can be realized
by numerical maximization of $e_p(U)$.
Such a procedure turned out to be successful
and allowed one to find the first such state for $d=6$
\cite{Rather_2022}.

{\color{\redCom}
Schmidt decomposition
of state
$|\Psi\rangle$
for the splitting $AB|CD$ implies that
any AME$(4,d)$ state, by definition maximally entangled
with respect to this partition,
is locally equivalent to the 
form (\ref{AME_U}). Thus 
it can be associated to a quantum orthogonal Latin square
determined by $U$ of order $d^2$ and its vectors
forming states $|\Psi_{ij}\rangle$
from eq. (\ref{eq:Unitarity}).
The link between combinatorial designs and quantum states \cite{CGSS05}
can be extended for a larger number of subsystems.}

To look for a state with $N=6$ systems
one needs to work with unitary matrices
of order $d^3$ and optimize 
multipartite entangling power \cite{Scott2003QECCentPow,LGMZ20}
in search of $3$-unitary matrices, 
which remain unitary after any of 
10 possible reordering of matrix entries \cite{Goyeneche2015}.
A simple construction of $3$-uniform states
AME(6,$d$) is provided by combinatorial
designs called mutually orthogonal {\sl Latin cubes} of 
order $d$,
being a natural generalization of the notion
of Latin squares.
Such a cube representing AME(6,4)
is provided in App.~\ref{app:AMEtoolbox}.

The theory of combinatorial designs
covers also $k$-dimensional mutually 
orthogonal hypercubes of order $d$.
If such a configuration exists
it allows one to construct 
$k$-unitary matrix of size $d^k$
and the corresponding $k$-uniform
state AME($2k,d$)
of $N=2k$ parties \cite{Goyeneche2015,Goyeneche_2018}.

\section{Classical codes and minimal support AME states}\label{sec:MinSupMDS}
Let us present here the simplest construction of AME states introducing relevant notation: The {\em support} of a state is the number of nonzero coefficients it has in a fixed product basis. A $k$-uniform state has at least support $d^k$, which is its Schmidt rank for any bipartition between $k$ and $N-k$ parties. If there exists a product basis expanding the state with support $d^k$, then we say it is a state of {\em minimal support}. Such states,
generated by any $k$-unitary permutation matrix $P_{d^k}$, 
are special because they can be constructed
by known schemes of classical error correction. 
In the case of a $N=4$ party system
such a state is obtained by plugging a 
$2$-unitary permutation matrix  $P_{d^2}$ into 
Eq. (\ref{AME_U}).

{\color{\redCom}To explain this connection, we let us sketch the key idea behind classical error correction. In its simplest form, the goal is to protect a classical bit from the potential error of a bit flip occurring with low probability. The simplest nontrivial example is so-called repetition code, where the bit $0$ is encoded into three repeated bits $000$, and the bit $1$ is encoded into $111$. Therefore, if a single bit flip error occurs with low probability $p$ and changes symbol $0$ into $1$, one can correct it by performing majority voting rule applied to the ``spoiled '' bit string e.g. $001$, which is registered instead of the expected string $000$. When the single error probability is very small $p \ll 1$, the probability of two bit flips $p^2$
can be neglected.

More generally,} a classical error correcting code $C[N,k,\delta]_d=\{\omega_j\}_{j=1}^{d^k}$ is a set of $d^k$ codewords of $N$ digits of a $d$-dimensional alphabet each, where every two codewords differ in at least $\delta$ digits. Classical codes with distance $\delta$ are able to identify $\delta-1$ bit (or {\em dit}) flips, namely classical errors. {\color{\redCom} For example, the repetition code is denoted as  $C[3,1,3]_2=\{000,111\}$
has $d^k=2$ codewords of Hamming distance $\delta=3$ composed of $N=3$ digits in an alphabet of dimension $d=2$, and can detect the presence of $\delta-1=2$ bit flips, while correcting at most one.} A classical code $C$ is {\em maximum distance separable (MDS)} if it saturates the Singleton bound \cite{Jo58,Si64},  $\delta\leq N-k+1$. 

MDS codes $C[N,k,\delta]_d$ can be used to construct $k$-uniform states of $N$ $d$-dimensional parties,
\begin{equation}
   \ket{\psi_{N,k,d}}=\frac{1}{\sqrt{d^k}}\sum_{j=1}^{d^k}\ket{\omega_j} ,
\end{equation}
which have minimal support~\cite{Goyeneche2015,helwig2013existApplic,bernal2018existence}. In the example of the repetition code above, this gives rise to the 1-uniform GHZ state $\ket{\text{GHZ}}=\big (\ket{000}+\ket{111}\big )/\sqrt{2}$. 

As a further example, the construction of minimal-support $k$-uniform AME states of $N=2k$ subsystems
from MDS codes is equivalent to their construction from mutually orthogonal Latin squares, cubes or hypercubes, and the associated multiunitary matrix is then a permutation matrix. Indeed, one can view each label as a vector $j=(i_1,...,i_k)$ and each codeword as $\omega_j=(i_1,...,i_k,T^{(1)}_{i_1,...,i_k}, ..., T^{(s)}_{i_1,...,i_k})$. Then $\{\omega_j\}$ form an MDS code if and only if $T^{(1)},...,T^{(s)}$ are mutually orthogonal Latin squares (cubes or hypercubes) with coordinates $i_1,...,i_k$. In turn, this occurs if and only if 
$$
\ket{\psi_{2k,k,d}}=
\frac{1}{\sqrt{d^k}}
\sum_{i_1,...,i_k=0}^{d-1}\ket{i_1,...,i_k}\ket{\Psi_{i_1,...,i_k}}
$$
is $k$-uniform~\cite{Rains1999QShadowEnum,
Scott2003QECCentPow}. For more
information see later contributions
\cite{helwig2013existApplic,
Goyeneche2014,bernal2018existence}. 
and the table of codes by Grassl
\cite{Grassl:codetables}.


\subsection{AME(4,3)}

As there are no states AME$(4,2)$~\cite{Higuchi_2000_twoCouples}, 
 the smallest local dimension of interest is $d=3$, states of four qutrits.
Due to the existence of OLS(3) (see Fig.~\ref{tab:ame_4_4}),
the construction of AME$(4,3)$ has been known for a while \cite{CGSS05,Goyeneche2014}. We can write a minimal support AME$(4,3)$ as
\begin{equation}
\label{eq:p9_state}
\begin{split}
\ket{\Psi_{P_9}}=\frac{1}{3}(&\ket{00\mathbf{00}}+\ket{01\mathbf{11}} +\ket{02\mathbf{22}} +\\& \ket{10\mathbf{12}} +\ket{11\mathbf{20}}+\ket{12\mathbf{01}} +\\&\ket{20\mathbf{21}}+\ket{21\mathbf{02}}+\ket{22\mathbf{10}}). 
\end{split}
\end{equation}

The subscript $P_9$ indicates that this state is equivalent to the 2-unitary permutation of size 9 that takes  $\{00,01,02,10,11,12,20,21,22\}$ to $\{00,12,21,22,01,10,11,20,02\}$.  This permutation represents the OLS(3) shown in Fig.~\ref{tab:ame_4_4}, where we read the first list as the entry of the cell and the second as the address. Hence the structure of
the state  $\ket{\Psi_{P_9}}$ is encoded 
in the Graeco-Latin square:
first two digits label the row and the column, while the latter two represent the rank and the suit of the corresponding card,
so each term of the state
is of the form $\ket{\text{Address}|\text{Entry}}$. 
The state (\ref{eq:p9_state}), distinguished by the AME property of maximal correlations for all three symmetric partitions,
{\color{\redCom} 
is also conjectured \cite{SG24}
to provide maximal geometric entanglement,
defined by the geodesic distance to the set of
separable pure states.
However, this property is not 
universal: it does not hold for 
AME$(4,d)$ with $d=4$ and $6$,
as in these cases  states with larger geometric entanglement were found \cite{SG24}.
}

Alternatively, to find an AME(4,3) state, one can use another tool
of combinatorial designs, namely orthogonal arrays.
An {\em orthogonal array} of $r$ rows, $c$ columns, dimension $d$, and strength $k$ -- written as OA($r,c,d,k$) -- is defined as a $r\times c$ arrangement of $d$ different elements such that every $r\times k$ subarray contains each $k$-tuple from the set $\{1,\cdots,d\}$ the same number of times~\cite{Rao_1946,Hedayat_1999}. 
Relation $r=\lambda d^k$ defines the notion of 
the {\sl index} $\lambda$  of the array.
Their main applications are statistics and the design of experiments. Here we preserve the notation where indices range from $1$ to $d$, consistently with the mathematical literature~\cite{Rao_1946,Hedayat_1999,ReedSol1960Codes,Sloane2007LibArrays}; but the indices shall be shifted to range from $0$ to $d-1$ if constructing quantum states, in concordance with the standard notation of the computational basis.

A special subset of OA are formed by \emph{irredundant} orthogonal arrays (IrOA), i.e., those for which every subset of $c-k$ columns contains no repeated rows~\cite{Goyeneche2014}.
Irredundant orthogonal arrays are useful in defining $k$-uniform states, as each IrOA($r,c,d,k$) of index unity leads
to an $k$-uniform state of $c$ qu$d$its, which is composed of $r$ superposed states 
of computational basis~\cite{Goyeneche2014}. 
Arrays of index unity, $\lambda=1$ 
so that $r=d^k$,
are always irredundant
and they generate  AME states with minimal support. This construction is equivalent
to the approach based on orthogonal Latin squares 
and MDS classical codes. 

In particular, the following array
\begin{equation}
    \text{OA}(9,4,3,2) =
    \begin{array}{cccc}
1&1&1&1\\
1&2&3&2\\
2&1&3&3\\
2&2&2&1\\
2&3&1&2\\
3&2&1&3\\
3&3&3&1\\
3&1&2&2\\
1&3&2&3\\
    \end{array}
\end{equation}
is irredundant \cite{Goyeneche2015} 
and, 
after a shift of labels
$i\to i-1$, 
leads to a minimal support AME(4,3) state
equivalent to Eq.~\eqref{eq:p9_state}.
The transformation takes each row $i,j,m,n$ and translates it into the state $\ket{ijmn}$ (after shifting indices). 
Subsequently, all of the states are summed and normalized, yielding the desired state
which by construction enjoys the AME property.

To see that this is the case
note that tracing out any $c-k = 4-2$ columns leaves a sum of projectors onto the remaining $k = 2$ columns. 
Thus, the resulting state is proportional to identity, so the original state is AME. 
The above discussion can be summarized as follows: All states related to OLS and orthogonal Latin cubes can be defined by IrOA.

One of the peculiarities of general AME$(4,3)$ states and the corresponding 2-unitaries is that they are all locally equivalent to  $\ket{\Psi_{P_9}}$ and to $P_9$
respectively \cite{Rather_2023}.
The proof relies on non-existence of universal entanglers for $d=3$. Such unitary matrices
of order $d^2$ entangle every bipartite product state. It is known that for
$d=3$ there are no universal entanglers \cite{Chen2008}. Thus for any unitary $U \in \mathcal{U}(9)$, there always exists a product state that remains a product under its action, and this restriction is sufficient to show
that the 2-unitary matrices of order $d^2=9$
are unique up to local rotations.


Therefore, every AME$(4,3)$ state has the form $(u_1\otimes u_2\otimes u_3\otimes u_4)\ket{\Psi_9}$
and every 2-unitary on $\mathcal{H}^3 \otimes \mathcal{H}^3$ has the form $U_9=(v_1\otimes v_2)\, P_9\, (v_3\otimes v_4)$, where each $u_i, v_i \in \mathcal{U}(3)$. 
If $U_9$ represents also the set of 9 dimensional 2-unitaries, the other peculiarity of AME$(4,3)$ is that $U_9$ seems to be isolated in the set of dual or T-dual unitaries. This is visually seen as a gap in the $e_p(U)-g_t(U)$ plot as $U$ goes over all possible two qutrit gates. 
 Figure \ref{fig:P9epgt} shows this in the vicinity of the point ($e_p=1$, $g_t=1/2$), marked $P_9$, which corresponds to the set of 2-unitary gates. 
 
 The figure is constructed from many sets of atypical highly entangled two-qutrit gates. Some of these are perturbed from $U_9$ at the extreme right and some from a one-parameter family of dual unitaries $U_s(\theta)$ and $SU_s(\theta)$, their T-dual cousins. The dual unitary $U_s(2 \pi/3)$ is an orthogonal matrix with entangling power $15/16=0.9375$ which is likely to be the highest possible value for dual unitaries that are not 2-unitary. Explicit form of $SU_s(\theta)$ is available in \cite{Suhail_Thesis}. Apart from these perturbations, Haar random unitaries as well as atypical matrices close to $U_9$ are subjected to few steps of an algorithm that converges to dual unitaries \cite{Rather_2020}, so that this reveals better the region of the gap, the rigorous existence of which is yet to be established.

 It is useful to contrast this with qubits ($d=2$), and ququarts ($d=4$), see Fig.~\ref{fig:P9epgt}. For two-qubit systems, the dual and T-dual lines do not meet and there are no $2$-unitary matrices of order 4. The maximum entangling power possible in this case
reads $e_p=2/3$, as known for a long time \cite{ZZF00}. This value is attained for entire family of gates including CNOT and DCNOT (double-CNOT, locally equivalent to a gate called iSWAP).
 On the other hand, for two-ququart system, AME$(4,4)$ state not only exists, but unlike AME$(4,3)$ it has an uncountable infinity of LU inequivalent (and also SLOCC inequivalent) realizations. There exists a continuous parameterization of dual and T-dual unitaries that limit to 2-unitaries \cite{Suhail_Thesis}  and hence there appears to be no excluded region in the vicinity of the $2-$unitary gate
 achieving the maximal value,  $e_p(U)=1$, and simply marked as $P_{16}$ in Fig.~\ref{fig:P9epgt}.
 For more details on construction of non-standard AME(4,4) states, see Appendix.~\ref{app:AMEtoolbox}.

\section{Conventional solution by graph/stabilizer states}\label{sec:Graphs}
    Absolute maximal entanglement is a global property of the state, meaning it is not a derivative of correlations between any given pair of subsystems. 
    Hence it is not surprising that there are no general schemes for constructing AME states.
    However, some of the constructions that proved useful for bipartite systems lead to multipartite AME states as well. 

    In particular, creation of a maximally entangled state of two qubits (e.g., the Bell state) is possible by an action of the controlled-Z (CZ) gate on $\ket{++} = \frac{1}{2}(\ket{00} + \ket{01} + \ket{10} + \ket{11})$ state,
    \begin{equation}
        \text{CZ} \ket{++} = \frac{1}{2}(\ket{00} + \ket{01} + \ket{10} - \ket{11}).
    \end{equation}
    Note that the matrix of coefficients of the transformed state 
    is proportional to the unitary Hadamard matrix, $H_2=(++;+-)$, and therefore the state above is locally equivalent to the
    maximally entangled  Bell state of two qubits.
    
    We can generalize this observation to the multipartite setting by denoting each qubit as a vertex in a graph and a two-qubit CZ gate by an edge connecting two corresponding vertices: An $N$-qubit state $\ket{\psi}$ is called a {\em graph state}
    ~\cite{Hein_2006},
    if it can be created via an action of CZ gates between pairs of qubits given by edges $E$ in a graph with qubits denoted by $N$ vertices $V$, each initiated in $\ket{+}$ state
        \begin{equation}\label{eq:graph_states}
            \ket{\psi} = \prod_E \text{\text{CZ}}_{E} \ket{+}^{\otimes N}.
        \end{equation}
    Significant examples of states that are locally equivalent (see Sec.~\ref{sec:lu_equivalence}) to graph states include Bell states and their generalized multiqubit versions, namely GHZ states. Entanglement in multiqubit states corresponding to a  graph 
    \cite{VandenNest_2004}
    was analyzed in~\cite{Hein_2006}
    and more recently characterized in \cite{Sharma_2025}.
 
    Notice that the order in which controlled-Z gates are applied
    is irrelevant since these operations do commute. For higher dimensions, this notion can be generalized by defining equally superposed states, $\ket{+_d}=\sum_{i=0}^{d-1}\ket{i}/\sqrt{d}$ and 
    using the basis related to 
    the Weyl-Heisenberg group~\cite{Teleport1993Bennett},
    \begin{equation}\label{eq_WeylHeis}
       X^k\ket{i}=\ket{i\oplus k}\!,\,\,\text{and}\,\, Z^m\ket{i}=\omega_d^{m}\ket{i}
    \end{equation}
    with $\omega_d=e^{\frac{2\pi\text{i}}{d}}$, where $\oplus$ denotes summation modulo $d$. 
    Hence the operator basis
    consist of $d^2$ operators,
       $X^k Z^{m}$ with $k,m=0,\dots,d-1$.
    Then qudit graph states are constructed analogously~\cite{gottesman1997stabilizer,Ashiknmin2001nonbinCodes}.
    \begin{figure}
     \centering
     \begin{subfigure}[b]{0.25\linewidth}
         \centering
         \includegraphics[width = \textwidth]{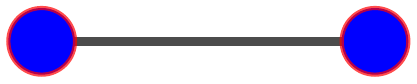}
         \caption{}
     \end{subfigure}
     \hfill
     \begin{subfigure}[b]{0.35\linewidth}
         \centering
         \includegraphics[width =\textwidth]{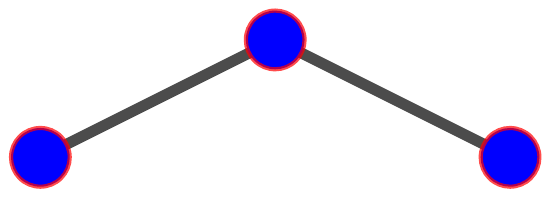}
         \caption{}
     \end{subfigure}
     \hfill
     \begin{subfigure}[b]{0.3\linewidth}
         \centering
         \includegraphics[width =\textwidth]{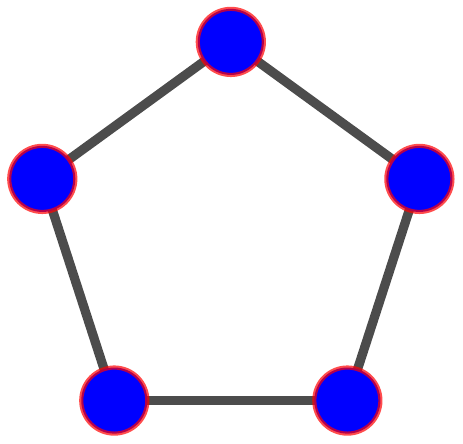}
         \caption{}
     \end{subfigure}
        \caption{Graphs corresponding to AME states
          of a) two, b) three and c) five qubits.
          Interestingly, a square graph with $4$ vertices
          does not represent AME(4,2) state, as it does not exist \cite{Higuchi_2000_twoCouples}. Further examples
           of AME graphs are provided in Appendix
           \ref{app:AMEtoolbox}.
           }
        \label{fig:graph_states_AME}
    \end{figure}

    Another approach is based on the notion of stabilizer states.
    For a given set of stabilizer operators $S_i$, we define the 
    {\em stabilizer states}~\cite{gottesman1997stabilizer,NielsenAndChuang,CSS1996Steane,CSS1996Goodexist}
     as those $\ket{\psi}$ that belong to the trivial eigenspace of each of the operators,
    $S_i\ket{\psi} = \ket{\psi}$ for all $i$.
    
    In this way one can provide an alternative definition of graph states \cite{Hein_2006}.
    For each vertex $i$ of the graph we construct an operator
    \begin{equation}
        S_i = \sigma_x^{(i)} \prod_{N(i)} \sigma_z^{(N)},
    \end{equation}
    where the product is taken over all the neighbors $N(i)$ of the vertex $i$.
    The graph state $\ket{\psi}$ is then defined as a state which is invariant under action of all operators $S_i$
    \begin{equation}
        S_i \ket{\psi} = \ket{\psi}.
    \end{equation}
    All graph states are stabilizer states  
    and all stabilizer states can be written as a graph state for some local unitaries~\cite{gottesman1997stabilizer}. Depending on the context, it might be more useful to apply the stabilizer definition of graph states, especially in the domain of quantum error correction.
    All stabilizer states and codes, and in particular graph states, are described by their parity-check (or check) matrix.

    For a wide range of qudit systems,  
    simple constructions of
    absolutely maximally entangled states are given by graph states~\cite{Helwig_2013}, as presented in Fig.~\ref{fig:graph_states_AME}.
    In particular, for graph states, the condition of maximal entanglement is in some cases easier to verify~\cite{Helwig_2013,Sharma_2025,Vande2025GraphFromMarg,Alsina_2021}.
    What is more, given a graph representation of a quantum state, it is straightforward to find a corresponding circuit, 
       determined by Eq.~\eqref{eq:graph_states}.
    Notably, graph AME states exist for any number $N$ of parties involved with certain local dimensions. Explicit connections between graph states, minimal support AME states, and classical codes are provided in Appendix~\ref{app:graph_states}.
    More examples of graphs leading to AME states are provided
    in Appendix~\ref{app:AMEtoolbox}. 
    
    We highlight that it has been shown that for some numbers of systems and local dimensions, AME states cannot be stabilizers~\cite{cha2026NoGoStab46,wojcik2026NoGoStabEvenDim}.

\section{Unconventional constructions: non-stabilizer states}\label{sec:unconventional}

Since MDS codes and stabilizer states do not suffice for the case of four quhexes, namely $N=4$ and $d=6$~\cite{cha2026NoGoStab46}, finding an AME state for this system size requires strategies beyond the techniques presented above.  In this section we will present 
non-stabilizer constructions of highly entangled states,
leading to examples of AME($4,6$) state.

\subsection{The Golden AME(4,6) state}
Negative solution of the Euler problem of 36 officers 
implies that there are no OLS$(6)$,
so the standard technique to create AME(4,6) does not apply.
Therefore any QOLS(6) that exists must be genuinely  quantum. 
    The search concerned the states of the form 
    \begin{equation}
        \ket{\psi_{ABCD}}=\frac{1}{6}\sum_{i,j=0}^5 \ket{ij}\otimes U \ket{ij},
    \end{equation}
    where $U \in \mathbb{U}(36)$ is 2-unitary, that is 
    $U$, $U^R$ and $U^{\Gamma}$ are all unitary. The lack of OLS(6) implies that there is no
    2-unitary permutation matrix.
    The hunt for the QOLS(6) is thus performed in an larger class of mathematical objects, wherein
    the discrete group of permutations
    is extended to continuous unitary group.
    If one restricts to permutations \cite{CGSS05},
the one that comes closest to being 2-unitary in terms of entangling power, is one for which $e_p(P_{36})=314/315\approx 0.996825$.
Thus the entire task can be seen as a search for an additional
increase of its entangling power by a missing fraction $1/315$.

    Using optimal permutation matrix $P_{36}$ and subspace rotation embellished versions of it, as seeds, modified Sinkhorn type algorithms reaches the final, genuinely quantum solution.
Non-permutation unitaries with higher entangling 
power were obtained using the polar decomposition, which finds the nearest unitary to any matrix.
This found a series of such `super-permutation unitaries' that were all orthogonal. One of them $A$ had $e_p(A)=419/420\approx 0.9976$, while the best that could be found $W$ had $e_p(W)\approx 0.99872$.  
    
    Starting from seed permutations (slighly perturbed to avoid 
    singulartities after reshuffling or partial transpose)
    that had compromised entangling power and moved gate-typicality closer to $0.5$,
    gave the first indication that 2-unitaries of size 36 do exist and hence there is an AME$(4,6)$ state.
    However, matrix $W$ of size $36$ remains the orthogonal matrix that has the highest entangling power known to us, all the others including the 2-unitary turn out to be complex.
    
    After a suitable permutation the 2-unitary matrix $U_{36}$ 
    obtained numerically was transformed into
    a block diagonal form with nine blocks of order four. 
    Assuming observed structure of their entries unitarity 
    conditions imposed for each block resulted in an analytical form for these elusive objects~\cite{Rather_2023,
    Burchardt_2022,Bruzda_2022,Rajchel-Mieldzioc_2022}.
     The matrix elements featured the golden mean $\varphi$ as a ratio
    of two moduli of its entries, while all complex phases 
    were found to be multiples of
    $20^{\text{th}}$ roots unity, $\omega=\omega_{20}=e^{i\pi/10}$. 
    
    Thus the orthonormal basis $\{\ket{\psi_{ij}}=U \ket{ij}, \; 0\leq i,j \leq 5\}$  of size $36$ forms a QOLS(6). Detailed forms of $U$ and also alternate representations are discussed provided in \cite{Rather_2022, Zyczkowski_2023}. An explicit form in the computational basis is provided in~\cite{Ex2021AME46}. We present here a somewhat different perspective suggesting that QOLS can be interpreted as a superposition of  classical combinatorial designs.
    
    It turns out that each bi-partite state $\ket{\psi_{ij}}$, 
    forming an entry of  the QOLS(6)
    can be written as a superposition of 4 states,
    \begin{equation}
    \begin{split}
\label{eq:AME46psiij}
\ket{\psi_{ij}}=(&\alpha_{ij}\ket{F_{ij}H_{ij}}+\beta_{ij}\ket{F_{ij}\tilde{H}_{ij}}+ \\ &\gamma_{ij}\ket{\tilde{F}_{ij}H_{ij}}+\delta_{ij}\ket{\tilde{F}_{ij}\tilde{H}_{ij}})/\sqrt{2}.
\end{split}
    \end{equation}
    Here $F,\tilde{F},H,\tilde{H}$, given below, are $6 \times 6$ {\em frequency} squares of 3 symbols. Frequency squares are such that each entry repeats a fixed number of times in each row and column~\cite{Hedayat_1999} and in these cases all entries repeat twice.
    A Latin square is a special case with frequency 1. The following examples are given in standard mathematical notation, where indices range between $1$ and $d=6$:
\begin{equation}
    F=\begin{bmatrix}
{\color{red}1} &{\color{red}1}&{\color{blue}5}&{\color{blue}5}&{\color{olive}3}&{\color{olive}3} \\
 {\color{red}5}&{\color{red}5}&{\color{blue}3}&{\color{blue}3}&{\color{olive}1} & {\color{olive} 1}\\
 {\color{olive}3}&{\color{olive}3} & {\color{red}1} &{\color{red}1}&{\color{blue}5}&{\color{blue}5}\\ 
{\color{olive}1} & {\color{olive} 1}&{\color{red}5}&{\color{red}5}&{\color{blue}3}&{\color{blue}3}\\
{\color{blue}5}&{\color{blue}5} &{\color{olive}3}&{\color{olive}3}&
{\color{red}1} &{\color{red}1} \\
{\color{blue}3}&{\color{blue}3}&{\color{olive}1} & {\color{olive} 1}&{\color{red}5}&{\color{red}5}
    \end{bmatrix},\,
    H=\begin{bmatrix}
        {\color{olive} 1}&{\color{olive} 3}&{\color{red}5}&{\color{red}1}&{\color{blue}3}&{\color{blue}5}\\
        {\color{olive} 1}&{\color{olive} 3}&{\color{red}5}&{\color{red}1}&{\color{blue}3}&{\color{blue}5}\\
        {\color{red}5}&{\color{red}1}&{\color{blue}3}&{\color{blue}5}&{\color{olive} 1}&{\color{olive} 3}\\
        {\color{red}5}&{\color{red}1}&{\color{blue}3}&{\color{blue}5}&{\color{olive} 1}&{\color{olive} 3}\\
        {\color{blue}3}&{\color{blue}5}&{\color{olive} 1}&{\color{olive} 3}&{\color{red}5}&{\color{red}1}\\
        {\color{blue}3}&{\color{blue}5}&{\color{olive} 1}&{\color{olive} 3}&{\color{red}5}&{\color{red}1}
    \end{bmatrix}
    \label{eq:FH}
\end{equation}
\begin{equation}
    \tilde{F}=\begin{bmatrix}
{\color{red}2}&{\color{red}2}&{\color{blue}6}&{\color{blue}6}&{\color{olive}4}&{\color{olive}4}\\
{\color{red}6}&{\color{red}6}&{\color{blue}4}&{\color{blue}4}&{\color{olive}2}&{\color{olive}2}\\
{\color{olive}4}&{\color{olive}4}&{\color{red}2}&{\color{red}2}&{\color{blue}6}&{\color{blue}6}\\
{\color{olive}2}&{\color{olive}2}&{\color{red}6}&{\color{red}6}&{\color{blue}4}&{\color{blue}4}\\
{\color{blue}6}&{\color{blue}6}&{\color{olive}4}&{\color{olive}4}&{\color{red}2}&{\color{red}2}\\
{\color{blue}4}&{\color{blue}4}&{\color{olive}2}&{\color{olive}2}&{\color{red}6}&{\color{red}6}
    \end{bmatrix}, \, \tilde{H}=\begin{bmatrix}
        {\color{olive}2}&{\color{olive}4}&{\color{red}6}&{\color{red}2}&{\color{blue}4}&{\color{blue}6}\\
        {\color{olive}2}&{\color{olive}4}&{\color{red}6}&{\color{red}2}&{\color{blue}4}&{\color{blue}6}\\
        {\color{red}6}&{\color{red}2}&{\color{blue}4}&{\color{blue}6}&{\color{olive}2}&{\color{olive}4}\\
       {\color{red}6}&{\color{red}2}&{\color{blue}4}&{\color{blue}6}&{\color{olive}2}&{\color{olive}4}\\
        {\color{blue}4}&{\color{blue}6}&{\color{olive}2}&{\color{olive}4}&{\color{red}6}&{\color{red}2}\\
        {\color{blue}4}&{\color{blue}6}&{\color{olive}2}&{\color{olive}4}&{\color{red}6}&{\color{red}2}
    \end{bmatrix}
    \label{eq:KpLp}
\end{equation}
    The array $FH$ superposing $F$ and $H$ contains 9 different symbols repeating $4$ times, and are examples of {\em orthogonal frequency squares} (OFS)~\cite{Hedayat_1999}.
    As marked with colors in matrices 
    all four patterns provide Latin squares of order three, if  symbols represent $2 \times 2$ blocks.
    The other combinations $F\tilde{H}$, $\tilde{F}H$ and $\tilde{F}\tilde{H}$ also have an identical structure, however the symbols are different in each and together they add up to the 36 symbols needed for states in $\mathcal{H}^6\otimes \mathcal{H}^6$. These four OLS further have the property that the repeating elements are at exactly the same positions in each of them. For example the elements at $\{(1,1), (4,2), (5,6), (6,3)\}$
    are the same in each of them, in $FH$ it is $11$, in $F\tilde{H}$ it is $12$, in $\tilde{F}H$ it is $21$, and in $\tilde{F}\tilde{H}$ it is $22$. See Appendix~\ref{app:ame46} for explicit forms of the 4 OFS and these elements are encircled.
    Thus the golden state QOLS(6) can be considered
    as a superposition of these four OLS.

\begin{figure}
    \centering
    \includegraphics[width=0.9\linewidth]{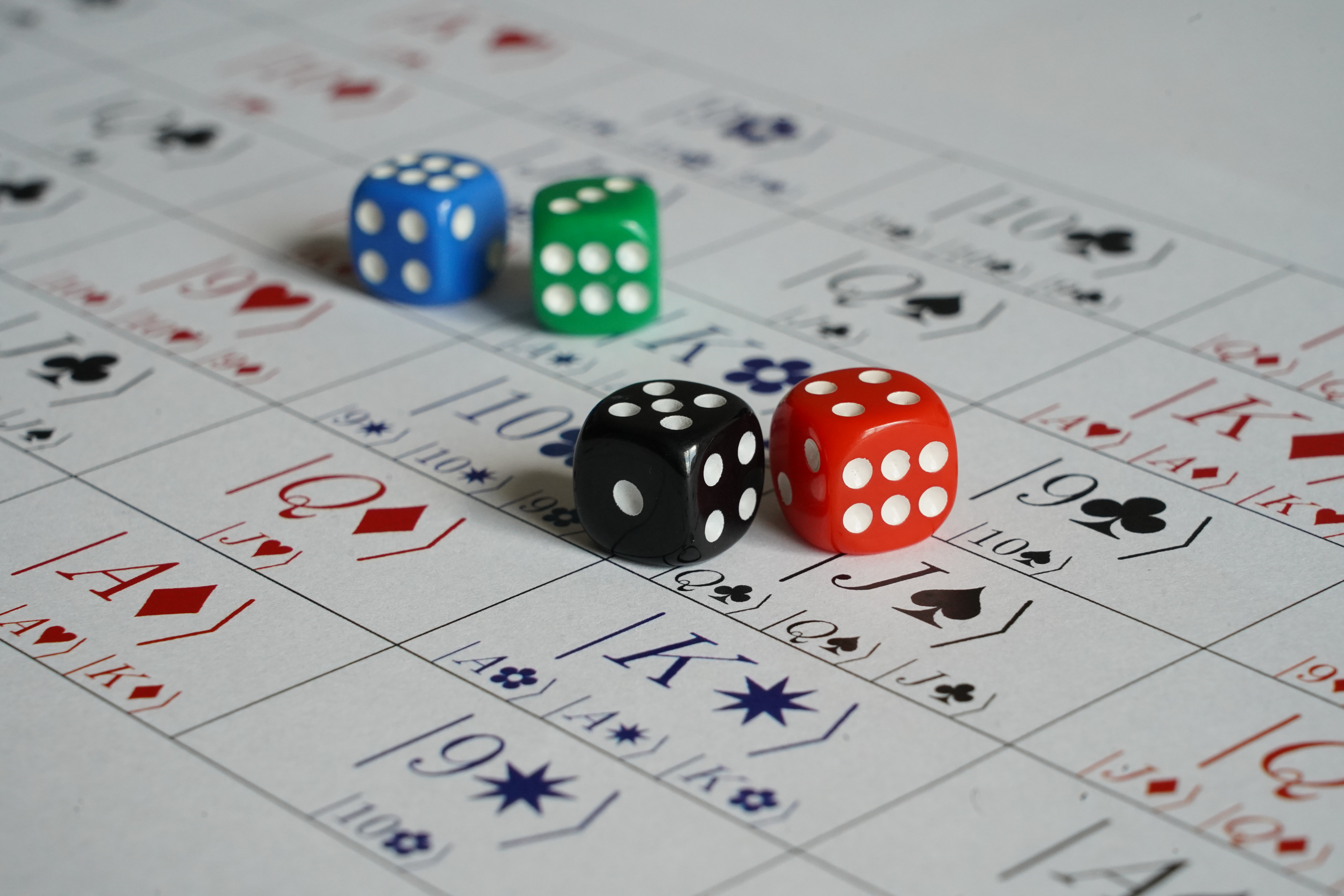}
    \caption{An artistic visualization of the golden state QOLS(6), using cards of 6 different ranks and suits. The outcome of any two dice prepared in such a state determines the outcome of the remaining
     two dice.
    Note that a classical solution to Euler's problem would correspond to the array with only one card in each entry.
    For a full figure created
     by Paulina Rajchel-Mieldzio\'c
    see the original paper~\cite{Rather_2022}.
      }
    \label{fig4}
\end{figure}

    The crucial inputs now are the coefficients in Eq.~(\ref{eq:AME46psiij}), which {\color{\redCom} can be combined (without the overall factor $1/\sqrt{2}$)  into matrices}:
    \begin{equation}
        M_{ij}=\begin{pmatrix}
            \alpha_{ij}& \beta_{ij}\\ \gamma_{ij}&\delta_{ij}
        \end{pmatrix}.
    \label{eq:Mij}
    \end{equation}
Remarkably enough, for the golden AME$(4,6)$ state~\cite{Rather_2022}, it turns out that all 36 matrices $M_{ij}$ are unitary matrices themselves. Hence each $\ket{\psi_{ij}}$ forms a two-qubit Bell state carrying 1 ebit of entanglement. The first four blocks $M_{ij}$ read
\begin{align}
    &M_{11}=\begin{pmatrix}
        0 &1\\1&0
    \end{pmatrix}, \;\,\quad\quad\quad M_{12}=\begin{pmatrix}
        -\omega^7 &0 \\0 & -\omega^9
    \end{pmatrix},\\
    &M_{13}=\begin{pmatrix}
        -a \omega^8 & b\omega^3 \\ b\omega^3 & -a \omega^8
    \end{pmatrix}, \; M_{14}=\begin{pmatrix}
        a \omega^3 & b \\ b & a \omega^7
    \end{pmatrix}.\nonumber
\end{align}
Here $b=\sqrt{\varphi}/5^{1/4}$,  $a=b/\varphi$, and $\varphi=(\sqrt{5}+1)/2$ is the golden mean. Thus, we can construct from the frequency squares and the $M_{ij}$ above, the following bipartite states,
which form QOLS(6),
\begin{align}
    \ket{\psi_{00}}&=\frac{1}{\sqrt{2}}\left( \ket{01}+\ket{10}\right)\label{eq:psi11}\\
    \ket{\psi_{01}}&=\frac{-1}{\sqrt{2}}\left( \omega^7\ket{02}+\omega^9\ket{13}\right)\nonumber\\
    \ket{\psi_{01}}&=\frac{1}{\sqrt{2}}\left( b \omega^3(\ket{45}+ \ket{54}) -a \omega^8(\ket{44}+\ket{55})\right)\nonumber\\
    \ket{\psi_{03}}&=\frac{1}{\sqrt{2}}\left( a \omega^3\ket{40}+b (\ket{41}+\ket{50})+a \omega^7\ket{51}\right).\nonumber
\end{align}
These can be also read off from the $2$-unitary matrix of size $36$ presented in~\cite{Rather_2022} (caution: $a$ and $b$ in that source  differs by a factor of $\sqrt{2}$ from the present usage). 

The fact that the repeating elements in the 4 OFS are at the same positions, (see Appendix~\ref{app:ame46}) implies that the condition of $\{\ket{\psi_{ij}}\}$ being orthonormal, translates to the corresponding $M_{ij}$.
Thus there are 9 sets of maximal (4) number of orthonormal matrices, for the example given above $\{M_{11}, M_{42}, M_{56},M_{63}\}$ form one set. The QOLS(6) found is then equivalent to a block unitary matrix with nine $4\times 4$ blocks.  The other two conditions for a QOLS to be satisfied, implies further constraints on the blocks $M_{ij}$.
The complete set of $M_{ij}$ is collected in the Appendix. This could be useful to elucidate the structure and derive general conditions for the OFS and the $M_{ij}$ to construct an QOLS$(d)$. Also see \cite{Zyczkowski_2023} for a listing of the QOLS(6) corresponding to the partial transpose $U^{\Gamma}$, and explicit forms of permutations that will take one from the block-diagonal form to the 2-unitary. In the approach presented here, these permutations are encoded in the 4 OFS. 

The same source is also a good reference for chess pieces subjected to QOLS(6) conditions. 
While the principles of quantum chess 
allow a given piece to be in a superposition state supported in two squares of the board, 
the additional rule ``No Double Occupancy'' 
does not allow for interference between 
different pieces~\cite{Ca19}.
However, this is not the case in
all solutions of the quantum version
of the Euler problem,
in which a single square  of the 
$6 \times 6$ chessboard 
is occupied by a superposition
of several different chess pieces -- see Fig.\ref{fig4}.
 Finally, the search for the golden AME(4,6) state uncovered much more than available in the main paper~\cite{Rather_2022}, see a
 superposition of three
 Ph.D.\ Theses~\cite{Burchardt_2022,Rajchel-Mieldzioc_2022,Bruzda_2022}.

\subsection{AME(4,6) from biunimodular vectors and complex Hadamard matrices}
An alternative 
simpler constructions of  AME$(4,6)$ states was achieved recently \cite{Rather_2024} by searching for 2-unitaries of the form 
\begin{equation}
    U= \sum_{a,b=0}^{d-1} \lambda_{a,b} \ket{\Phi_{ab}}\bra{\Phi_{ab}},
    \label{eq:U_Diag_Max_Ent}
\end{equation}
where the $\lambda_{a,b}$s are phases; $\lambda_{a,b} \in \mathbb{U}(1)$,   $\{ \ket{\Phi_{ab}}:= \ket{Z^{a}X^{-b}},\, 0 \leq a,b \leq d-1 \}$ is a maximally entangled basis obtained from vectorizing Weyl-Heisenberg operator basis, and $X$ and $Z$ are as defined in Eq.~\eqref{eq_WeylHeis}.


By construction $U$ is unitary, 
and it is also $2-$unitary if it satisfies 
the following additional requirements \cite{Rather_2024}:
  \begin{equation}
  \begin{split}
  & \sum_{a,b=0}^{d-1} \lambda_{a,b} \lambda_{a+k,b+l}^*=0,\\
       & \sum_{a,b=0}^{d-1} \omega^{al-bk} \lambda_{a,b} \lambda_{a+k,b+l}^*=0, \; \forall (k,l) \neq (0,0).
        \end{split}
        \label{eq:lam_T_dual}
    \end{equation}
The first part assures that $U$ is dual-unitary \cite{Yu2024hierarchical} and the second that it is T-dual. If such a sequence of phases exist, they were dubbed ``perfectly perfect'' in \cite{Rather_2024} and have been called ``doubly perfect'' in \cite{Gross_2025}. Note that sequences that satisfy the dual-unitary condition alone (having zero autocorrelation) have been called perfect in the literature. They were found based on the literature concerning biunimodular vectors:
unimodular (phase) vectors that remain unimodular under
the action of $F_6 \otimes F_6$, where $F_6$ is the 6 dimensional discrete Fourier transform. 

It is by no means obvious that for $d=6$
such doubly perfect sequences $\lambda_{a,b}$ exist.
Taking recourse to numerical algorithms 
several solutions were obtained, and some of them 
displayed remarkably simple form \cite{Rather_2024}. For instance, 
one solution consisted of cubic roots of unity: $\lambda_{a,b}=\exp(2\pi i \phi_{a,b}/3)$ and the set of phases $\phi$ as a vector is
\begin{equation}
\label{phases}
\begin{split}
\{ &0, 2, 2, 0, 0, 1, 0, 1, 1, 1, 2, 1, 0, 2, 0, 2, 2, 2, \\&2, 0, 2, 2, 
    2, 1, 1, 1, 2, 0, 2, 2, 0, 1, 2, 2, 1, 0 \}.
\end{split}
\end{equation}
Thus, among all solutions known up to date,
 the one provided \cite{Rather_2024}
 involves the smallest, third, root of unity. 
    
 In a more recent work  
 on AME$(4,6)$ states an `artisanal' construction, 
 not involving any numerical search was 
   provided  \cite{Gross_2025}. 
 Building on the earlier work, doubly perfect sequences have been constructed using  sophisticated algebraic and number theoretical tools. 
 While the golden state QOLS(6) has a $36=9 \times 4$ structure as elucidated above, the most recent
 approach  exploits the structure  $36=3^3+3^2$, 
 so the space of 2 quhexes is decomposed into a direct sum of three qutrits and another two qutrits. 
 A $2$-unitary matrix $V_{36}$ is then constructed by acting with Clifford unitaries separately on the two sectors.
 This implies that the resulting AME state 
 can be considered as a superposition of two
stabilizer states.
Hence, this latest construction   \cite{Gross_2025}
stands out as a first fully analytical solution to the AME$(4,6)$ problem.
This solutions posses an elegant property, all eigenvalues of corresponding two-unitary matrix are multiples of $\omega_3 = e^{2 \pi i/3}$, thus the order of this matrix is equal to three as $V_{36}^3 = \mathbb{I}$. 
 
 
    Furthermore, a complex Hadamard construction of a $2$-unitary matrix $H_{36}$ with all
    entries of the same modulus and phases being multiples of sixth root of identity, 
    $\omega_6=e^{2\pi i/6}$,
    was found  \cite{Bruzda_2024Two-Had}. 
    This solution, based on numerical search,
    can be generalized to a 19-parameter family of AME(4,6) states. The connections to 
    doubly perfect sequences have been noted both in that work and subsequently in \cite{Gross_2025}. 

\subsection{Beyond AME(4,6)}

    While AME$(4,6)$ was the smallest number of particles that posed a serious  challenge, non-standard constructions, apart from graph states or OLS, for other cases is also of natural interest. As AME$(4,3)$ has a unique solution, up to local unitaries, AME$(4,4)$ is of interest. By an exhaustive numerical search it has been shown that all AME$(4,4)$ obtained from the 6912 OLS(4), are in fact LU equivalent \cite{Rather_2022_v2}. Enphasing 2-unitaries obtained from OLS(4) was shown to be sufficient to lead to LU inequivalent 2-unitaries and hence AME states \cite{Rather_2023}. However, numerical algorithms have also found other 
    structures that are not LU equivalent to any permutation. One such orthogonal matrix with entries that are simply $0$ or $\pm 1/2$ is shown in the Appendix~\ref{app:AMEtoolbox}.
    
    The search of non-standard constructions of AME states led the authors of~\cite{bistron_2023} to quantum convolutional channels, resembling the half-quantum, half-classical orthogonal Latin squares. Within this construction, by means of a modified Sinkhorn-type algorithm, the authors found continuous classes AME(4,7) and AME(4,9) states with 2 and 4 free nonlocal parameters. Moreover, in the latter case, the obtained family happened to connect previously known minimal-support AME states.

    {\color{\redCom} Finally, there have also been approaches to construct new AME, or $k$-uniform, states as combinations of AME states with a fewer number of subsystems \cite{Pozsgay2024, Raissi2020_2}. Although states obtained in these constructions were minimal support states or stabiliser states, respectively, they can be non-equivalent to standard constructions.}

\section{Entanglement in subsystems of AME states}
\label{Ent_AME}

The defining property of AME states maximizes entanglement between any number of particles and its complementary set, and hence from this point of view all AME$(N,d)$ states are alike. However, there is a less explored aspect of how much entanglement is present among subsystems that are not complementary, as the resulting states are mixed. For simplicity we consider the case when $N$, the number of particles, is even. Given any labeling of the $N$ particles, a general AME$(N,d)$ state can be written as
\begin{equation}\label{eq:AMEbip}
    \ket{\psi}=\frac{1}{d^{N/4}}\sum_{i_1 \cdots i_{N/2}=0}^{d-1} \ket{i_1 \cdots i_{N/2}}\ket{\phi_{i_1 \cdots i_{N/2}}},
\end{equation}
where $\{ \ket{\phi_{i_1 \cdots i_{N/2}}} \}$ forms an orthonormal set of $d^{N/2}$ states of $N/2$ particles whose labels are $N/2+1, \cdots N$. 

We wish to find the entanglement of any $n$ particles with another $m$ particles, when $m\neq N-n$.  If $n+m\leq N/2$, there is no entanglement as the reduced density matrix of the $n+m$ particles, $\rho_{n+m}$, is proportional to Identity and hence trivially separable. Let $l=N-n-m$. The nontrivial case is when $n+m>N/2$, when $\rho_{n+m}$ has $d^{l}$ non-zero eigenvalues (all equal to $1/d^l$), and the rest $d^{N-l} -d^{l}$, are $0$.  The rank-deficiency of the reduced density matrix allows for it to be negative under partial transpose (NPT) and hence for entanglement to exist.

Let the $l$ particles to be traced out have labels $1, \cdots, l$, then we have 
\begin{equation}
    \rho_{{\color{\redCom} n}+m}=\frac{1}{d^{l}} \sum_{i_1 \cdots i_l=0}^{d-1} \hat{P}_{i_1 \cdots i_l},
\end{equation}
where 
\begin{equation}
    \hat{P}_{i_1 \cdots i_l}= \ket{\Phi_{i_1 \cdots i_l}}
    \bra{\Phi_{i_1 \cdots i_l}},
\end{equation}
are orthogonal rank-1 projectors, and 
\begin{equation}
    \ket{\Phi_{i_1 \cdots i_l}}=\sum_{i_{l+1} \cdots i_{\frac{N}{2}}=0}^{d-1} \frac{\ket{i_{l+1} \cdots i_{\frac{N}{2}}}\ket{\phi_{i_1 \cdots i_{\frac{N}{2}}}}}{d^{\frac{N/2-l}{2}}}
\label{eqA:Phi}
\end{equation}
is a state of $N-l=n+m$ subsystems.

Without loss of generality we take $n \leq m$, and examine $\hat{P}^{T_n}_{i_1 \cdots i_l}$, the projector's partial transpose with respect to $n$ qudits. Consider two cases separately.

\subsubsection*{\texorpdfstring{{\rm Case (a):} $n+m>N/2$, 
{\rm and} $m \geq N/2$}{}}
In this case we can choose the labels $l+1, \cdots l+n$ for the $n$ particles. As $l+n=N-m \leq N/2$, the $n$ particles are part of the labels in the first ket of Eq.~(\ref{eqA:Phi}). 


Observe that from the orthonormality of the $\ket{\phi}$, it follows that  $\hat{P}^{T_n}_{i_1 \cdots i_l} \hat{P}^{T_n}_{j_1 \cdots j_l}=0$, unless $i_1 =j_1, \cdots,  i_l= j_l$. In other words,
the projectors span orthogonal subspaces also after partial transposition, hence we get
\begin{align}
    \|\rho_{n+m}^{T_n}\|_1&=\|\hat{P}^{T_n}_{i_1 \cdots i_l}\|_1 \quad \text{and}\nonumber \\ \mathcal{N}(\rho_{n+m})&=\mathcal{N}(\hat{P}_{i_1 \cdots i_l})
\end{align}
for any index set $i_1, \cdots,i_l$. As $\ket{\Phi_{i_1\cdots i_l}}$ in Eq.~(\ref{eqA:Phi}) is
already in the Schmidt decomposed form of a maximally entangled state of $n$ particles with $m$, it follows that 
\begin{equation}
    \mathcal{N}(\rho_{n+m})=\frac{d^n-1}{2}.
    \label{eq:negativity}
\end{equation}
This implies that if the subsystems are sufficiently large, more precisely one contains at least half of the particles,
they are maximally entangled. The negativity is the maximal possible value, as if the $n+m$ particle state is pure, and is simply determined by the smaller number of particles (here $n$). 

Thus in all AME$(4,d)$ states, the $1:2$ split is the only nontrivial one and in this case the negativity is $(d-1)/2$.
This is the same value as the $1:3$ split, 
while for the $2:2$ split the negativity is $(d^2-1)/2$. In particular these cases do not distinguish one AME$(4,d)$ state from another.
\begin{figure}
    \centering
    \includegraphics[width=0.35\linewidth]{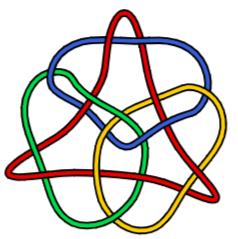}
     \includegraphics[width=0.35\linewidth]{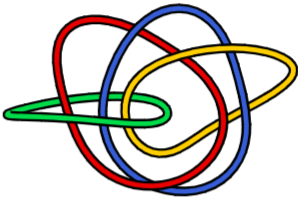}
    \caption{Link structure illustrating 1-resistant 4-party states
    (left) and  2-resistant state (right) borrowed from 
     \cite{Burchardt_2022}.  Removing any single ring
     from the left configuration, renders the others in a Borromean configuration \cite{Karol-GeoQstates2006}. This is analogous to AME$(4,d)$ states being 1-resistant: after removing (tracing away) any single subsystem, the remaining three are still entangled. However, if 
     any two subsystems are traced out, the other two become separable. In contrast, a typical 4-party state is 2-resistant, as performing partial trace over any two particles produces an entangled state of the remaining two, analogous to the link structure on the right.}
    \label{fig:Borrom}
\end{figure}
The notion of ``$\ell$-resistance'' was introduced in \cite{Quinta2019}: An entangled state of $N$ parties is called $\ell-$resistant if:
\begin{itemize}
\item  It remains entangled as any $\ell$ of its $N$ subsystems
are traced away;
\item  It becomes separable if a partial trace is performed
over an arbitrary set of $\ell + 1$ subsystems.
\end{itemize}
In particular, $\ell$-resistant states were constructed based on a mapping to the link structure of $N$ rings. It has been appreciated that the three loop Borromean knot, and their Brunnian generalizations with $n$ loops such that cutting any one will unlink all,
are analogs of the 3-qubit GHZ and generalized GHZ states of $N$ qubits. If any one qubit is erased, or traced out, the others become separable and thus they are $0-$resistant in the terminology of~\cite{Quinta2019}. 

It follows from Eq.~(\ref{eq:negativity}) that all  AME$(4,d)$ states ($d\ge 3$)
are $1$-resistant. The reduced density matrices of any 3 particles in such states have a Borromean configuration, they are entangled maximally as shown above, but any further particle loss or tracing out leaves separable states of two particles. This is illustrated by the link diagram in the left part of Fig.~\ref{fig:Borrom}. Further, the above analysis of $m+n$ particle entanglement for $m\geq N/2$, implies that AME$(N,d)$ is $\left( N /2-1\right)$-resistant for even values of $N$. 

For $N=4$, typical (Haar random) states are 2-resistant, as follows from results in \cite{Aubrun2012,Bhosale2012}. The fact that AME$(4,d)$ are 1-resistant implies that the entanglement is more multipartite in nature than in typical states, or reflects
a decrease in its monogamy. Reduced density matrix obtained
by tracing out a single subsystem in a typical 4-party state does not have the Borromean configuration and hence these states have a different link structure shown in the right part of Fig.~\ref{fig:Borrom}. 
\subsubsection*{\texorpdfstring{{\rm Case(b):} $n+m>N/2$, 
{\rm and} $m < N/2$}{}}

The smallest even value of $N$ for which this occurs is $6$, when $n=m=2$. There are now less than $n$ particles in the $\ket{i_{l+1}  \cdots i_{N/2}}$ part of Eq.~(\ref{eqA:Phi}), and hence
analyzing partial transpose
it is essential to consider correlated states of $N/2$ particles $\ket{\phi_{i_1 \cdots i_{N/2}}}$. Their entanglement properties with respect to the partition $m| N/2-m$ potentially play a role, 
opening possibility to apply them to distinguish among AME$(N,d)$ states. A recent work uses negativity to study the effects of AME states subjected to noisy channels \cite{Stawska2025}, although the partitions analyzed there are complementary ones as $m=N-n$.

\section{Local equivalence}\label{sec:lu_equivalence}
    Quantum entanglement is the distinctive characteristic of quantum mechanics that differentiates it from any classical correlations.
    Therefore, any operations that are classical or local will not increase properly defined entanglement. 
    This gives rise to so called {\em local operations and classical communications} (LOCC) and corresponding equivalence classes~\cite{Chitambar_2014}. 

    However, from the operational perspective, that might be too little, as it gives rise to one-way operations only. 
    To change the state back to its original form, one of the extensions includes probabilistic interconversion, i.e. stochastic LOCC (or SLOCC for short)~\cite{Zhang_2016}. 
    In this setup, we say that two states are equivalent if we can convert one to the other with non-zero probability of success, and vice versa. 

    In this scheme, one may lose the resource \emph{on average}, as the probability of success is usually not 1. 
    Here we introduce the special case of local unitary (LU) equivalence classes, which specifies that two pure states $\ket{\psi}$ and $\ket{\phi}$ are in the same class if and only if
    \begin{equation}
        \ket{\psi} = \left(U_1 \otimes ...\otimes U_N \right)\ket{\phi}.
    \end{equation}
    In the above, each $U_i$ is a local unitary operator acting on subsystem $\mathcal{H}_i$. 
    Straightforwardly, due to the invertibility of the unitary operations, also $\ket{\phi}$ can be obtained from $\ket{\psi}$ via local unitaries $U_i^\dagger$. 
    The probability of success in both directions is thus 1. 
    Although, in the general case, for a given state $\ket{
    \psi}$ its orbit of SLOCC-equivalent states is strictly larger than its LU orbit, these sets are equal when we restrict the orbit to AME states~\cite{Burchardt_2020}.
    This results from the application of Kempf-Ness theorem~\cite{Kempf_1979,Gour_2011} for 1-uniform states.
    Consequently, since LU/LOCC/SLOCC sets coincide for AME states, for the remainder of this section, we shall mention the LU class alone.

    Already for general pure states of two qubits, there is an infinite number of LU-equivalence classes, each characterized by different Schmidt coefficients.
    An open problem of multipartite entanglement for AME states is their classification in LU classes, although for multi-quibit system
    the answer is known    \cite{Kraus_2010,Liu_2012}. 
    In particular, even for the cases for which the existence of AME states with local dimension $d\ge 3$ is proven, 
    it is not known in general,
    how many classes of entanglement are they grouped in. 
    Furthermore, verifying whether two given AME states of several qudits are equivalent up to local unitaries is far from trivial. Although in the case of AME stabilizer and graph qubit states the problem can be simplified by restricting to the Clifford group, identifying LU-equivalence has been mostly done for each case individually~\cite{Burchardt_2022}. 
    The hardness stems from the fact that reduced density matrices, which provide an entire local description of AME states, are maximally mixed.

\begin{figure}[b]
    \centering
    \includegraphics[width=0.99\linewidth]{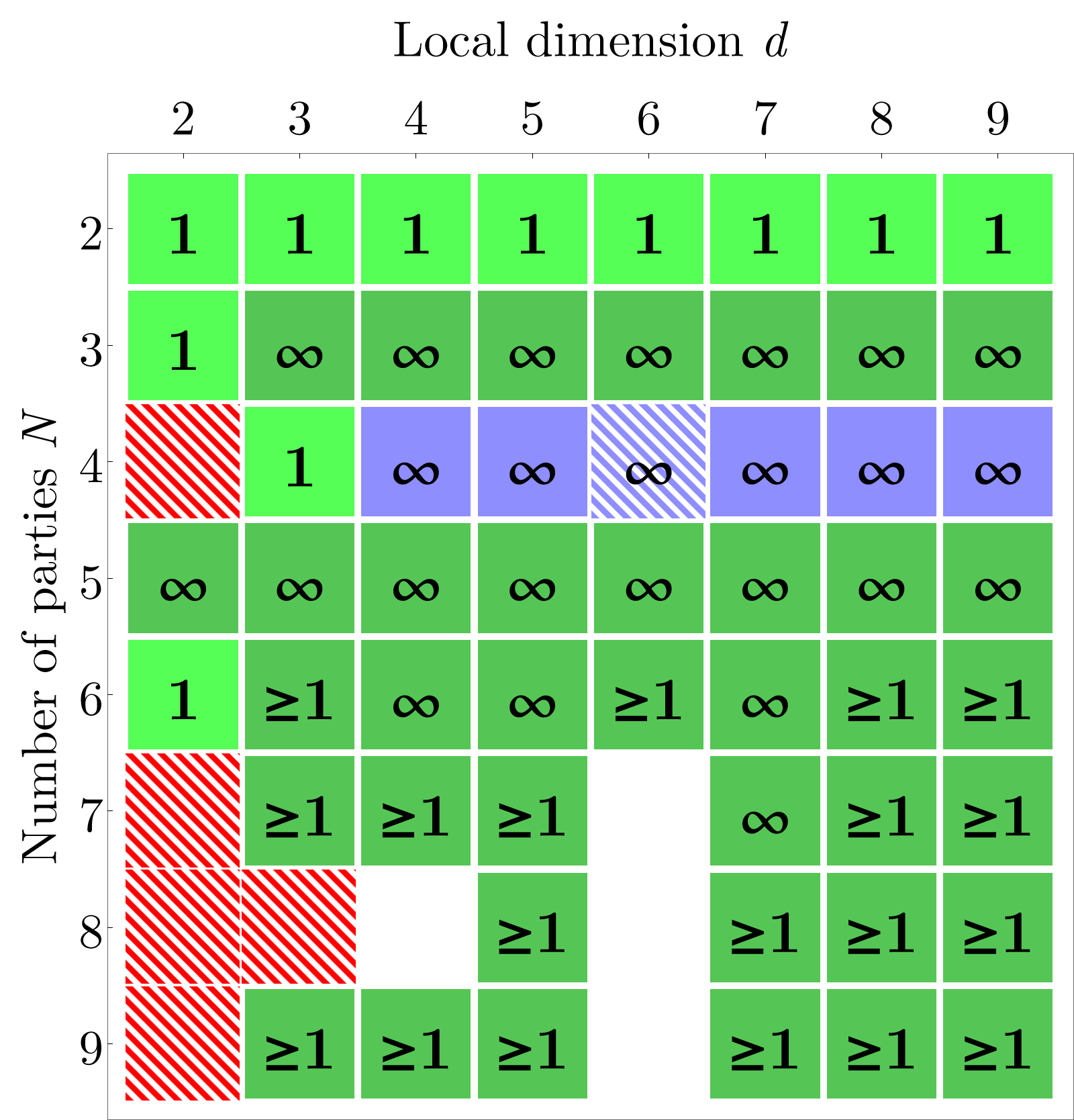}
    \caption{
    Existence of AME states for different local dimension $d$ and number of parties $N$. Red stripes denote that the corresponding state is known {\em not} to exist.
    For blank positions the AME existence problem remains unsolved.
    Light green squares denote
    cases for which all known
    AME states are equivalent to graph states. 
    Blue squares denote the setups for which a non-graph AME state is known. The special case AME$(4,6)$ is depicted with blue stripes, as in this case there is no
    AME graph state~\cite{Rather_2023}.
    Dark green squares represent cases where there exists a graph state, but nothing is known about non-graph states.
    Numbers inside a square describe the known numbers of LU/SLOCC equivalence classes discussed in Sec.~\ref{sec:lu_equivalence}.}
    \label{fig:AME_table}
\end{figure}
    
    To the rescue comes the theory of polynomial invariants~\cite{Rains_2000,Grassl_1998}. 
    This type of local-unitary invariants are very useful both in theory and experiments to classify~\cite{VES11,LL13}, quantify~\cite{EBOS12,RA2016invariant,ES14} and detect~\cite{EKHVKDCKPZV20,NCVKEDCZVKK21,RH24,IWKG21,AR25mixed} entanglement through randomized measurements~\cite{EFHKPVZ22,CIDGKLWMPV24}. For AME states, polynomial invariants have been used to identify and construct infinitely many equivalence classes for AME(5,$d$)~\cite{Ramadas_2024}, as well as AME(4,$d$) states from $d$-dimensional quantum Orthogonal Latin squares~\cite{Rather_2023} whenever $d\geq 4$.  
    Here, the true \emph{quantum} Latin squares are necessary in the case of $d=6$, as there are no orthogonal Latin squares of size $6$.  
    More generally, using quantum orthogonal arrays~\cite{MV16, Goyeneche_2018, Ramadas_2024}, one can show the existence of infinitely many equivalence classes for AME(3,$d$) states for all $d>2$ as well as for AME(5,$d$) for all $d$~\cite{Ramadas_2024,Tan_2025}.

    Not all cases, however, admit infinitely many equivalence classes. 
    Obviously, for every bipartite system, there is only one equivalence class with a generalized Bell state as a representative.
    Surprisingly, for the case of AME(4,3), due to a lack of the corresponding universal entangler~\cite{Chen2008}, there is only one equivalence class~\cite{Rather_2023}, with the representative given by a state of minimal support, as presented in Eq.~\eqref{eq:p9_state}.
    Similarly, three and six qubits AME states admit a single LU-equivalence class~\cite{Dur_2000,Rains_1999}, in the case of 3 qubits, given by the GHZ state.
    A systematic framework to construct AME states
    belonging to different equivalence classes
    was provided in Ref. \cite{RBB22}.
    
    We summarize the above discussion in \com{Fig.}~\ref{fig:AME_table}, in which the number of known equivalence classes is written for different numbers of subsystems and local dimensions.
    {\color{black}
    This table can be considered as an extension
    of earlier data collected in \cite{TableOfAME,Huber_2018Shadow,
    Ramadas_2024,casas2025circuits}.
    }
    A recent work \cite{Ramadas_2025} uses local unitaries to minimize the decomposition entropy   
    of multipartite states \cite{EPZ15} 
    and compares AME states with typical ones.

As a further remark, we consider local-unitary equivalence of four-partite AME states by analyzing their defining isometries. Consider two AME states $\ket{\psi}=(\one\otimes U)\ket{\phi_{CD|AB}}$ and $\ket{\varphi}=(\one\otimes V)\ket{\phi_{CD|AB}}$, where $U$ and $V$ are 2-unitary matrices and $\ket{\phi_{CD|AB}}=\sum_{i,j=0}^{d-1}\ket{ij}\ket{ij}$ is the maximally entangled state between $CD$ and $AB$. The relation between both states is then $\ket{\psi}=(\one\otimes W)\ket{\varphi}$, where $W=UV^\dag$.

If the unitary transition matrix $W$ is local, $W=W_A\otimes W_B$, then the states $\ket{\psi}$ and $\ket{\varphi}$ are LU-equivalent. This condition can be tested through the operator Schmidt decomposition of $W$, $W=\sum_{i=1}^r\lambda_i O^A_i\otimes Q^B_i$ where $\lambda_i\geq 0$ and $O^A_i$ and $Q^B_i$ are matrices: the matrix $W$ is local, if and only if $r=1$, which occurs if and only if $\sum_{i=1}^r\lambda_i\log \lambda_i=0$. Notice however that the converse is not true, since the unitary matrix mapping $\ket{\psi}$ to $\ket{\varphi}$ is not unique. 
A similar property holds also for two $3$-uniform AME states
of six parties,
provided the transition matrix $W$
is local, so it 
has the structure of tensor product
of three local unitaries.

\section{Maximal entanglement and quantum Error correction}\label{sec:QEC}
In Section~\ref{sec:MinSupMDS}, we have considered constructions of AME states from classical codes $C[N,k,\delta]_d=\{\omega_j\}_{j=1}^{d^k}$ involving $d^k$ codewords of $N$ digits of a $d$-dimensional alphabet each with distance $\delta$, which can identify $\delta-1$ classical errors. The constructions given in Section~\ref{sec:Graphs} can be seen as a quantum extension of the classical ones, since stabilizer states and codes are a quantum generalization of classical codes~\cite{CSS1996Goodexist,CSS1996Steane, Grassl:codetables}.  Here we will sketch a more general connection between AME states and quantum error correcting codes~\cite{Lidar_Brun_2013}. 

{\color{\redCom} Quantum error correction is of fundamentally different nature from classical error correction, in the sense that it does not suffice to correct bit flips. Quantum bits can suffer from arbitrary unitary errors such as phase gates, and correcting them requires 
to detect which error occurred and 
to correct strings of Pauli operator errors. Therefore, one may naturally expect that a richer structure arises by considering constructions of AME states from quantum error correcting codes.}

\subsection{Maximally entangled states and code subspaces}
Let us start briefly introducing necessary notation~\cite{NielsenAndChuang,gottesman1997stabilizer}: A {\em quantum error correcting code} 
$C[\![N,k,\delta]\!]_d=\{\ket{\omega_i}\in\mathcal{H}_d^{\otimes N}\}_{i=1}^{d^k}$ is a subspace spanned by $d^k$ orthogonal states $\ket{\omega_i}$ encoding $k$ logical qudits into an $N$-partite $d$-dimensional quantum system. We use this notation to clearly distinguish quantum from classical codes, although the notation $C((N,d^k,\delta))_d$
is sometimes used in parallel for the same object -- see~\cite{NielsenAndChuang,Huber2020QMDScodesAME,Shor1995SchemeQECC,CSS1996Goodexist,gottesman1997stabilizer}. 

Quantum errors are operators acting on $\mathcal{H}_d^{\otimes N}$, and an error $M$ has {\em weight} $\text{wt}(M)=t$ if it acts locally on $t$ subsystems and trivially elsewhere (namely $M=E_1\otimes...\otimes E_{t}\otimes\one^{\otimes N-t}$ and permutations thereof). In analogy to classical codes, a quantum code $C$ with {\em distance} $\delta$ can be used to identify errors $M$ of weight $\text{wt}(M)=\delta-1$ or less~\cite{NielsenAndChuang}. 

The Knill-Laflamme condition~\cite{KnillLaflammeATheoryQECC} states that a code $C$ has distance $\delta$ if and only if
\begin{equation}\label{eq:KnillLaflammeThm}
\begin{gathered}
\bra{\omega_i}M^\dag_a M_b\ket{\omega_j}=\delta_{i,j}\zeta_{a,b} 
\end{gathered}
\end{equation}
for all errors $M_a$ and $M_b$ of weight $\text{wt}(M_a^\dag M_b)\leq\delta-1$, where $\zeta_{a,b}\in\C$ is independent of the codewords $\ket{\omega_i}$ and $\ket{\omega_j}$. The code $C$ is called {\em pure} if $\zeta_{a,b}=\tr(M_a^\dag M_b)/d^N$ for all errors with $\text{wt}(M_a^\dag M_b)\leq\delta-1$~\cite{Gour2007QECentSubs,Huber2020QMDScodesAME}. 

By expanding projector $\ket{\omega_i}\bra{\omega_i}$ into the orthonormal error basis made by tensor products of the Weyl-Heisenberg operators of Eq.~\eqref{eq_WeylHeis} (Pauli strings for qubits), one finds that this occurs if and only if $\tr_{N-\delta+1}\ket{\omega_i}\bra{\omega_i}\propto\one$. This is because all Weyl-Heisenberg operators except for the identity are traceless (see~\cite{preskill1998QInfoAndCompu,Scott2003QECCentPow} for a full derivation). In analogy to the classical case, the {\em quantum Singleton bound}~\cite{Cerf1997QSBound}
establishes that $\delta\leq(N-k)/2+1$, and codes saturating it are called {\em quantum maximum distance separable (QMDS) codes}~\cite{Huber2020QMDScodesAME}. As a result, the $d\times k$ codewords $\ket{\omega_i}$ of a pure QMDS code $C[\![N,k,\floor{(N-k)/2}+1]\!]_d$ satisfy 
relation
$\tr_{\ceil{(N+k)/2}}\ket{\omega_i}\bra{\omega_i}\propto\one$. 

A special case is that of {\em self-dual} codes, for which $k=0$ and thus consist of a single codeword $\ket{\omega}$ and encode a one-dimensional subspace. Although these are unnatural in quantum error correction since no information can be encoded, it is clear from the connections sketched above that a self-dual pure quantum code $C[\![N,0,\floor{N/2}+1]\!]_d$ is equivalent to an AME$(N,d)$ state. These can be seen as maximally resourceful at defining maximally entangled subspaces, since the existence of a pure quantum code $C[\![N,k,\delta]\!]_d$ implies the existence of a pure quantum code $C[\![N-1,k+1,\delta-1]\!]_d$ but the converse is in general not true~\cite{Rains1998,Huber2020QMDScodesAME,Raissi2020Modifying} --see Proposition 7 and Fig. 1 in~\cite{Huber2020QMDScodesAME} for positive examples and counterexamples, respectively.  A simple example where the construction works in both directions is the AME($6,2$) state $\ket{\psi}=(\ket{0}\ket{0_L}+\ket{1}\ket{1_L})/\sqrt{2}$ with~\cite{Ramadas_2024,Goyeneche2014}
\begin{align}
\ket{0_L} &= \frac{1}{4}(\ket{00000} + \ket{00111} - \ket{01010} + \ket{01101}\nonumber \\
&- \ket{10001} - \ket{10110} - \ket{11011} + \ket{11100})
\label{eq:0L} \\
\ket{1_L} &= \frac{1}{4} (\ket{00011} + \ket{00100} - \ket{01001} + \ket{01110}\nonumber \\
&+ \ket{10010} + \ket{10101} + \ket{11000} - \ket{11111} )\label{eq:01L},
\end{align}
which is a one-dimensional (thus self-dual) pure code $C[\![6,0,4]\!]_2$. This defines a pure code $C[\![5,1,3]\!]_2=\{\ket{0_L},\ket{1_L}\}$~\cite{laflamme1996AperfectQECC(5-2)}, which in turn corresponds to the one-parameter two-dimensional subspace of AME($5,2$) defined by~\cite{Ramadas_2024}
\begin{equation}
\ket{\psi(\theta)} = \cos(\theta)\ket{0_L} + \sin(\theta)\ket{1_L}\,.
\end{equation}

A higher-dimensional paradigmatic example is the three-qutrit code $[\![3,1,2]\!]_3$ with codewords
\begin{align}
\ket{0_L}&=\frac{1}{\sqrt{3}}(\ket{000}+\ket{111} +\ket{222})\label{eq:0Lf} \\
\ket{1_L}&=
\frac{1}{\sqrt{3}}(\ket{012} +\ket{120}+\ket{201})\label{eq:1L} \\
\ket{2_L}&=\frac{1}{\sqrt{3}}(\ket{021}+\ket{102}+\ket{210}), \label{eq:2L}
\end{align}
constructed from the code $C[\![4,0,3]\!]_3$, which corresponds to 
the AME(4,3) state~(\ref{eq:p9_state}): similarly as above, the codewords are obtained by writing the full state as  $\ket{\Psi_{P_9}}=\sum_{j=0}^2\ket{j}\ket{j_L}/\sqrt{3}$. 
Notice that indeed, the subspace spanned by the codewords $\{\ket{0_L},\ket{1_L},\ket{2_L}\}$ is maximally entangled. In fact, this encoding can be used for so-called {\em quantum secret sharing}~\cite{Hillery1999QSS}: a message can be encoded nonlocally among multiple users, in such a way that none of them is able to recover the information individually with local operations. AME states are particularly useful for this type of schemes~\cite{Helwig2012}.

To sum up, the link between AME states and quantum codes has numerous applications in quantum information processing, such as multipartite teleportation and quantum secret sharing~\cite{Helwig2012,Helwig_2013}. These combine two main features of AME states: the fact that two complementary subsets $A$ and $B$ of parties sharing an AME state can produce a high-dimensional Bell state by a unitary acting on $A$ or $B$, which then enables to perform different forms of teleportation~\cite{helwig2013existApplic}; and the fact that an AME state is composed by subspaces where all states are in turn AME states of smaller system sizes, which allows for nonlocal encodings of information~\cite{Helwig_2013}.

\subsection{Existence of Absolutely Maximal Entanglement}
Besides practical applications, the connection between AME states and quantum codes above is very useful to determine whether a given system size $(N,d)$ admits the existence of an AME($N,d$) state~\cite{Rains1998,Huber_2017,Huber2020QMDScodesAME}: if an AME($N,d$) 
exists, then a $d$-dimensional subspace of AME($N-1,d$) 
exists as well (and the equivalent statement in the opposite direction can be obtained by negating both existences). An explicit construction of optimal quantum error correcting codes from subspaces of absolutely maximally entangled states is given in~\cite{Raissi2018}.

Using this connection, an extensive machinery determining the existence of quantum codes applies to determine that certain AME states cannot exist~\cite{Rains1998,Rains_1999,Rains1999QShadowEnum,Scott2003QECCentPow,Huber_2018Shadow,Huber2020QMDScodesAME,angles2024sdpbounds}, besides case-by-case study~\cite{Higuchi_2000_twoCouples,Huber_2017,Rather_2022}. Up to date, these techniques can be divided as follows:

{\em Quantum weight enumerators}~\cite{Shor-Lafflamme1997Weight,Rains1998,Rains_1999,Huber2020QMDScodesAME,angles2024sdpbounds}. Given a code $C$ with projector $\Pi_C$ onto the code subspace, the Shor-Laflamme~\cite{Shor-Lafflamme1997Weight} weight enumerators are
\begin{align}
A_j(\Pi_C)&=\sum_{E:\text{wt}(E)=j}\tr(E\Pi_C)\tr(E^{\dag}\Pi_C)\\
B_j(\Pi_C)&=\sum_{E:\text{wt}(E)=j}\tr(E\Pi_C E^{\dag}\Pi_C)\,,
\end{align}
where the summation is performed over the elements of an orthonormal error basis with certain weight. Note that the weight enumerators are basis independent. The code $C$ exists only if $A_j(\Pi_C)\geq 0$, $B_j(\Pi_C)\geq 0$, $A_0(\Pi_C)=d^{2k}$ and
$d^k B_j(\Pi_C)\geq A_j(\Pi_C)$. AME states (i.e. codes with maximum distance and $k=0$) satisfy relation $A_j=B_j$. This leads to the following necessary condition for an AME$(N,d)$ to exist~\cite{Scott2003QECCentPow},
\begin{equation}\label{eq:ScottBound}
\begin{gathered}
N\leq\begin{cases}
    2(d^2-1)\quad \forall N\in\mathbb{N}_{\text{even}}\\
    2d(d+1)-1\quad \forall N\in\mathbb{N}_{\text{odd}}
\end{cases}.
\end{gathered}
\end{equation}

{\em Quantum Shadow enumerators}~\cite{Rains1999QShadowEnum,Huber_2018Shadow,Huber2020QMDScodesAME}. Although the weight enumerators above determine the non-existence of the majority of cases in~\cite{TableOfAME}, further bound is found as follows~\cite{Huber_2018Shadow}. Let $S$ be a subset of $k\leq N$ parties and $S^c$ its complementary subset. Define 
\begin{equation}
S_j(\Pi_C)=\sum_{\substack{T,S\subseteq\mathcal{N}\\|T|=j}}(-1)^{|S\cap T^c|}\tr_{S}\big ((\tr_{S^c}\Pi_C)^2 \big ),
\end{equation}
summing over all the subsets $T$ and $S$ of $\mathcal{N}=\{1,...,N\}$ where $T$ has cardinality $j$, and $T^c$ and $S^c$ are the complementary sets of $T$ and $S$. The scalars $S_j(\Pi_C)$ are the coefficients of the 
{\em Shadow enumerator polynomial}, and they must be nonnegative if a hypothetical quantum error correcting code (or AME state in particular) with projector $\Pi_C$ exists. In the case of AME states $\Pi_C=\ket{\psi}\bra{\psi}$, this condition can be imposed with linear programming~\cite{Huber_2018Shadow}. This linear program can detect  non-existence of AME states that weight enumerators cannot detect, such as AME($4,2$) state~\cite{Higuchi_2000_twoCouples}. Recently, it has been generalized to a semidefinite programming hierarchy using state-polynomial optimization, leading to a more powerful machinery~\cite{angles2024sdpbounds}. In particular, this extension detects the non-existence of codes and AME states undetectable with the techniques above.

{\em The marginal problem for pure states}~\cite{Hig2003QubitMarginal,Klyachko_2006QuantumMarginal,YSWNG21}. 
Determining whether an AME$(N,d)$ state exists is in fact a case of the quantum marginal problem~\cite{Hig2003QubitMarginal,Klyachko_2006QuantumMarginal}. Namely, determining whether there exists a global state $\rho$ compatible with certain prescribed marginals $\{\rho_S\}_S$ belonging to subsets $S$ of $|S|$ parties such that $\tr_{S^c}\rho=\rho_S$. For the case of AME states, one needs to impose that the global state is pure,
$\rho=\ket{\psi}\bra{\psi}$, and the marginals are maximally mixed, $\rho_S=\tr_{S^c}\ket{\psi}\bra{\psi}=\one_S/d^{|S|}$. 

This problem was tackled with a semidefinite programming hierarchy~\cite{YSWNG21}, so that each level approximates two copies of the hypothetical pure state to be found. Using this technique, the existence problem of an \com{AME$(N,d)$} state is reformulated to the separability problem of a corresponding mixed state in extend
 dimension,
 $d^N \times d^N$. This proved to be a powerful tool to determine the existence of a range of AME states and pure quantum codes.

{\color{\redCom}
\section{Other applications}
 Up to date, the main known applications of AME states come from their property of maximal bipartite entanglement between any pair of complementary subsystems. Therefore, any multipartite protocol with a priori unknown target pairs that uses entanglement will see benefit from using AME states.
 Here we sketch these applications introduced in Ref.~\cite{helwig2013existApplic}, following the lines of Refs.~\cite{helwig2013existApplic,Helwig2012,Cleve1999HowToSharaeQSS}.
 \subsection{Parallel teleportation}
 Standard quantum teleportation~\cite{Teleport1993Bennett} consists of transferring an unknown state $\ket{\psi}$ from a system $A$ to a system $B$, by means of performing local operations on a maximally entangled state $\ket{\phi}_{AB}$ and classical communication. 
 If $A$ and $B$ consist of subparts themselves, 
 AME states allow to perform this protocol in parallel, with each user of a subset $S$ of size $|S|=\floor{N/2}$ sending an unknown quantum state to a user in the complementary set $S^c$ of size $|S^c|=\ceil{N/2}$. To understand this protocol, it is handy to present the $N$-partite shared AME state $\ket{\psi}$ as in Eq.~\eqref{eq:AMEbip} in the bipartition $S|S^c$, namely
 \begin{equation}
    \label{eq:AMEbipU}\ket{\psi}=\frac{1}{\sqrt{d^{k}}}\sum_{i_1,...,i_k=0}^{d-1}\ket{i_1,...,i_k}U\ket{i_1,...,i_k},
\end{equation}
The transformation applied by senders is the joint unitary $U_{S|S^c}$, 
which modifies each possible state of the receivers into
\begin{equation}\label{eq:teleportUnitaryOperation}
\begin{aligned}
& U_{S|S^c}^{\dag}\ket{\psi_{i_1,...,i_k}}=\ket{i_1}\cdots\ket{i_k}\ket{0}\quad (N \text{ odd})\\
& U_{S|S^c}^{\dag}\ket{\psi_{i_1,...,i_k}}=\ket{i_1}\cdots\ket{i_k}\quad (N \text{ even}),
\end{aligned}    
\end{equation}
depending on whether $N$ is even or odd. Thus we obtain the state $\ket{\psi'}=\big (\tfrac{1}{\sqrt{d}}\sum_{i=0}^{d-1}\ket{i,i}\big )^{\otimes\floor{N/2}}$,  
where each party in $S$ shares a Bell state with a party in $S^c$, transforming it to the standard teleportation protocol. Note that the entire reasoning can be extended to $k$-uniform states with a lower-dimensional state being teleported. Similar strategies allow for other teleportation protocols such as open-destination, where an unknown state is teleported to an arbitrary local party~\cite{helwig2013existApplic,Helwig2012}.

\subsection{Entanglement swapping}
Entanglement swapping~\cite{ExperimentalSwappingPan1998} is a protocol used to entangle non-interacting local systems, utilizing entanglement with an intermediate system. Notably, it can be applied for multipartite systems using AME states shared among an even number $N=2k$ of parties. The work~\cite{helwig2013existApplic} presents a scenario enabling this possibility: consider $M+1$ parties, 
organized in subsystems $\{1,...,2k\},\{k+1,...,3k\},...,\{mk+1,...,(m+1)k\},...,\{Mk+1,...,(M+1)k\}$, sharing each an AME$(2k,d)$ state, defined as in Eq.~\eqref{eq:AMEbipU} with unitary matrices $U_M$. Furthermore, we require that the state obtained by applying $M$-th power $U^M\ket{i_1,...,i_k}$ is LU-equivalent to $U\ket{i_1,...,i_k}$ for all elements of the computational basis $\{\ket{i_1,...,i_k}\}$ up to permutations of local parties. If each of the intermediate parties $\{k+1,...,Mk\}$ performs a Bell measurement on two qudits they share, then the outermost parties $\{1,...,k,Mk+ 1,...,(M+ 1)k\}$ will share an AME$(2k,d)$ state.

As a simple example, consider three AME(4,$d$) states from Eq.~\eqref{eq:AMEbipU}, shared between eight parties $\{A,B,C,D\}$, $\{C,D,E,F\}$, and $\{E,F,G,H\}$.
If the Bell measurements will be performed by four parties that share two qudits: $C,D,E$, and $F$, then the state shared between parties $\{A,B,G,H\}$ will be AME(4,$d$) state.

\subsection{Quantum secret sharing}
Quantum secret sharing~\cite{Hillery1999QSS} is a quantum cryptography protocol designed to distribute a secret among multiple parties, 
such that no local party can recover the state without the collaboration of other users. This method inherently uses entanglement, namely the nonlocal spread of information stored in quantum systems. 
This makes AME states a perfect platform for this purpose~\cite{Helwig2012}. 

Below we present the simplest implementation of this protocol based on the four-qutrit AME state of Eq.~\eqref{eq:p9_state}. Following the lines of Ref.~\cite{Cleve1999HowToSharaeQSS}, we will consider a three-dimensional quantum message encoded in a qutrit,
\begin{equation}
    \ket{\phi}=\alpha\ket{0}+\beta\ket{1}+\gamma\ket{2},
\end{equation}
with $\alpha,\beta,\gamma\in\C$ and $|\alpha|^2+|\beta|^2+|\gamma|^2=1$. Let us store the information of this qutrit in three parties $A$, $B$ and $C$ as $\ket{0}\rightarrow\ket{0_L}$,  $\ket{1}\rightarrow\ket{1_L}$ and $\ket{2}\rightarrow\ket{2_L}$ through Eqs.~(\ref{eq:0Lf}-\ref{eq:2L}), into 
\begin{equation}
    \ket{\phi_L}=\alpha\ket{0_L}+\beta\ket{1_L}+\gamma\ket{2_L}.
\end{equation}
No information can be obtained by local parties as all one-body reductions are maximally mixed, independently of the coefficients $\alpha$, $\beta$, and $\gamma$. However, the state $\ket{\phi}$ can be obtained from the collaboration of any two local parties: 
by applying two-qudit operations~\cite{Helwig2012} on their reduced state $\varrho_{AB}=\tr_C\dyad{\phi_L}$, these can obtain the separable state
\begin{align}
    \sigma_{AB}=&\big (\alpha\ket{0}+\beta\ket{1}+\gamma\ket{2}\big )\big (\alpha\bra{0}+\beta\bra{1}+\gamma\bra{2}\big )\nonumber\\
    &\otimes\big (\ket{0}+\ket{1}+\ket{2}\big )\big (\bra{0}+\bra{1}+\bra{2}\big ),
\end{align}
which recovers the original message $\ket{\phi}\bra{\phi}$. 
}

\section{Perfect tensors and boundary-bulk correspondence}
\label{sec:holo}

Beyond practical applications in quantum error correction, AME states found their way in simulating one of the best-known conjectures of theoretical physics:  AdS/CFT correspondence and holographic duality.
The general notion for this correspondence is that certain physical theory defined on \textit{bulk}, usually consisting of anti-de Sitter space (AdS), can be reformulated into a conformal field theory (CFT) defined on the \textit{boundary} of the bulk. In case of original construction, the theories of interest were type IIB strings given by type IIB supergravity in the 
bulk, and super-Yang–Mills theory, occupying 10 and 4-dimensional spaces respectively \cite{Maldacena1999}. 

The main premise of AdS/CFT correspondence is that properties of one theory, which is difficult to tackle due to strong coupling, can be translated into properties of another one with weak coupling, thus allowing one to ``bypass'' demanding or impossible calculations. On the other hand,
this very nature makes the 
correspondence difficult to study, since important scenarios of theories of interest
are out of reach,
resulting in a demand for simplified toy models.

Seminal work on this frontier \cite{Pastawski2015} provided the so-called HaPPY code, which constituted a mapping between the time slice of 2+1 dimensional AdS space -- Poincar\'e disk, and its 1-dimensional boundary using a network of perfect tensors. 
To outline this construction we recall that each AME state 
$|\psi\rangle$ can be interpreted as a perfect tensor of its amplitudes \eqref{eq:General_Nd_state}, or, for given bi-partition $A \cup B= [N]$ as an $d^{|A|} \times d^{|B|}$ isometry.

\begin{figure}[ht]
\includegraphics[width=0.9\columnwidth]{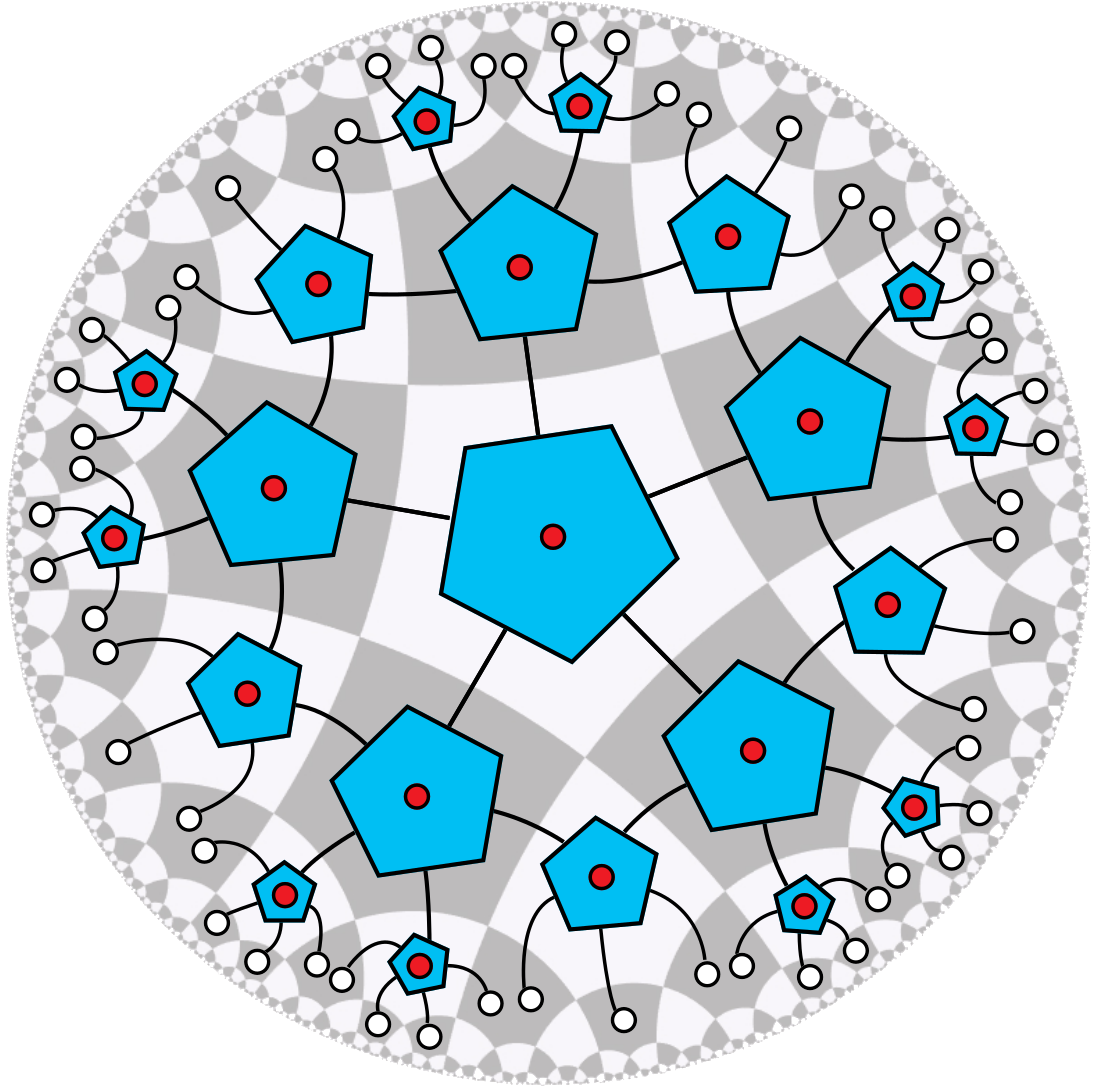}
\caption{ HaPPY code tensor network placed on tiling of Poincar\'e disk. Each blue pentagon represents one perfect tensor of 6 qubits describing an AME$(6,2)$ state and a $[\![5,1,3]\!]_2$ error correction code. Red dots correspond to bulk (logical) indices, whereas the white dots correspond to boundary (physical) ones. Figure borrowed from  \cite{Mazurek2020}, 
inspired by the original figure in \cite{Pastawski2015}. }
\label{HaPPY_plot}
\end{figure}

The authors of~\cite{Pastawski2015} decomposed the Poincar\'e disk into regular pentagons, such that four of them meet at each vertex, and associated each tile with a copy of AME(6,2) state
described by a
perfect tensor related to a $[\![5,1,3]\!]_2$ quantum error correction code \cite{Laflamme1996, Bennett1996}, as presented in Fig.~\ref{HaPPY_plot}. In such a way, each logical qubit index of a tensor corresponds to a bulk degree of freedom associated with a proper tile. The remaining indices are either contracted with neighboring tensors or, in the case of boundary tiles, become boundary degrees of freedom. Such a \textit{tensor network} can be constructed starting from the central tile and then attaching new ones layer by layer. Then each new tile can be interpreted as an isometry from the bulk index and the contracted indices to the remaining ones, due to the perfectness of the tensor. Thus the resulting network is a large isometry from the bulk into the boundary indices.

The HaPPY code quickly gained attention, not only due to its properties desired for modeling AdS/CFT like entropy scaling \cite{Ryu2006, Pastawski2015}, but also due to startling error-correcting properties. 
For example, the work \cite{Pastawski2017} has proven that the HaPPY code exhibits \textit{\"uberholography}, which means that in the limit of a large network, only a fractal subset of the boundary indices is sufficient to reconstruct a localised bulk operator.
Furthermore, the reformulation of HaPPY code as Majorana dimer code \cite{Jahn2019} enabled efficient construction of the tensor network and description of boundary states. 
Using these methods, many properties of boundary states, like correlation functions, were studied \cite{Jahn2019_2, Jahn2020}, bridging the gap between toy-modes of AdS/CFT correspondence and the physical properties of field theories of interest. The next milestone on the venture to derive physical properties of AdS/CFT models was the calculation of central charges corresponding to this model, and similar tensor networks \cite{Jahn2020}.

The standard generalisations of HaPPY codes were based on modification of Poincar\'e disk tilling by introducing different perfect tensors or \textit{block-perfect} tensors, which are maximally entangled with respect to bipartitions into adjacent sets of indices \cite{Harris2018}. The extension of perfect tensor networks into block perfect ones allowed to employ other well-known error correction codes, such as Calderbank–Shor–Steane (CSS) code
 \cite{CSS1996Steane,CSS1996Goodexist}.

During the course of further investigation, the work \cite{Cao2022} proved that ``{\em when the network consists of a single type of tensor [...] it cannot be both locally contractible and sustain power-law correlation functions}''. This implies, for the HaPPY code, that the crucial physical properties of boundary states -- power-law decay of correlation functions -- cannot be realised by localised bulk operators.
The first bypass for this theorem
was based on a network consisting of two types of tensors~\cite{Cao2022}.

However, at the moment, another model -- hyper-invariant tensor network (HTN) -- devised by Evenbly \cite{Evenbly2017, Steinberg2022} gained traction. The core idea of this approach was to introduce two types of tensors, one placed on the tiles of Poincar\'e disk and another one on edges between them, and with such a network demand minimal isometry conditions for blocks of tensors in the network. Although a perfect tensor network constitutes a simple example of HTN, in general, their
properties can be substantially different.
Motivated by HTN, the work \cite{Bistro2025} explicitly demonstrated the power-law decay of correlation functions for bulk to boundary mapping tensor network. 
Furthermore, another extension of HTN incorporating bulk degrees of freedom \cite{Steinberg2023} led to a new family of powerful error correcting codes: Evenbly codes \cite{Steinberg:2024ack}.

This overview would not be complete without mentioning related approaches to holographic tensor networks, like the ones based on small distortions of perfect tensors \cite{bhattacharyya2016, bhattacharyya2018} or random tensor networks \cite{Hayden2016}.
An alternative approach is to simulate AdS/CFT correspondence which aims to create a discrete version of CFT -- Quasi-CFT -- directly on the tensor network \cite{Jahn2022}. Finally we point out to 
a recent construction of quantum circuits devised to implement the HaPPY code \cite{AnglsMunn2024}.  

A vast amount of research on holographic tensor network models has been performed recently, and therefore a comprehensive discussion of this field is beyond the scope of this work. For a better understanding of these problems and the motivations behind them, we encourage the
readers to consult the recent
reviews \cite{Jahn2021, Morsalani2025}.

\section{Concluding Remarks}

Absolutely maximally entangled (AME) states of several subsystems
exhibit maximal correlations between 
results of measurements
performed by any selected parties.
These particular states can be used 
to gauge the quality of
emerging quantum processors: 
they can serve as benchmarks not easy to create~\cite{Paraskevopoulos_2025}, 
but their entanglement is rather robust against the
 noise \cite{CerveraLierta_2019,casas2025circuits}.

AME states are required to execute certain tasks of quantum information processing.
For instance, a four-party GHZ state allows one to teleport
a {\em single qudit} between any two chosen parties of the system,
while the corresponding AME$(4,d)$ state, which exists for any $d\ge 3$,
enables teleportation of {\em two qudits} from any selected 
pair of users to the remaining two of them.

In this work we presented the current state of
research on these highly entangled states.
Special attention was paid to the newly discovered
class of AME states, which do not belong to the
known class of stabilizer and graph states.
Although the first such example 
of the golden AME state \cite{Rather_2022}
was shown to be related to the problem of 36 officers of Euler
\cite{Zyczkowski_2023},
several other non-stabilizer 
AME states were recently found for $d=6$ 
\cite{Rather_2024,Gross_2025} 
and other dimensions \cite{bistron_2023,Bruzda_2024Two-Had}.
Note that some of these recent non-standard AME solutions can be interpreted as a superposition of two (or more) 
classical combinatorial designs or stabilizer states.

The structure of the set of AME states of various dimensions
is rather intricate. In this contribution 
we managed to improve our understanding of key properties of AME states
closing several gaps concerning their features.
We showed that certain 2-uniform AME states of $N$ parties 
can be created as a superposition of $1$-uniform states
and  analyzed entanglement of AME states reduced 
to $N-1$ subsystems. Furthermore, in some cases
we found solutions of AME states
and extended the study on their 
local equivalence \cite{Burchardt_2020,Ramadas_2024},
where it was demonstrated that
for five parties there are infinite number of 
AME$(5,d)$ equivalence classes for any $d\ge 2$.

As the question concerning existence of AME states for different
number of parties $N$ and local dimension $d$
was first integrated in an online table by Huber and Wyderka \cite{TableOfAME},
following \cite{Ramadas_2024,casas2025circuits}
we gathered in Fig. \ref{fig:AME_table}
the data concerning the number of known
non-equivalent AME states.

{\color{black}
As mentioned in the introduction, it is also possible to analyze AME states for heterogeneous systems with different local dimensions \cite{GBZ16,Huber_2018Shadow,shen2021heterogeneous}. Pure-state entanglement in 
$2 \times L \times M \times N$
 systems was classified in \cite{Sun15}, while the advantages of planning experiments with heterogeneous systems were advocated in \cite{Krenn2016}. The AME problem for such systems was posed in \cite{GBZ16} and solved for the $2 \times 2 \times 3 $ and $2 \times 3 \times 3$ systems, for which $1$-uniform states were constructed. The former one is generated using the notion of mixed orthogonal arrays \cite{Hedayat_1999}, linked to  
 quantum error correction codes over mixed
alphabets \cite{Mazurek2020,ball2025AMEhetero}.
AME states do not exist for four-party systems 
 $2 \times 2 \times 2 \times  3$ 
 and
 $2 \times 2 \times 3 \times  3$, 
but such a state 
$|AME\rangle \in {\cal H}_{54}$
was constructed for
$2 \times 3 \times 3 \times  3$ system  
~\cite{Huber_2018Shadow}.
 For several other heterogeneous systems such an existence problem remains open.
}

In spite of numerous spectacular results on
multipartite strongly entangled quantum states achieved recently,
our comprehension of these issues is by far not complete.
{\color{\redCom} For instance, the work on fermionic AME states has only just started~\cite{Giorgadze_2025}, with potential applications extending to quantum chemistry.
Although some schemes to create AME states experimentally
were proposed \cite{CerveraLierta_2019,ZBE24,casas2025circuits},
so far, limited cases have been achieved~\cite{miller2024weightEnumExperiment}.}
Therefore, this work will be concluded with a list of 
some relevant open questions concerning 
both theoretical and experimental physics.
\\\\

\hskip 1.0cm (T) Theory

\smallskip 
\begin{enumerate}[label=(T\arabic*).]
    \item 
    Erase white spots in the table above:
    Are there states AME(8,4) and AME(7,6)?
    \item 
    Is there an AME(4,6) state with real
    coefficients? Equivalently, is there an
orthogonal $2$-unitary matrix of order $36$?
\item 
What is the simplest case of an AME$(N,d)$ state
    which does not belong to the stabilizer class?
\item {\color{\redCom} For a given state AME$(N,d)$ find the minimal
number of two-qudit gates
necessary to realize it by a quantum circuit 
and the minimal
number of layers
(circuit depth).}
    
    \item 
    Are there parameters $N$ and $d$
     for which the number of non-equivalent  
     AME$(N,d)$ states is larger than one but finite?
     \item 
     Given $N$ and $d$ for which the number of non-equivalent  
     AME$(N,d)$ states is infinite, classify them
     according to their 
     (i) internal entanglement structure (ii) robustness to noise and
     (iii) potential usage in an experiment.
     \item 
Are there dual/T-dual gates in $U(9)$ arbitrarily close to the 2-unitary permutation matrix $P_9$  related to  the state AME$(4,3)$? 
For  which other 
dimensions, are there
analogs of dual unitary matrices of size 
$d^{\floor{N/2}}$
arbitrarily close to the multi-unitary matrix
defining  the state  AME$(N,d)$? 
\\
\end{enumerate}

{\color{\redCom}
\hskip 1.0cm (E) Experiment 

\smallskip

\begin{enumerate}[label=(E\arabic*).]
    \setcounter{enumi}{7}
    \item A six-qubit AME graph state was implemented in an ion trap~\cite{miller2024weightEnumExperiment}, but noise effects were considerable in comparison to less entangled states. What implementation remains noise-robust in the presence of strong correlations?
    
\item Which experimental platform is the most suitable for implementing qudit non-stabilizer AME states?

\item What is the optimal strategy to protect these highly correlated states from decoherence? 
\end{enumerate}

}

\medskip

{\bf Acknowledgments}.
K.{\.Z}. is thankful to Ryszard Horodecki for his constant support during the last quarter-century.
It is also a pleasure to thank 
Wojciech Bruzda,  Adam Burchardt and Suhail Rather
for a long collaboration in hunt for new
AME states and for
several valuable remarks concerning this contribution. 
We are also grateful to Jakub Czartowski, 
Paolo Facchi, Masoud Gharahi,
Dardo Goyeneche,
David Gross,  Otfried G{\"u}hne, 
Felix Huber,
Pawe{\l} Mazurek, Gerard Anglès Munné, Ion Nechita, Saverio Pascazio, 
Balazs Pozgay,
Zahra Raissi and N. Ramadas 
for sharing with us data and figures, numerous fruitful discussions and valuable comments on this work.

Financial support by the European Union under ERC
Advanced Grant {\em TAtypic}, Project No. 101142236,
is gratefully acknowledged. 
GR-M acknowledges funding from the European Innovation Council accelerator grant COMFTQUA, no. 190183782.
RB acknowledges support by the National Science Center, Poland, under the contract number 2023/50/E/ST2/00472. 
AR acknowledges 
financial support
by the Deutsche Forschungsgemeinschaft (DFG, German
Research Foundation), project number 563437167 and Project BeRyQC, Grant No.
13N17292), and 
funding from Spanish MICIN (PID2022:141283NBI00; 139099NBI00), FEDER funds, the Spanish Government with funding
from European Union NextGenerationEU (PRTR-C17.I1), the Generalitat de Catalunya,
the Ministry for Digital Transformation and of Civil Service of the Spanish Government through the QUANTUM ENIA project - Quantum Spain Project - through the Recovery, Transformation and Resilience Plan NextGeneration EU within the framework
of the Digital Spain 2026 Agenda. 
AL gratefully acknowledges the hospitality of the Institute of Theoretical Physics, Jagiellonian University, Krak\'ow, during the writing of this manuscript.



\begin{widetext}
\appendix
\section{Details of the QOLS(6) corresponding to the golden AME(4,6) state}
\label{app:ame46}
The four OFS (\ref{eq:FH},~\ref{eq:KpLp})
are explicitly given below and that the entries repeat 4 times, in exactly the same positions in each of them, is also illustrated in one case. Additional circles with blue digits  
close to square $KL$ mark position of the two
distinguished entries on the torus and reveal
a four-knight structure.

\begin{tikzpicture}
\node at (-5.15,-2.2) {\Circle{\color{blue}11}};

\node at (-6,-0.95) {\Circle{\color{blue}11}};

\node at (0,0) {
$
KL = 
\begin{bmatrix}
\Circle{11} & 13 & 55 & 51 & 33 & 35 \\
51 & 53 & 35 & \Circlee{31} & 13 & 15 \\
35 & 31 & \Circlee{13} & 15 & \Circlee{51} & 53 \\
15 & \Circle{11} & 53 & 55 & 31 & 33 \\
53 & 55 & 31 & 33 & 15 & \Circle{11} \\
33 & 35 & \Circle{11} & 13 & 55 & 51
\end{bmatrix},
\qquad
K\tilde{L} =
\begin{bmatrix}
\Circle{12} & 14 & 56 & 52 & 34 & 36 \\
52 & 54 & 36 & \Circlee{32} & 14 & 16 \\
36 & 32 & 14 & 16 & \Circlee{52} & 54 \\
16 & \Circle{12} & 54 & 56 & 32 & 34 \\
54 & 56 & 32 & 34 & 16 & \Circle{12} \\
34 & 36 & \Circle{12} & 14 & 56 & 52
\end{bmatrix}
$};
\end{tikzpicture}
\vskip 0.15cm
\begin{tikzpicture}
\node at (0,0) {$
\tilde{K}L =
\begin{bmatrix}
\Circle{21} & 23 & 65 & 61 & 43 & 45 \\
61 & 63 & 45 & \Circlee{41} & 23 & 25 \\
45 & 41 & 23 & 25 & \Circlee{61} & 63 \\
25 & \Circle{21} & 63 & 65 & 41 & 43 \\
63 & 65 & 41 & 43 & 25 & \Circle{21} \\
43 & 45 & \Circle{21} & 23 & 65 & 61
\end{bmatrix},
\qquad
\tilde{K}\tilde{L} =
\begin{bmatrix}
\Circle{22} & 24 & 66 & 62 & 44 & 46 \\
62 & 64 & 46 & \Circlee{42} & 24 & 26 \\
46 & 42 & 24 & 26 & \Circlee{62} & 64 \\
26 & \Circle{22} & 64 & 66 & 42 & 44 \\
64 & 66 & 42 & 44 & 26 & \Circle{22} \\
44 & 46 & \Circle{22} & 24 & 66 & 62
\end{bmatrix}
$};
\end{tikzpicture}
\begin{figure}[H]
\begin{center}
    \includegraphics[scale=.26]{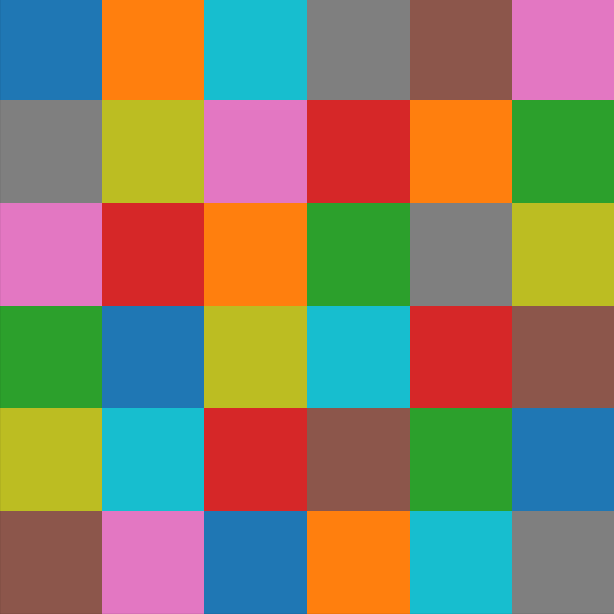}
    \caption{The elements of any one of the four OFS written above are colored to reveal the same pattern in all of them. This square of size 6 has a structure similar to a Latin square, as each color 
    occurs once in each row and column. However, there are 9 colors instead of 6, each appearing in the pattern four times,  which visualizes the title of Ref.~\cite{Zyczkowski_2023}.
    Note that four red fields are connected by a closed path of a chess knight. The same property holds for any other color provided the knight jumps on the torus.
    Boundaries between any pair of colors appear exactly twice on the torus, and are vertical (for instance between red and orange) or horizontal (between red and yellow). 
    }
    \end{center}
\end{figure}

All the 36 unitary matrices $M_{ij}$ 
or order two
-- see Eq.~(\ref{eq:Mij}) -- that construct the golden AME$(4,6)$ from the 4 frequency squares~(\ref{eq:FH},~\ref{eq:KpLp}) are listed below. They are organized into 9 quartets, each of which is an orthonormal set ensuring the unitarity of the QOLS(6). The first quartet corresponds to the entries circled in the OFS above.

\begin{align*}
M_{11} &= \begin{pmatrix} 0 & 1 \\ 1 & 0 \end{pmatrix}, &
M_{42} &= \begin{pmatrix} -b & a\omega^5 \\ -a\omega^{5} & b \end{pmatrix}, &
M_{56} &= \begin{pmatrix} 1 & 0 \\ 0 & 1 \end{pmatrix}, &
M_{63} &= \begin{pmatrix} -a & -b\omega^{5} \\ b\omega^5 & a \end{pmatrix}, \\
M_{12} &= \begin{pmatrix} -\omega^{7} & 0 \\ 0 & -\omega^{9} \end{pmatrix}, &
M_{25} &= \begin{pmatrix} a\omega^2 & b\omega \\ b\omega^5 & -a\omega^{4} \end{pmatrix}, &
M_{33} &= \begin{pmatrix} 0 & \omega^5 \\ -\omega^{9} & 0 \end{pmatrix}, &
M_{64} &= \begin{pmatrix} -b\omega^{4} & a\omega^3 \\ a\omega^7 & b\omega^6 \end{pmatrix}, \\
M_{13} &= \begin{pmatrix} -a\omega^{8} & b\omega^3 \\ b\omega^3 & -a\omega^{8} \end{pmatrix}, &
M_{44} &= \begin{pmatrix} -b & a\omega^5 \\ a\omega^5 & -b \end{pmatrix}, &
M_{52} &= \begin{pmatrix} 0 & 1 \\ -1 & 0 \end{pmatrix}, &
M_{65} &= \begin{pmatrix} \omega & 0 \\ 0 & -\omega \end{pmatrix}, \\
M_{14} &= \begin{pmatrix} a\omega^3 & b \\ b & a\omega^7 \end{pmatrix}, &
M_{21} &= \begin{pmatrix} 0 & 1 \\ -1 & 0 \end{pmatrix}, &
M_{35} &= \begin{pmatrix} -\omega^{3} & 0 \\ 0 & \omega^7 \end{pmatrix}, &
M_{66} &= \begin{pmatrix} b\omega^9 & -a\omega^6 \\ -a\omega^{6} & -b\omega^{3} \end{pmatrix}, \\
M_{15} &= \begin{pmatrix} b & a\omega^5 \\ a\omega^5 & b \end{pmatrix}, &
M_{46} &= \begin{pmatrix} a & -b\omega^{5} \\ -b\omega^{5} & a \end{pmatrix}, &
M_{54} &= \begin{pmatrix} 1 & 0 \\ 0 & -1 \end{pmatrix}, &
M_{61} &= \begin{pmatrix} 0 & 1 \\ -1 & 0 \end{pmatrix}, \\
M_{16} &= \begin{pmatrix} -b\omega^{4} & a\omega \\ -a\omega^5 & -b\omega^2 \end{pmatrix}, &
M_{23} &= \begin{pmatrix} a\omega^2 & -b\omega^{9} \\ -b\omega^{3} & a \end{pmatrix}, &
M_{31} &= \begin{pmatrix} \omega^8 & 0 \\ 0 & -\omega^{6} \end{pmatrix}, &
M_{62} &= \begin{pmatrix} 0 & -\omega^{6} \\ 1 & 0 \end{pmatrix}, \\
M_{22} &= \begin{pmatrix} \omega^7 & 0 \\ 0 & -\omega^{9} \end{pmatrix}, &
M_{36} &= \begin{pmatrix} -a\omega^{2} & -b\omega^{5} \\ b\omega & -a\omega^4 \end{pmatrix}, &
M_{43} &= \begin{pmatrix} -b\omega^{4} & a\omega^7 \\ -a\omega^{3} & -b\omega^{6} \end{pmatrix}, &
M_{51} &= \begin{pmatrix} 0 & -\omega^4 \\ -1 & 0 \end{pmatrix}, \\
M_{24} &= \begin{pmatrix} -1 & 0 \\ 0 & \omega^6 \end{pmatrix}, &
M_{32} &= \begin{pmatrix} 0 & \omega^2 \\ \omega^2 & 0 \end{pmatrix}, &
M_{45} &= \begin{pmatrix} a\omega^4 & -b\omega^{7} \\ b\omega^7 & -a \end{pmatrix}, &
M_{53} &= \begin{pmatrix} b\omega^7 & -a \\ a & -b\omega^{3} \end{pmatrix}, \\
M_{26} &= \begin{pmatrix} b\omega^4 & a\omega^3 \\ a\omega^9 & -b\omega^{8} \end{pmatrix}, &
M_{34} &= \begin{pmatrix} a\omega & -b \\ b\omega^4 & a\omega^3 \end{pmatrix}, &
M_{41} &= \begin{pmatrix} a\omega & -b\omega^{4} \\ b\omega^6 & -a\omega^{9} \end{pmatrix}, &
M_{55} &= \begin{pmatrix} b\omega^2 & a\omega^5 \\ -a\omega^5 & b\omega^8 \end{pmatrix}.
\end{align*}


\section{Translation between classical codes, minimal support states and graph states}\label{app:graph_states}


In this Appendix, we present the correspondence between minimal support states obtained from classical codes and graph states.
Although we are mostly focused on AME states, corresponding to the classical maximal distance separable codes, we, in fact, use relations between any classical linear code, stabilizer states and graph states.
The methods used in this section were introduced in \cite{Helwig_2013}.

To make our discussion grounded, we focus on an exquisite example of AME$(4,5)$ state, presented in \eqref{eq:AME_45_def}.
First, let us notice that this state, as many other minimal support states, is based on a simple \textit{generator matrix of classical linear code} $G$, which is maximal distance separable $[4,2,3]_5$:
\begin{equation}
\label{eq:AME_45_generator_def}
|\Psi\rangle := |\text{AME}(4,5) \rangle = \frac{1}{5} \sum_{\mathbf{x} \in \mathbb{Z}_5^2} |G \mathbf{x}\rangle \text{ where } \mathbf{x} = (i,j)\,,  G^{\top} = \begin{pmatrix}
1&1&1&1\\
0&1&2&4
\end{pmatrix}
\end{equation}
with all operations performed modulo $5$.



For every code $G$ one can define a \textit{parity check} matrix $H$ such that rows of $H$ are linearly independent and orthogonal to the columns of $G$: $HG = 0$. 
In our case
\begin{equation}
H = \begin{pmatrix}
1&3&1&0\\
3&1&0&1
\end{pmatrix}~,
\end{equation}
which can be found by solving a simple system of linear equations. For classical error correction, columns of $G$ serve to encode the vector $\mathbf{x}$ as in \eqref{eq:AME_45_generator_def}, whereas rows of $H$ are used to detect errors.
Let us denote $X^\mathbf{z} = X^{z_0} \otimes X^{z_1} \otimes \cdots$ and analogously for $Z$. Then one can check by a direct computation that the generator matrix $G$ encodes $X$-stabilizers of $|\Psi\rangle$, namely for all $\mathbf{y} \in \mathbb{Z}_5^2$
\begin{equation*}
X^{G\mathbf{y}}|\Psi\rangle = \frac{1}{5} \sum_{x \in Z_5^2}|G \mathbf{x} + G \mathbf{y}\rangle = \frac{1}{5}  \sum_{x \in Z_5^2}|G \mathbf{x} \rangle = |\Psi\rangle~.
\end{equation*}
Moreover, parity check matrix $H$ encodes $Z$-stabilizers of $|\Psi\rangle$, for all $\mathbf{z} \in \mathbf{Z}_5^2$
\begin{equation*}
Z^{H^\top\mathbf{z}}|\Psi\rangle = \frac{1}{5} \sum_{x \in Z_5^2}\omega^{\mathbf{z}^\top HG \mathbf{x}} |G \mathbf{x} \rangle = \frac{1}{5}  \sum_{x \in Z_5^2}|G \mathbf{x} \rangle = |\Psi\rangle~.
\end{equation*}
Thus, we have a full set of stabilisers for our state, which is usually encoded into the so-called \textit{generator matrix of stabilizer state}:

\begin{equation*}
M = \left(\begin{array}{c|c} 
G^\top &0  \\
0 & H \\
\end{array}\right) = 
\left(\begin{array}{cccc|cccc}
1&1&1&1 &0&0&0&0 \\
0&1&2&4 &0&0&0&0 \\
0&0&0&0 &1&3&1&0\\
0&0&0&0& 3&1&0&1
\end{array}\right)~,
\end{equation*}
with the left block corresponding to $X$ operators and the right block to $Z$ operators.

All graph states can be represented using a stabilizer state generator matrix as well. However, for the graph states, the left block is an identity, whereas the right block is an adjacency matrix for the graph, with the numbers representing the multiplicity of $CZ$ operator on a given edge. To find the graph representation of the considered state, we must modify it using local operations. 

In \cite{bahramgiri2006graph}, the authors characterized the action of local \textit{Clifford} operators by linear operations on the generator matrix of stabilizer state. Each such action can be represented as $M' = U M Y$ where $U$ and $Y$ are invertible and $Y$ is a block diagonal matrix satisfying:
\begin{equation}
\label{clif_eq}
\begin{aligned}
&\hspace{6 cm} Y = \
\begin{pmatrix}
E& F\\
E' & F' \\
\end{pmatrix} \\
&E = \text{diag}(e_1,\cdots e_n), \;\; F = \text{diag}(f_1,\cdots f_n),\;\;
E' = \text{diag}(e_1',\cdots e_n'), \; F' = \text{diag}(f_1',\cdots f_n'), \\
&\hspace{5 cm} \text{and } \; e_i f_i' - e_i' f_i = 1 \; \text{ for all } i. 
\end{aligned}
\end{equation}
First we choose $U =\one$ and for $Y$ we set $E  = F'= \one$, $F = 0$ which ensures that condition \eqref{clif_eq} is satisfied, and $E' =\text{diag}(0,0,1,1)$ to make first block of $M$ nondegenerate
\begin{equation*}
M \to MY = 
\left(\begin{array}{cccc|cccc}
1&1&1&1 &0&0&0&0 \\
0&1&2&4 &0&0&0&0 \\
0&0&1&0 &1&3&1&0\\
0&0&0&1 &3&1&0&1
\end{array}\right)~.
\end{equation*}
Next we set $Y = \one$ and $U$ equal the inverse of first block of $M$ (in the field $\mathbb{Z}_5$), to obtain desired identity matrix
\begin{equation*}
\begin{aligned}
&\hspace{2 cm} U = \begin{pmatrix}
1&4&1&3\\
0&1&3&1\\
0&0&1&0\\
0&0&0&1\\
\end{pmatrix} ,
&M \to UM = 
\left(\begin{array}{cccc|cccc}
1&0&0&0 &0&1&1&3 \\
0&1&0&0 &1&0&3&1 \\
0&0&1&0 &1&3&1&0\\
0&0&0&1 &3&1&0&1
\end{array}\right)~.
\end{aligned}
\end{equation*}
Thus, we almost obtained the proper form. The final step is to eliminate the nonzero elements on the diagonal of the second block by the matrix $Y$ with blocks: $E = F' = \one$, $E'= 0$, $F = \text{diag}(0,0,-1,-1)$:
\begin{equation*}
M \to MY = 
\left(\begin{array}{cccc|cccc}
1&0&0&0 &0&1&1&3 \\
0&1&0&0 &1&0&3&1 \\
0&0&1&0 &1&3&0&0\\
0&0&0&1 &3&1&0&0
\end{array}\right)~.
\end{equation*}
The graph for the obtained graph state is presented in the Figure \ref{fig:graph_4_5}.

\section{Toolbox for exemplary AME states}
\label{app:AMEtoolbox}

\subsection{1-uniform states}
\begin{enumerate}
	\item Bell state $\ket{\mathrm{B}} = \frac{1}{\sqrt{2}}\big( \ket{00} + \ket{11} \big)$, which is LU-equivalent to graph states of the form $A - B$
	\item generalized Bell state $\ket{\mathrm{B}_d} = \frac{1}{\sqrt{d}}\sum_{j=0}^{d-1} \ket{jj}$, which is LU-equivalent to graph states of the form $A - B$
	\item $n$-qubit GHZ state $\ket{\mathrm{GHZ}_n} = \frac{1}{\sqrt 2}\big( \ket{0}^{\otimes n} + \ket{1}^{\otimes n} \big)$, which is LU-equivalent to graph states of the full graph form, which in turn are LU-equivalent to a star graph via local complementation~\cite{VandenNest_2004}
	\item $n$-qu$d$it GHZ state $\ket{\mathrm{GHZ}_n^d} = \frac{1}{\sqrt d}\sum_{j=0}^{d-1} \ket{j\cdots j}$, all of them are graph states, so LU-equivalent to stabilizer states
\end{enumerate}


\begin{figure}[ht]
\centering
\begin{tikzpicture}

\node at (0,0) {\includegraphics[width=0.2\linewidth]{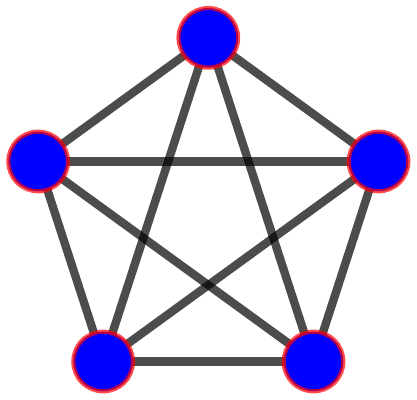}};

\node at (3.5,0) {\scalebox{2}{$\equiv$}};

\node at (7,0) {\includegraphics[width=0.16\linewidth]{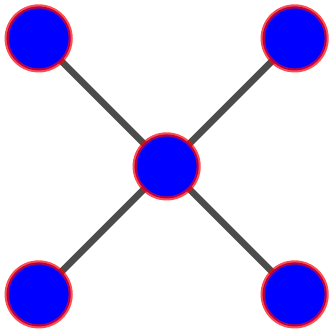}};

\end{tikzpicture}
\caption{The graphs corresponding to GHZ state of $N = 5$ parties. Every edge represents a qudit CZ gate 
Eq.~\eqref{eq:qudit_cz}
between corresponding qubits. $ N$-partite GHZ state is equivalent to two graph states -- the full graph and the star graph \cite{VandenNest_2004,Hein_2006}.}
\label{fig:graph_5_3}
\end{figure}

In the case of three systems, AME states are exactly the 1-uniform states.
As an example of an absolutely maximally entangled state of 3 subsystems with $d$ levels each, consider
\begin{equation}\label{ame3d}
	\ket{\mathrm{AME}(3,d)} = \frac{1}{d}\sum_{i,j=0}^{d-1} \ket{i,j,i\oplus j},
\end{equation}
where $\oplus$ denotes the addition modulo $d$.

Another noteworthy examples of one-uniform states are AME$(3,3)$ with maximally antysymetric structure
\begin{equation*}
    |\text{AME}(3,3)\rangle = \frac{1}{\sqrt{6}}(|012\rangle+|201\rangle+|120\rangle- |102\rangle- |201\rangle- |021\rangle )
\end{equation*}
as well as a special type of AME$(3,4)$ with higher measure of geometric entanglement than for the standard one \eqref{ame3d}
\begin{equation*}
    |\text{AME}(3,4)\rangle = \frac{1}{\sqrt{8}}(|022\rangle + |033\rangle + |120\rangle + |131\rangle + |212\rangle + |203\rangle + |310\rangle + |301\rangle).
\end{equation*}
Both of these states were derived in \cite{SG24}.

\subsection{2-uniform states}
{\color{\redCom} 
A $2$-uniform state can be related to classical or quantum orthogonal Latin squares. Two OLS($d$) determine the state AME(4,$d$),
while three OLS($d$) lead to the state AME(5,$d$). Four OLS(5)
yield a state of six subsystems of size $5$ that is $2$-uniform but not 3-uniform, thus not an AME state.}
In the subsequent constructions, we will use the qudit controlled Z gate, which, depending on the dimension $d$, means the following unitary operation
\begin{equation}\label{eq:qudit_cz}
    \text{CZ}_d = \sum_{i = 0}^{d-1}\ket{i}\bra{i}_k\otimes Z^i_l,
\end{equation}
acting between qudits $k$ and $l$.
Here, the qudit $Z$ gate is the standard clock operator, which in the computational basis $\ket{k}$ reads
\begin{equation}
    Z \ket{k} = \omega^{k}\ket{k},
\end{equation}
where $\omega = e^{2\pi i/d}$ is the $d$-th root of unity.
We shall start with the list of known constructions of AME(4,$d$).  

\begin{itemize}
	\item AME(4,2) state -- in this case, as already proven by Higuchi and Sudbery in 2000~\cite{Higuchi_2000_twoCouples}, absolutely maximally entangled states of four qubits do not exist. Some alternative proofs
    are recently provided in \cite{Huber_2025}.
    Equivalently, there are no $2$-unitary matrices of order $4$. 
    {\color{\redCom} However, strongly entangled $4$-qubit states, which do not satisfy AME property but maximize other entanglement measures were also analyzed in Refs.~\cite{VDMV02,Brown05,Borras07,GW10,EWZ16,GMO20,SG24,OT25}. }
    
	\item AME(4,3) state --
    appeared in unpublished notes by Sandu Popescu, while its construction
     was analyzed in 
    \cite{Helwig2012,Helwig_2013}.
    The standard solution related to
       OLS(3) reads \cite{helwig2013existApplic,
       Goyeneche2014},
    \begin{equation}
    \label{AME43}
    \ket{\mathrm{AME}(4,3)} = \frac{1}{3}\sum_{i,j = 0}^2 \ket{i,j,i\oplus j,i \oplus 2j}, 
	\end{equation} 
    where $\oplus$ denotes addition modulo $d=3$.

The corresponding generator matrix, defined  in Appendix~\ref{app:graph_states}, reads
\begin{equation}
\label{eq:AME43_gen}
	G_{4,3}^{\top} = \left[
	\begin{array}{cccc}
	1 & 0 & 1 & 1 \\
	0 & 1 & 1 & 2
	\end{array}
	\right],
\end{equation}
which explains the structure of expression
 (\ref{AME43}).

The state AME(4,3) is related to the classical ternary Hamming code $[4,2,3]_3$~\cite{Helwig_2013}
 and to the
quantum error correction codes $[\![4,0,3]\!]_3$ and $[\![3,1,2]\!]_3$, see ~\cite{Scott2003QECCentPow,Alsina_2021}. 
It forms a graph state associated 
with the graph presented in Fig.~\ref{fig:graph_4_3}.
Furthermore, this state is associated
to the magic square of size three with 
entries summing to $12$ in each row and column,
which written in ternary basis implies the following pair of orthogonal Latin squares, 
\begin{equation}
	\begin{bmatrix}
	0 & 5 & 7 \\ 
	4 & 6 & 2 \\ 
	8 & 1 & 3
	\end{bmatrix} = \begin{bmatrix}
	00 & 12 & 21 \\ 
	11 & 20 & 02 \\ 
	22 & 01 & 10
	\end{bmatrix}
      = 2\text{OLS}(3).
\end{equation}
This configuration, plotted with nine cards in Fig.  
\ref{tab:ame_4_4},
can be obtained from an orthogonal array OA(9,4,3,2).

\begin{figure}[ht]
\centering
\begin{tikzpicture}

\node at (0,0) {\includegraphics[width=0.18\linewidth]{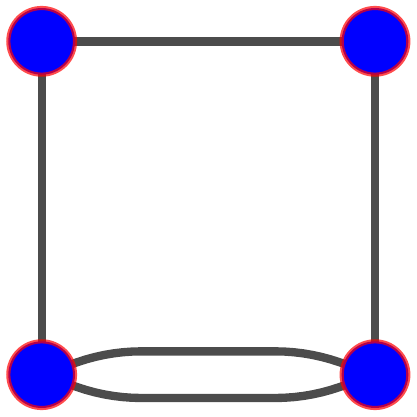}};

\node at (3.5,0) {\scalebox{2}{$\equiv$}};

\node at (7,0) {\includegraphics[width=0.18\linewidth]{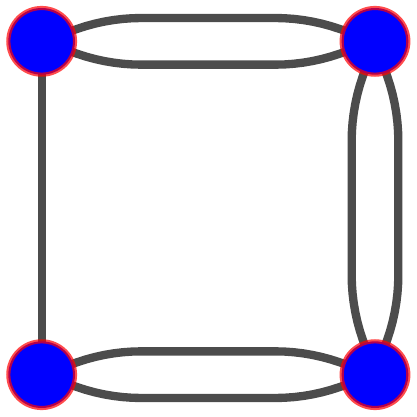}};

\end{tikzpicture}
\caption{Graphs corresponding to graph AME(4,3) state. Every edge represents a qutrit CZ$_3$ gate 
Eq.~\eqref{eq:qudit_cz}
between corresponding qubits, while the double edge 
represents the double usage of the gate, qutrit $(\text{CZ}_3)^2$. 
Both graphs are equivalent. The same graph represents also AME$(4,p)$ state for a prime dimension $p$. For $p=5$
we obtain AME state not equivalent to the one
associated with the graph 
shown in Fig. \ref{fig:graph_4_5}.}
\label{fig:graph_4_3}
\end{figure}

Finally, AME(4,3) state can be constructed using a 2-unitary permutation matrix $P_9$ \cite{Goyeneche2015} that satisfies strong Sudoku conditions 
\begin{equation}
	P_9 = \left[
	\begin{array}{ccc|ccc|ccc}
	1 &   &   &   &   &   &   &   &   \\
	  &   &   &   &   &   &   &   & 1 \\
	  &   &   &   & 1 &   &   &   &   \\
	\hline
	  &   &   &   &   & 1 &   &   &   \\
	  & 1 &   &   &   &   &   &   &   \\
	  &   &   &   &   &   & 1 &   &   \\
	\hline
	  &   &   &   &   &   &   & 1 &   \\
	  &   &   & 1 &   &   &   &   &   \\
	  &   & 1 &   &   &   &   &   &   \\
	\end{array}
	\right],
\end{equation}
where every empty position corresponds to a 0,
out of which the state can be created via 
\begin{equation}
	\ket{\mathrm{AME}(4,3)} = 
    {\color{\redCom}\frac{1}{3}}
    \sum_{i,j=0}^{d-1} \ket{ij}\otimes P_9 \ket{ij}.
\end{equation}
Since this  state,
determined by a $2-$unitary permutation matrix $P_9$,
contains a superposition of 9 states of computational basis, it belongs to the class of minimal support AME states.
It is conjectured 
that this state provides maximal
geometric entanglement among all states of the
four-qutrit system \cite{SG24},
which is not the case in other dimensions.

As proven by Rather et al.~\cite{Rather_2023}, all absolutely maximally entangled states of four qutrits are LU-equivalent to the one written above.

	\item AME(4,4) state -- in this case, there exist a standard solution for a pair of OLS(4), given by a stabilizer state depicted in Fig.~\ref{fig:graph_4_4}, found by Helwig~\cite{Helwig_2013}.
This minimal support state can be obtained by a 2-unitary permutation $P_{16}$  from Appendix D, item (iii) in \cite{Goyeneche2015}. 

For four ququarts, the minimal support constructions -- i.e., based on the 6912 possible pairs of orthogonal Latin squares -- are all equivalent up to local unitaries. An example of an AME$(4,4)$ state of minimal support is~\cite{Helwig_2013}
\begin{equation}\label{eq:AMEmin4-4}
\ket{\text{AME}(4,4)}=\frac{1}{4} \bigg(\sum_{i=0}^3 \ket{iiii} \,\,+ \sum_{\pi\in A_4} \ket{\pi(0)\pi(1)\pi(2)\pi(3)}\bigg),
\end{equation}
where $A_4$ denotes the set of even permutations of the digits $\{0,1,2,3\}$. This state is also locally equivalent to the eight-qubit graph state presented in Fig.~\ref{fig:graph_4_4}, considering local operations in the four-ququart system as two-qubit operations in the eight-qubit system.  

\begin{figure}[h]
\includegraphics[width=0.2\linewidth]{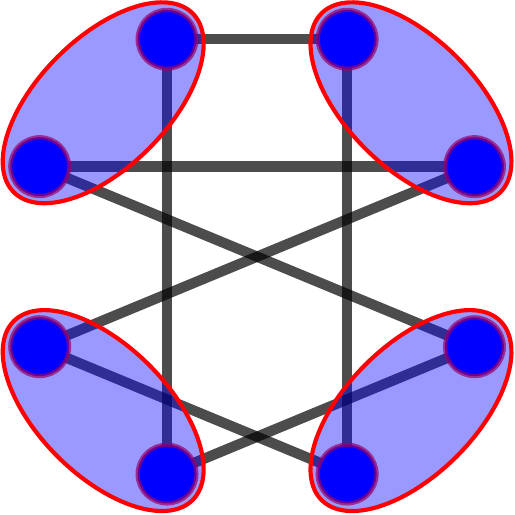}
\caption{
The graph corresponding to an eight-qubit state \cite{Helwig_2013}, which can be interpreted as AME$(4,4)$ state from \eqref{eq:AMEmin4-4}. Pairs of qubits, each grouped into ququart, are indicated by blue nodes in ellipses.
Every edge corresponds to a CZ gate between the corresponding qubits.  Note that this is not AME$(8,2)$ state, if each qudit is regarded as a single party.}
\label{fig:graph_4_4}
\end{figure}

A non-minimal-support construction that is not equivalent to the above one, is given by the two-unitary matrix $O_{16}$ that is also an orthogonal matrix
\begin{equation}
O_{16}=\frac{1}{2}
\begin{pmatrix}
\begin{array}{cccc|cccc|cccc|cccc}
1 & . & . & . & . & 1 & . & . & . & . & -1 & . & . & . & . & -1 \\
. & 1 & . & . & -1 & . & . & . & . & . & . & -1 & . & . &-1 & . \\
. & . & -1 & . & . & . & . & 1 & -1 & . & . & . & . & -1 & . & . \\
. & . & . & -1 & . & . & 1 & . & . & 1 & . & . & 1 & . & . & .  \\
\hline
. & 1 & . & . & -1 & . & . & . & . & . & . & 1 & . & . & 1 & . \\
-1 & . & . & . & . & 1 & . & . & . & . & 1 & . & . & . & . & -1 \\
. & . & . & -1 & . & . & 1 & . & . & -1 & . & . & -1 & . & . & . \\
. & . & -1 & . & . & . & . & -1 & 1 & . & . & . & . & -1 & . & . \\
\hline
. & . & -1 & . & . & . & . & -1 & -1 & . & . & . & . & 1 & . & . \\
. & . & . & 1 & . & . & 1 & . & . & -1 & . & . & 1 & . & . & . \\
-1 & . & . & . & . & -1 & . & . & . & . & -1 & . & . & . & . & -1 \\
. & -1 & . & . & -1 & . & . & . & . & . & . & -1 & . & . & 1 & . \\
\hline
. & . & . & -1 & . & . & -1 & . & . & -1 & . & . & 1 & . & . & . \\
. & . & 1 & . & . & . & . & -1 & -1 & . & . & . & . & -1 & . & . \\
. & 1 & . & . & 1 & . & . & . & . & . & . & -1 & . & . & 1 & . \\
1 & . & . & . & . & -1 & . & . & . & . & 1 & . & . & . & . & -1
\end{array}
\end{pmatrix}\,.
\label{eq:O16mat}
\end{equation}
To obtain the AME state, it suffices to use the standard construction with 2-unitary matrices
\begin{equation}
        \ket{\text{AME}(4,4)'} =
         {\color{\redCom}\frac{1}{4}}
        \sum_{i,j=0}^3 \ket{ij}\otimes O_{16} \ket{ij}.
    \end{equation}
Another infinite family of non-stabilizer states was found by Rather et al.~\cite{Rather_2022_v2}. These states correspond to a family of 2-unitary matrices $U_{16}$.

	\item AME(4,5) state -- in this case, due to variety of equivalent OLS, there are multiple minimal support AME states.
    One of them arises from the following orthogonal Latin squares
%
%
%
%

\begin{equation}
   \label{panOLS}
	\text{OLS}(5) = \left[
	\begin{array}{ccccc}
	{\color{blue}11} & 35 & 54 & 23 & 42 \\
	53 & {\color{blue}22} & 41 & 15 & 34 \\
	45 & 14 & {\color{blue}33} & 52 & 21 \\
	32 & 51 & 25 & {\color{blue}44} & 13 \\
	24 & 43 & 12 & 31 & {\color{blue}55} \\
	\end{array}
	\right] \mapsto
	\left[
	\begin{array}{ccccc}
	{\color{blue}00} & 24 & 43 & 12 & 31 \\
	42 & {\color{blue}11} & 30 & 04 & 23 \\
	34 & 03 & {\color{blue}22} & 41 & 10 \\
	21 & 40 & 14 & {\color{blue}33} & 02 \\
	13 & 32 & 01 & 20 & {\color{blue}44} \\
	\end{array}
	\right],
\end{equation}
where the two pairs of OLS are related by a translation of both indices $i \mapsto i-1$ and $j \mapsto j-1$. The latter form, treated as numbers
in quinary (pental) system yield a diabolic square  
of order five,
\begin{equation}
	\left[
	\begin{array}{ccccc}
	0 & 14 & 23 & 7 & 16 \\
	22 & 6 & 15 & 4 & 13 \\
	19 & 3 & 12 & 21 & 5 \\
	11 & 20 & 9 & 18 & 2 \\
	8 & 17 & 1 & 10 & 24 \\
	\end{array}
	\right]
\end{equation}
with peculiar properties:
Not only sums in each row and column
are equal 60, 
but this is also the case for any sum
along its any left and right (generalized) diagonals.
This property is thus inherited by the
OLS (\ref{panOLS}), which are called 
{\em pandiagonal}, as
there is no repetition of any symbols
in each row, each column, and
its 5 left and 5 right generalized diagonals.
Hence  the superposition
of five states taken along any 
generalized diagonal of the square 
(treated as a torus) 
is unitarily similar to the superposition
along the main color diagonal,
which forms the GHZ state,
$|11\rangle + \dots + |55\rangle$.

Since entire square can be 
stratified into five generalized diagonals,
this pandiagonal AME(4,5) state can be 
written as 
\begin{equation}
\label{eq:AME_45_def}
	\ket{\mathrm{AME}(4,5)} = \frac{1}{5}\sum_{i,j=0}^4 \ket{i,i\oplus j,i\oplus 2j, i \oplus 4j} = \frac{1}{5}\sum_j \big(\mathbb{I} \otimes X^j \otimes X^{2j} \otimes X^{4j}\big)\ket{\text{GHZ}_5^4},
\end{equation}
with additions $\oplus$ modulo 5, where $X$ is the cyclic permutation matrix of order five, 
$X|i\rangle=|i \oplus 1\rangle$.
For instance,
the term  $j=0$ in the expression on the right
hand side corresponds to the main diagonal
and the original $|GHZ_5^4\rangle$ state,
while the term $j=1$ 
describes states $|2153\rangle +|3214\rangle +
|4325\rangle \dots$, 
determined by the
  generalized subdiagonal of the left square in
 (\ref{panOLS}). 
Hence this $2$-uniform state AME(4,5)
can be  formed as a superposition of five
$1$-uniform states, which are locally equivalent to the generalized GHZ state, $\ket{\text{GHZ}_5^4} = \frac{1}{\sqrt{2}}\sum_{i=0}^4 \ket{iiii}$.
The corresponding generator $G_{4,5}$,
discussed in Appendix~\ref{app:graph_states},
reads
\begin{equation}
\label{G45}
	G_{4,5}^\top = \left[
	\begin{array}{cccc}
	1 & 1 & 1 & 1 \\
	0 & 1 & 2 & 4\\
	\end{array}
	\right],
\end{equation}




while the graph which defines this AME state is depicted in Fig.~\ref{fig:graph_4_5}.

\begin{figure}[ht]
\centering
\includegraphics[width=0.2\linewidth]{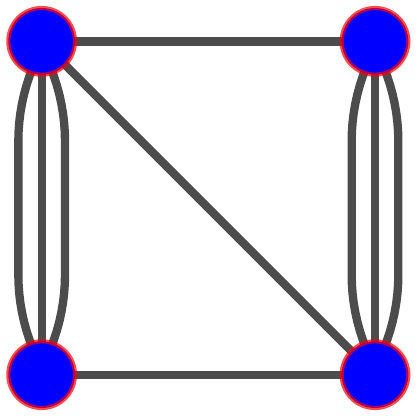}
\caption{The graph corresponding to the state AME(4,5) from \eqref{eq:AME_45_def}.
As previously, the multiplicity of edges corresponds to the power of the relevant $\text{CZ}_5$ gate, defined by Eq.~\eqref{eq:qudit_cz}}
\label{fig:graph_4_5}
\end{figure}


	\item AME(4,6) state -- this case was already discussed in details in the main body of the text, see Sec.~\ref{sec:unconventional}.
    Here, we provide the references for the original golden state~\cite{Rather_2022,Zyczkowski_2023}, the subsequent 
    solutions based on biunimodular sequences \cite{Rather_2024},
    Hadamard $H_{36}$ solution~\cite{Bruzda_2024Two-Had}, as well as the recent artisanal solution~\cite{Gross_2025}.
   The latter work provides 
an equivalent reformulation of the existence
of AME$(4,d)$ in
terms of certain quasi-orthogonal decompositions of matrix algebras.   To demonstrate similarities and differences
between various solutions we show 
in Fig.~\ref{fig:ame_4_6_visualization}
building blocks of the corresponding 2-unitary matrices.

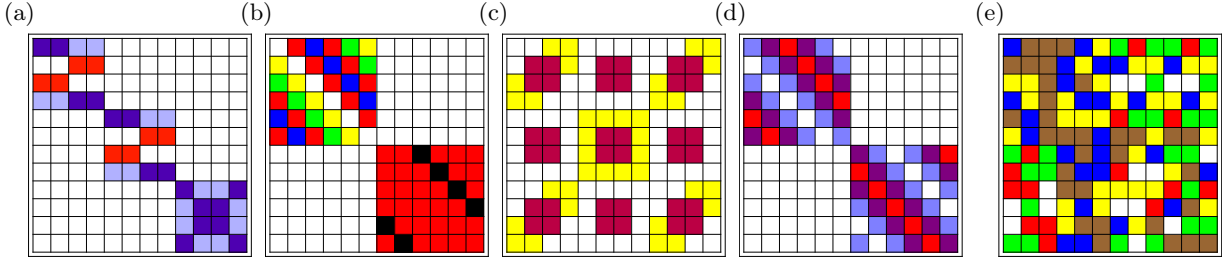
\begin{figure}[h]
 \begin{tikzpicture}
    \newcommand\x{3.15}
    \newcommand\z{0.18\linewidth}
    \newcommand\xx{-1.6}
    \newcommand\y{1.75}
       \node at (\xx, \y) {(a)};
       \node at (\xx+\x, \y) {(b)};
       \node at (\xx+2*\x, \y) {(c)};
       \node at (\xx+3*\x, \y) {(d)};
       \node at (\xx+4*\x+0.3, \y) {(e)};
       \node at (0,0) {\includegraphics[width=\z]{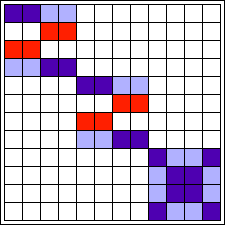}};
       \node at (\x,0) {\includegraphics[width=\z]{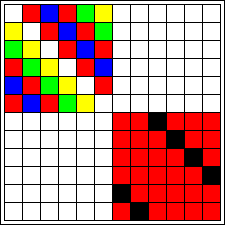}};
       \node at (2*\x,0) {\includegraphics[width=\z]{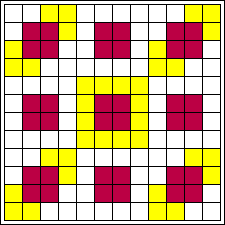}};
       \node at (3*\x,0) {\includegraphics[width=\z]{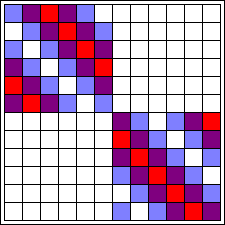}};
       \node at (4*\x+0.3,0) {\includegraphics[width=\z]{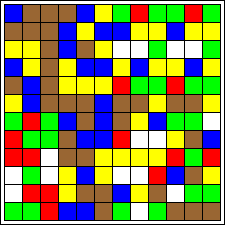}};
    \end{tikzpicture}
\caption{
Visualization of exemplary solutions of $2$-unitary matrices $U_{36}$
representing non-equivalent states AME$(4,6)$ -- to save the
place only a diagonal block of size $12$ of each matrix is shown.
(a) original golden state~\cite{Rather_2022,Zyczkowski_2023}
  obtained by a suitable permutation of block diagonal matrix
  with 9 blocks of order four (only $3$ such blocks are shown here),  with three colors representing three different amplitudes, while phases are multiples of $\omega_{20}$
   with $\omega_n=\exp(i 2 \pi/n)$;
(b) solution of the form (\ref{eq:U_Diag_Max_Ent})
based on bi-unimodular sequence \cite{Rather_2024} 
  with six blocks of order six,
  phases (\ref{phases})  being multiples  of $\omega_3$ 
  and five amplitudes marked by colors;
  (c) solution from~\cite{Bruzda_2024Two-Had} with two different amplitudes;
(d) artisanal solution constructed analytically in~\cite{Gross_2025}
with three amplitudes and phases being multiples of $\omega_{3}$;
and
(e) complex Hadamard matrix~\cite{Bruzda_2024Two-Had}
with all amplitudes equal
and phases being multiples of $\omega_6$, in this panel
represented by different colors.
}
\label{fig:ame_4_6_visualization}
\end{figure}

	\item AME(4,7) state -- this state can be obtained in a standard way from 2 OLS(7), which
     leads to a $2$--unitary permutation matrix
     $P_{49}$. Furthermore, a non-stabilizer solutions were found recently \cite{bistron_2023},
     using 2-unitary matrix $U_{49}$ that is not locally equivalent to  $P_{49}$.
     It defines a $2$-parameter locally non-equivalent family of AME states via
    \begin{equation}
    \label{eq:ame_47_constr}
        \ket{\text{AME}(4,7)} =
         {\color{\redCom}\frac{1}{7}}
        \sum_{i,j=0}^6 \ket{ij}\otimes U_{49} \ket{ij}.
    \end{equation}
    The building blocks of the above 2-unitary matrix are shown in Fig.~\ref{fig:dimension_7}(a).
    \item AME(4,8) state -- as in all cases of four-party AME states apart from qubits and quhexes, we can construct the AME states using orthogonal Latin squares. 
    
    \item  AME(4,9) state -- there exist several minimal support solutions generated by various permutation matrices $P_{81}$, derived from OLS$(9)$, among them tensor product $P_{81} = P_{9}\otimes P_{9}$ with $P_9$ defining AME$(4,3)$ state.
    Non-stabilizer solutions can be constructed using 2-unitary $U_{81}$ from~\cite{bistron_2023}, which provides $4$-parameter locally non-equivalent family of AME states, interpolating between minimal-support solutions, constructed as in equation \eqref{eq:ame_47_constr}.
    The building blocks of  $U_{81}$ are shown in Fig.~\ref{fig:dimension_7}(b).

\begin{figure}[h!]
    \centering
    \begin{tikzpicture}
    \newcommand\x{2.9}
    \newcommand\z{1.9}
    \newcommand\xx{-0.3}
    \newcommand\y{2.8}
    \newcommand\s{0.11\linewidth}

       \node at (-1.7,1.45) {(a)};
       \node at (-1.7,1.45-0.3-\y) {(b)};

       \node at (6.32,0) {\includegraphics[width=0.92\linewidth]{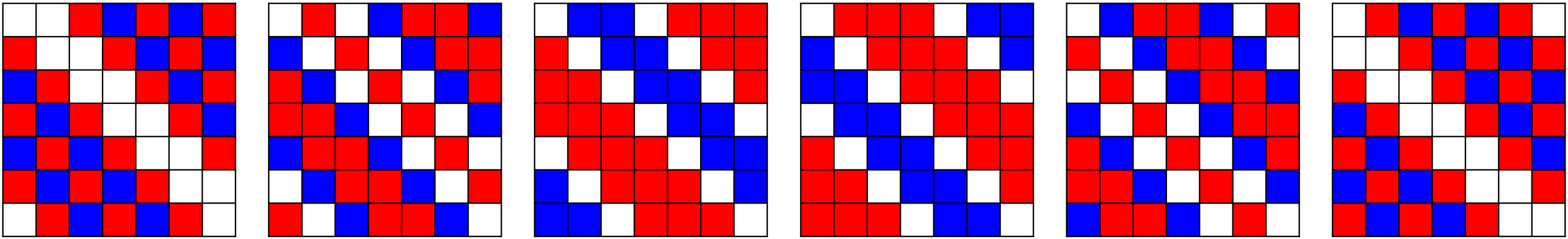}};

       \node at (\xx,-\y) {\includegraphics[width=\s]{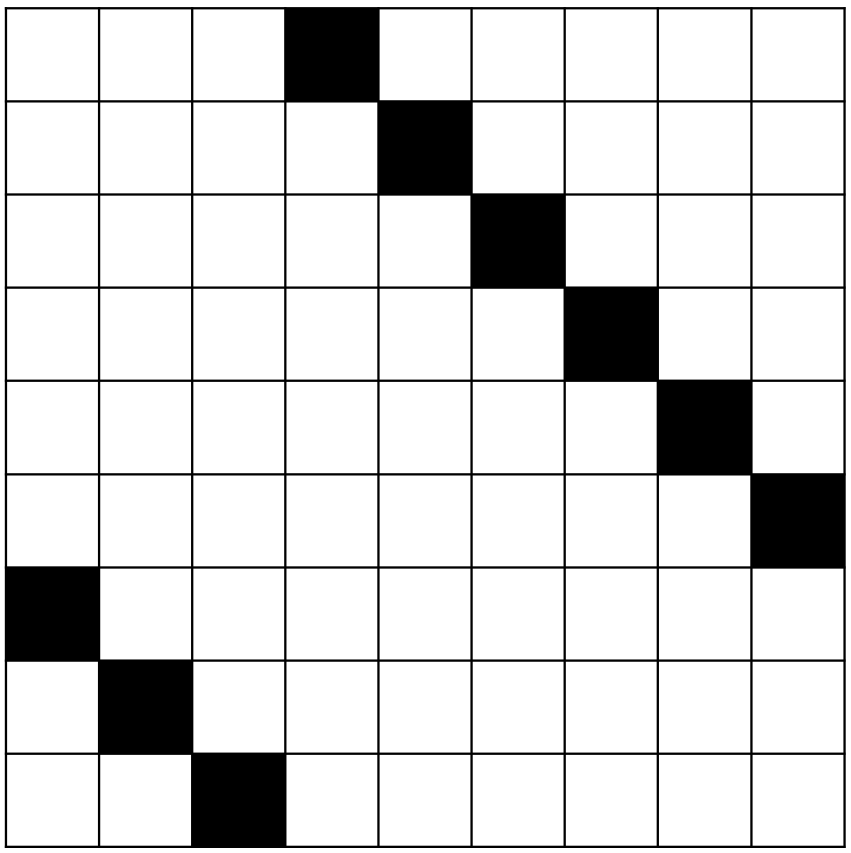}};
       \node at (\z+\xx,-\y) {\includegraphics[width=\s]{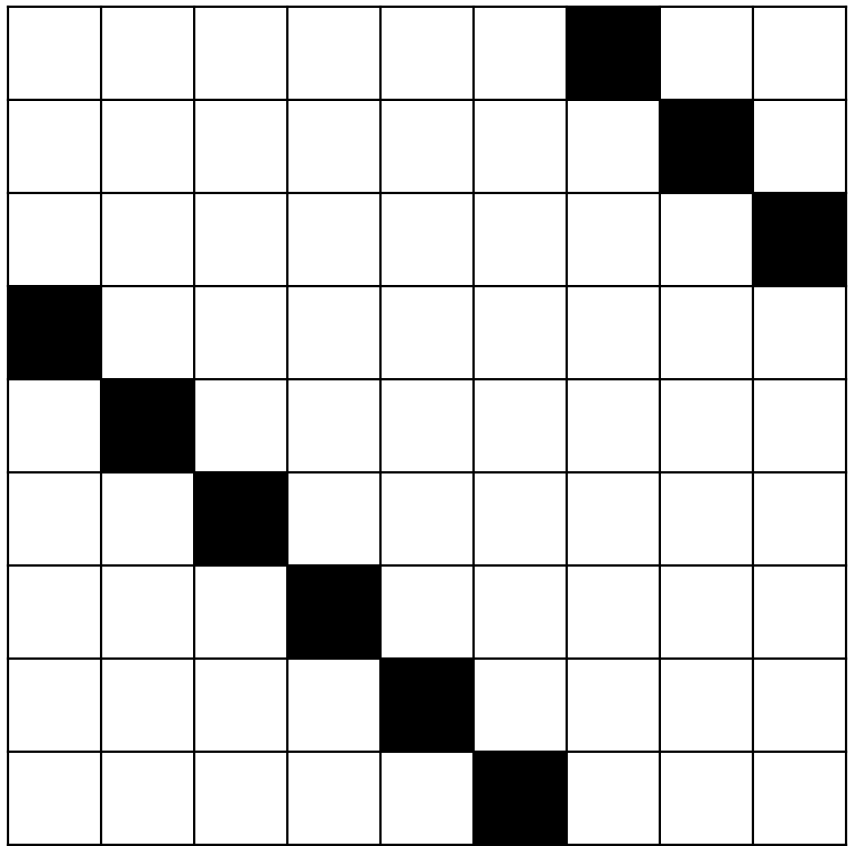}};
       \node at (2*\z+\xx,-\y) {\includegraphics[width=\s]{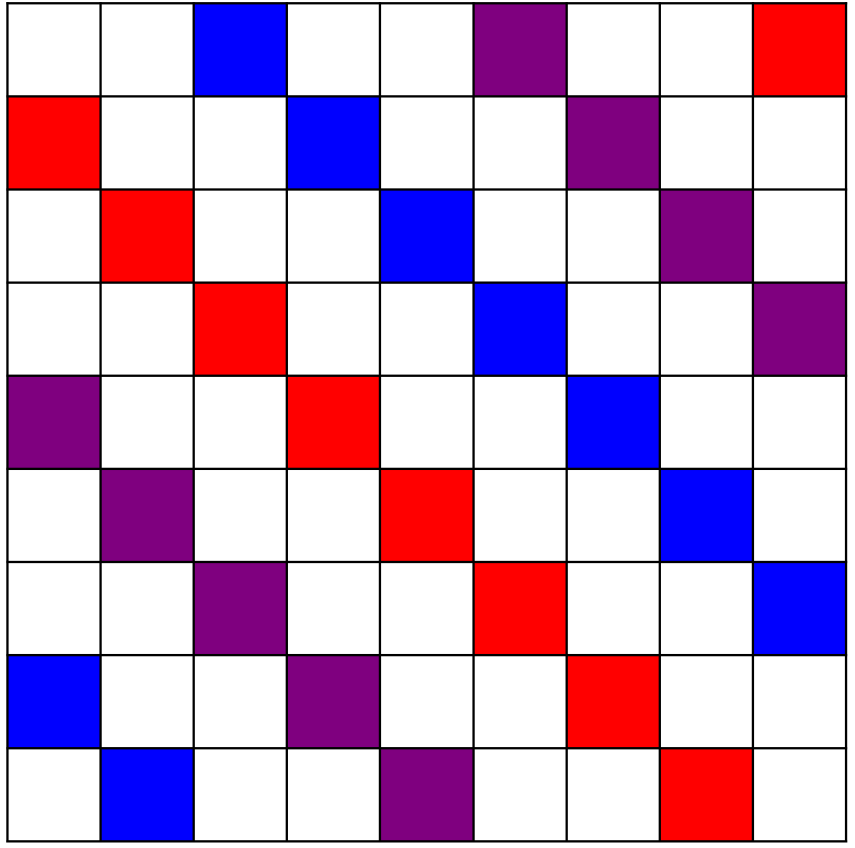}};
       \node at (3*\z+\xx,-\y) {\includegraphics[width=\s]{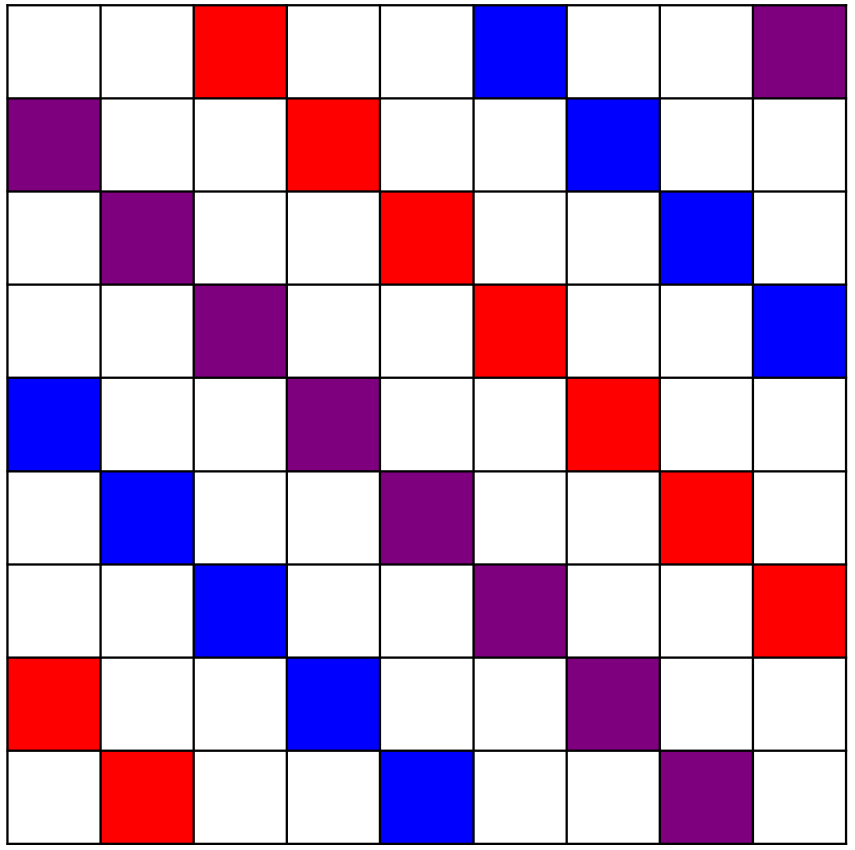}};
       \node at (4*\z+\xx,-\y) {\includegraphics[width=\s]{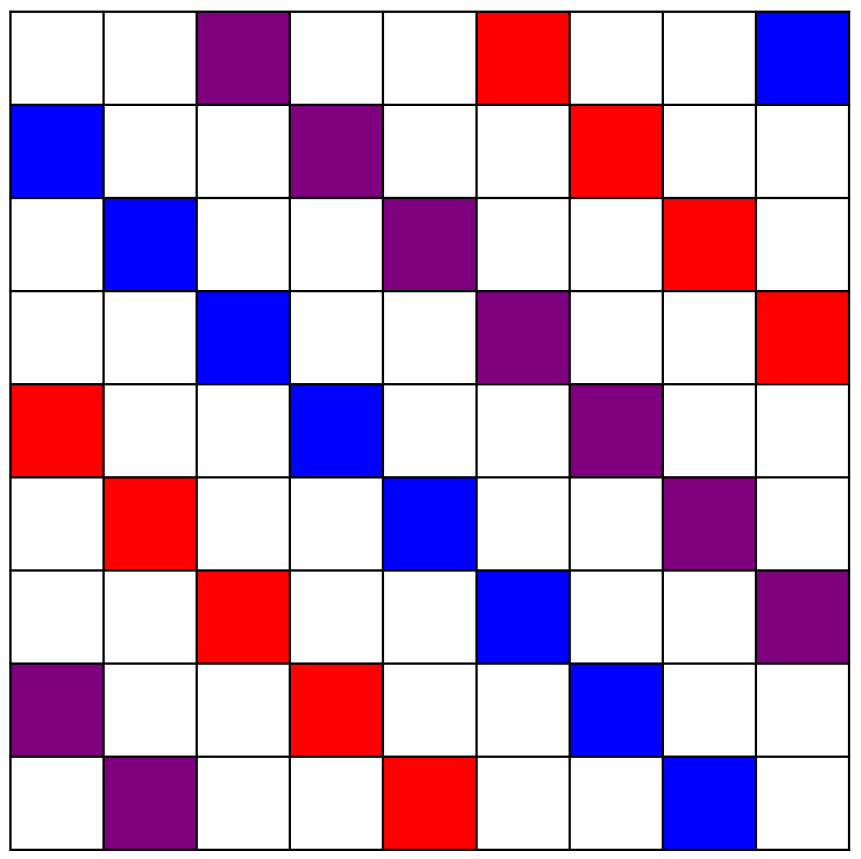}};
       \node at (5*\z+\xx,-\y) {\includegraphics[width=\s]{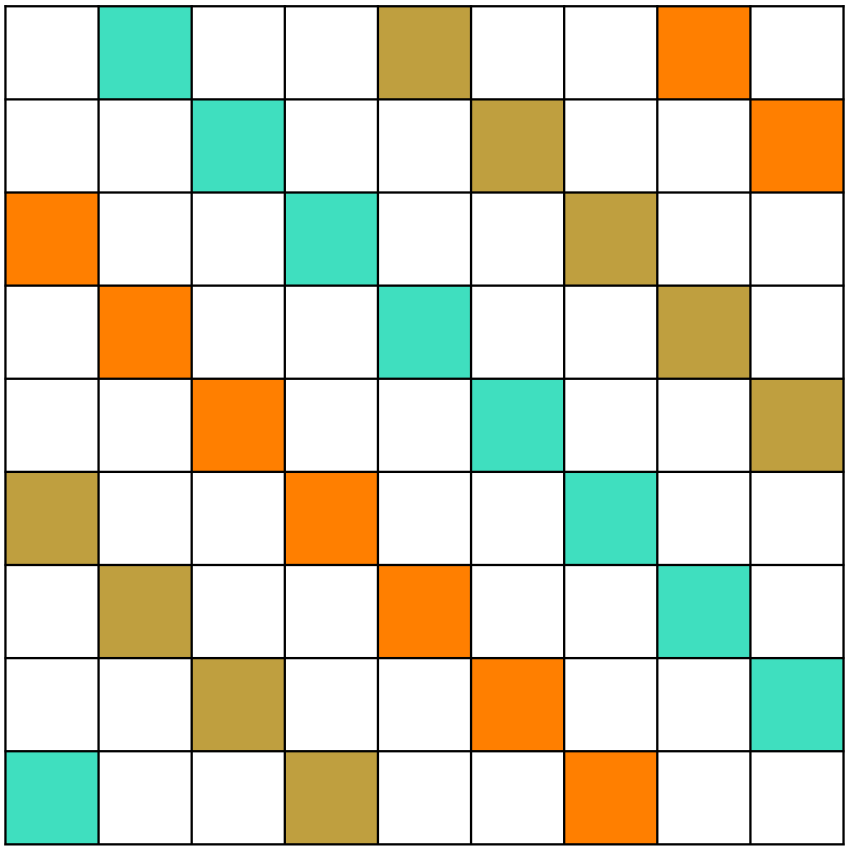}};
       \node at (6*\z+\xx,-\y) {\includegraphics[width=\s]
       {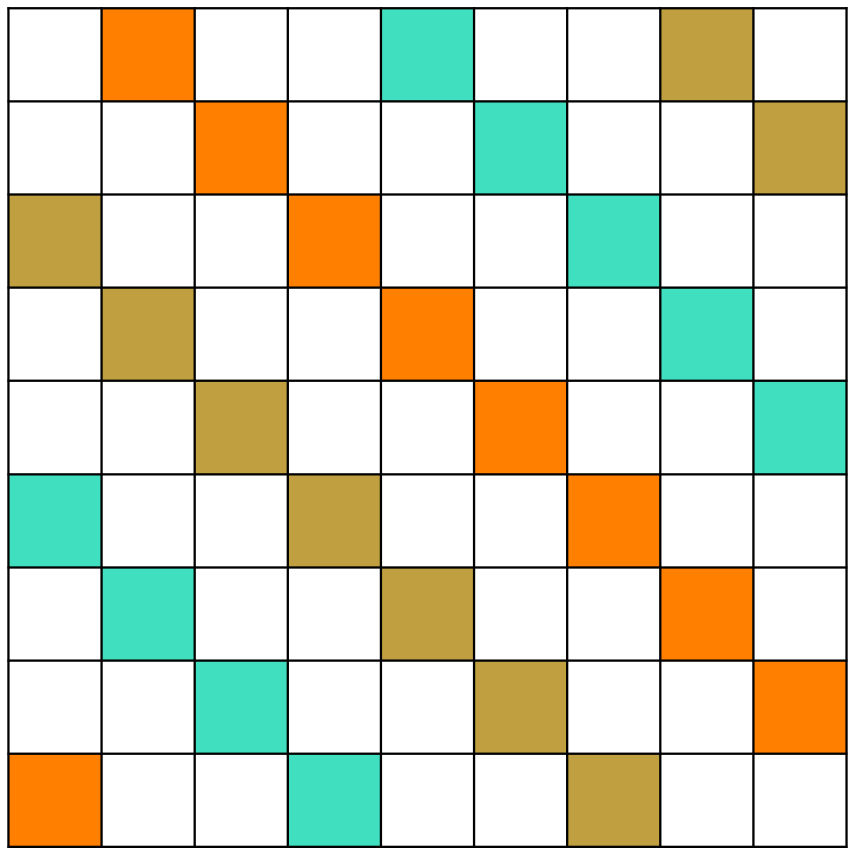}};
       \node at (7*\z+\xx,-\y) {\includegraphics[width=\s]{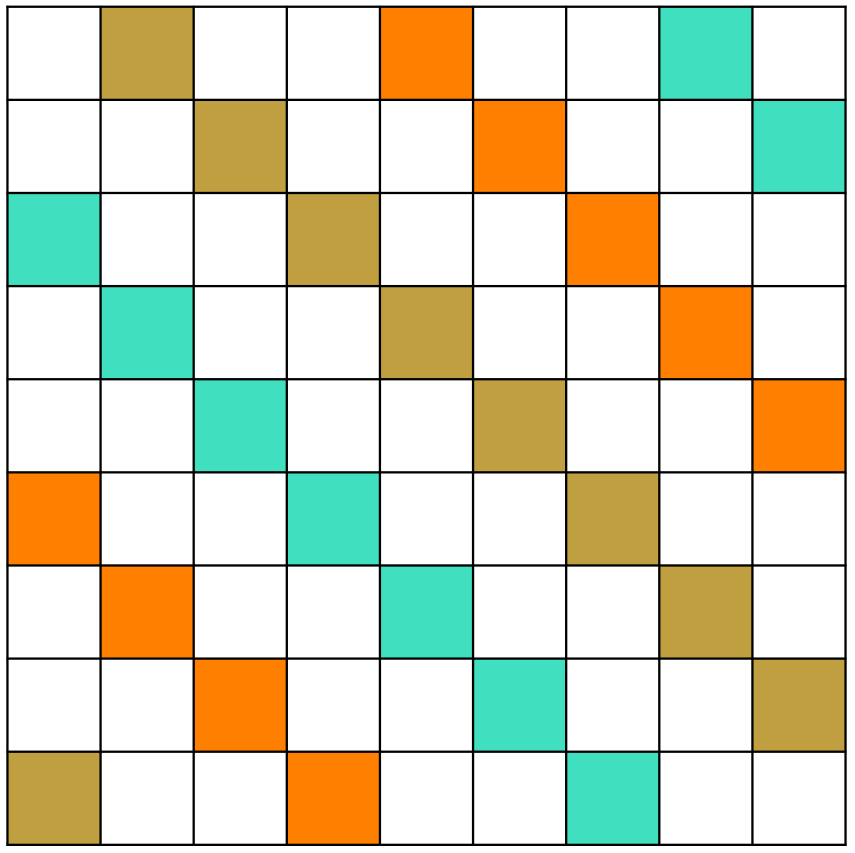}};
    \end{tikzpicture}
    \caption{
    Visual representation of the structure  
    of 2-unitary matrices corresponding to 
      four-party non-stabiliser    AME states.
    (a) Amplitudes describing  six building blocks of $U_{49}$,
    each of size $7$, 
    determining AME$(4,7)$ state, with the first block, proportional to identity, omitted.
    (b) Amplitudes of eight building blocks of order $d=9$
    (identity omitted) which allow one to construct $2$-unitary matrix $U_{81}$
    and AME$(4,9)$ state as shown in \cite{bistron_2023}.
    }
    \label{fig:dimension_7}
\end{figure}


\item AME(4,$d$) state -- 
in the general case of $d$-dimensional systems, we can create AME(4,$d$) states of the stabilizer class via generator matrix \eqref{eq:AME43_gen}:
\begin{equation}
    \ket{\text{AME} (4,d)}= \frac{1}{d}\sum_{i,j = 0}^{d-1} \ket{i,j,i\oplus j,i \oplus 2j},
\end{equation}
with addition modulo $d$,
which implies a $2-$unitary permutation matrix
  $P_{d^2}$, whenever $d\neq 2,6$.
The above state defines a quantum error correcting code $[\![4,0,3]\!]_d$ and a quantum error detection code $[\![3,1,2]\!]_d$ also
written $((3,d,2))_d$~\cite{Alsina_2021}.

As mentioned above, 
non-stabilizer $AME(4,d)$ states were constructed 
for $d=4,6,7$ and $9$. Furthermore,
techniques to construct 
$2$-unitary Hadamard matrices
developed in ~\cite{Bruzda_2024Two-Had}
where extended  in~\cite{Gross_2025}.
It was shown there that for any dimension $d$ 
a doubly perfect
sequence gives rise to a $2$-unitary complex
Hadamard matrix
of dimension $d^2$. This approach allows one to construct 
explicit examples of such matrices for all $d$,
except possibly those of the form $d=2m$, 
where $m$ is neither divisible by $2$ nor by $3$,
so the first cases not covered are $d = 10, 14, 22$.

\end{itemize}




\vspace{0.5cm}

Let us move on to $2$-uniform  AME states of $N=5$ subsystems.

\vspace{0.5cm}
\begin{itemize}
    \item 

AME(5,2) state -- was already analyzed 
in an early work by  Laflamme et al. \cite{Laflamme1996} and later in \cite{Brown05}.

In this  case it is useful to define stabilizer states using quantum orthogonal arrays (QOA)~\cite{Goyeneche_2018}, which extend the standard notion of orthogonal arrays, already introduced in Sec.~\ref{sec:MinSupMDS}. 
As their generalization, a quantum orthogonal array QOA($r,c+q,d,s$) is an arrangement of $r$ rows, $c$ classical columns, $q$ quantum 'entangled' columns, dimension $d$, and strength $s$.
The difference from the classical case is that now some of the columns are quantum, meaning they form a state from $q$-dimensional Hilbert space.
Therefore, the definition of the QOA is slightly more involved, requiring partial traces over the subsystems.
For the exact definition, we refer the interested reader to~\cite{Goyeneche_2018}, while here we shall use QOA to define AME states.
As an example, consider 
a quantum array
\begin{equation}\label{eq:QOA}
	 \text{QOA}(4,3_c+2_q,2,2) = \left[
	\begin{array}{ccc|cc}
	0 & 0 & 0 & \ket{\phi^+} \\
	0 & 1 & 1 & \ket{\psi^+} \\
    1 & 0 & 1 & \ket{\psi^-} \\
    1 & 1 & 0 & \ket{\phi^-} \end{array}
	\right],
\end{equation}
where two 'quantum' columns are formed
by four maximally entangled states of the Bell basis, 
\begin{equation}
        \ket{\psi^{\pm}} = \frac{1}{\sqrt{2}}\big(\ket{01}\pm \ket{10}\big) \quad \text{and} \quad 
        \ket{\phi^{\pm}} = \frac{1}{\sqrt{2}}\big(\ket{00}\pm \ket{11}\big).
\end{equation}
Observe that the first three ``classical'' columns correspond to an orthogonal array OA(4,3,2,2) \cite{Pa59}.

Finally, to construct AME(5,2) from the above quantum orthogonal array, we apply the same algorithm as in Sec.~\ref{sec:MinSupMDS}, namely, read out each row separately
\begin{equation}\label{eq:ame_5_2}
    \ket{\text{AME}(5,2)} = \frac{1}{2}\big(\ket{000}\otimes\ket{\phi^+} + \ket{011}\otimes\ket{\psi^+} + \ket{101}\otimes\ket{\psi^-} + \ket{110}\otimes\ket{\phi^-} \big)
\end{equation}
with the corresponding cycle graph depicted in
Fig.~\ref{fig:graph_states_AME}c.
Furthermore, a new family of AME(5,2) states not equivalent to to the standard solution
was recently found~\cite{Ramadas_2024},
\begin{equation}
    \ket{\text{AME}(5,2)'} =\frac{1}{\sqrt{2}}\rl{ \cos\theta \ket{\tilde{0}} + \sin\theta \ket{\tilde{1}} },
\end{equation}
where 
\begin{equation}
    \begin{split}
        \ket{\tilde{0}}  =& \frac{1}{\sqrt{8}} \left( \ket{00000} + \ket{00111} - \ket{01010} + 
 \ket{01101} \right.\\& \left. - \ket{10001} - \ket{10110} - 
 \ket{11011} + \ket{11100} \right)
    \end{split}
\end{equation}
and 
\begin{equation}
    \begin{split}
        \ket{\tilde{1}}  =&\frac{1}{\sqrt{8}} \left( \ket{00011} + \ket{00100} - \ket{01001} + 
 \ket{01110}\right.\\& \left.+ \ket{10010} + \ket{10101}+ 
 \ket{11000} - \ket{11111} \right).
    \end{split}
\end{equation}

\item AME(5,4) state -- in this case, an AME state of five ququarts can be defined using 3 mutually orthogonal Latin squares of size $4$~\cite{Goyeneche2015}, which form a full set of orthogonal Latin squares in this dimension, see Fig.~\ref{fig:OLS_AME_4_4}, panels (b) and (c). The corresponding state reads
\begin{equation}
    \begin{split}
|\text{AME}(5,4)\rangle=\frac{1}{4}\big (&|00000\rangle+|10312\rangle+|20231\rangle+|30123\rangle+\nonumber\\
&|01111\rangle+|11203\rangle+|21320\rangle+|31032\rangle+\nonumber\\
&|02222\rangle+|12130\rangle+|22013\rangle+|32301\rangle+\nonumber\\
&|03333\rangle+|13021\rangle+|23102\rangle+|33210\rangle\big )\nonumber.
    \end{split}
\end{equation}

\begin{figure}[h]
\centering
\begin{tikzpicture}

\node at (-2.8,1.) {(a)};
\node at (0,0) {\includegraphics[width=0.3\linewidth]{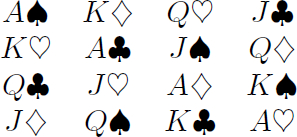}};


\node at (3.4,1.) {(b)};
\node at (6.2,0) {\includegraphics[width=0.3\linewidth]{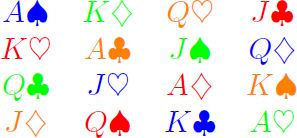}};


\node at (9.9,1.) {(c)};
\node at (11.6,0)
{\includegraphics[width=0.155\linewidth]{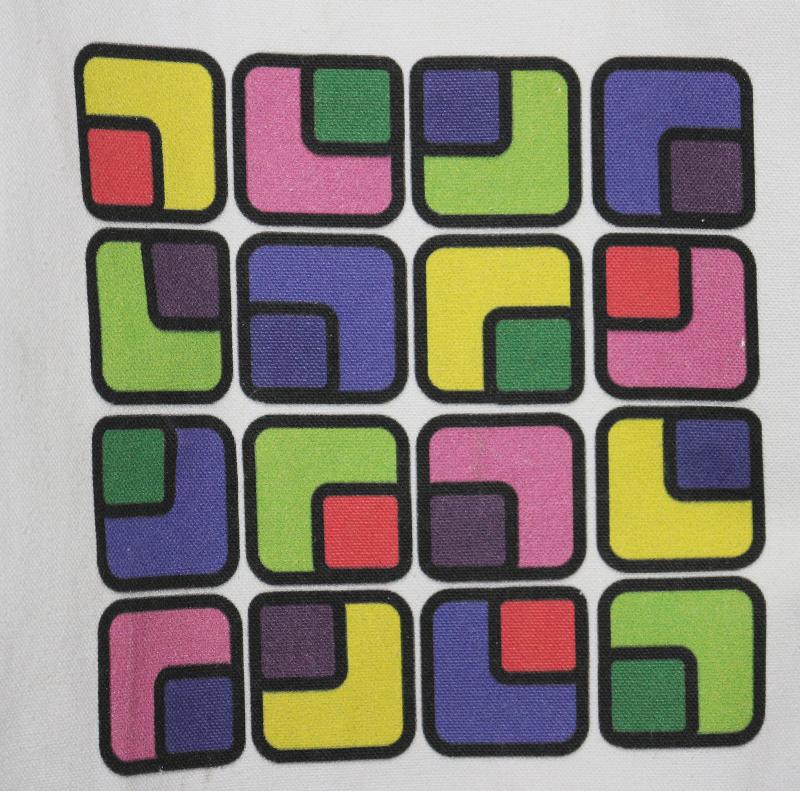}};

\end{tikzpicture}
\caption{
Squares of size $4$ containing $16$
objects with: (a) two features form 2OLS(4)
and AME(4,4) state; (b) three features form 3OLS(4);
(c) pattern decorating a bag offered to participants of a meeting of {\sl American Mathematical Society} also implies 3OLS(4) and a state AME(5,4).
    In parts (b) and (c), three Latin squares are encoded, respectively, in: (1) the rank of the card/orientation of the inset, (2) the suit of the card/color of the inset, and (3) the color of the entire card/large element.
    }
\label{fig:OLS_AME_4_4}
\end{figure}
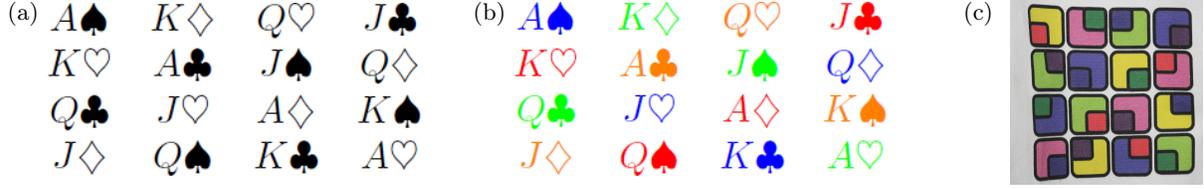

Equivalently, this state is constructed from the homogeneous superposition between the codewords of an MDS $C[5,2,4]_4$ code of the type {\em Reed-Solomon}~\cite{ReedSol1960Codes}, explicitly given in~\cite{Sloane2007LibArrays}. and is a self-dual pure quantum code $C[\![5,0,3]\!]_4$. In turn, it defines a quantum code $C'[\![4,1,2]\!]_4$ code with codewords
\begin{align}
\ket{0_L}&=\frac{1}{2}\big (
|0000\rangle
+|1111\rangle
+|2222\rangle
+|3333\rangle\big )\\
\ket{1_L}&=\frac{1}{2}\big (
|0312\rangle
+|1203\rangle
+|2130\rangle
+|3021\rangle\big )\\
\ket{2_L}&=\frac{1}{2}\big (
|0231\rangle
+|1320\rangle
+|2013\rangle
+|3102\rangle\big )\\
\ket{3_L}&=\frac{1}{2}\big (
|0123\rangle
+|1032\rangle
+|2301\rangle
+|3210\rangle\big ).
\end{align}
One verifies that all the codewords above are 1-uniform states, which follows from the fact that they define a code $C'$ with quantum distance $2$.

\item
AME(5,$d$) state -- for a more general case of five qu$d$its, one can use QOA($d^2,3_c+2_q,d,2)$ to construct an AME state, see Eq.~(36) in~\cite{Goyeneche_2018}.
Furthermore, a similar construction as for 4 parties holds: the state~\cite{rico2020absolutely}
\begin{equation}\label{eq:AMEmin(5,d)odd}
\ket{\text{AME}(5,d)}=\frac{1}{d}\sum_{i,j=0}^{d-1}\ket{i}\ket{j}\bigotimes_{s=1}^{3}\ket{i\oplus   n_s\cdot j},
\end{equation}
with $n_s\in\mathbb{N}\setminus 0$ is an AME state when $\{n_s\}$ and any subtraction between them $mod(d)$ are all coprime with $d$. There exists a suitable choice of $\{n_s\}$ fulfilling this condition when $d$ is not a multiple of 2 nor 3.
In particular, there exist two families of AME(5,$d$) states
\begin{equation}
    \ket{\text{AME}'(5,d)} =\dfrac{1}{d} \sum_{i,j=0}^{d-1} \ket{i,j,i\oplus j,2i\oplus j,3i\oplus j}, 
\end{equation}


which is absolutely maximally entangled whenever $d$ is prime.
The second family, which is AME for all dimensions $d$, reads
\begin{equation}
\ket{\text{AME}''(5,d)} =\dfrac{1}{\sqrt{d^3}} \sum_{i,j,k=0}^{d-1} \omega^{(3i+j) k} 
\ket{i,j,i\oplus j,2i\oplus j \oplus k,k}.
\end{equation}
In both of the above equations, $\oplus$ denotes addition modulo $d$.
Interestingly, these two families are not LU-equivalent~\cite{Burchardt_2020}.
\end{itemize}

\vspace{0.5cm}


Before discussing 3-uniform AME states of six parties,
as a warm up we shall 
discuss
2-uniform states of these systems. Let us start 
presenting a family of 
 six-qudit 
 two-uniform 
 states that are not necessarily AME states~\cite{rico2020absolutely}
\begin{equation}\label{eq:AMEmin(6,d)odd}
\begin{gathered}
\ket{\psi(6,d)}=\frac{1}{\sqrt{d^3}}\sum_{i,j,k=0}^{d-1}\ket{ijk}\bigotimes_{s=1}^{3}\ket{\alpha_s i\oplus\beta_s j\oplus\gamma_s k}.
\end{gathered}
\end{equation}
Positive integer 
constants $\alpha_s,\beta_s,\gamma_s$
are defined by 
the matrix
\begin{equation}
O=\begin{pmatrix}
\alpha_1 & \beta_1 & \gamma_1\\
\alpha_2 & \beta_2 & \gamma_2\\
\alpha_3 & \beta_3 & \gamma_3
\end{pmatrix},
\end{equation}
whose determinant is nonzero and coprime with $d$. Note that $O$ is a minor of the generator matrix $G^\top = (\mathbb{I}_3|O)$. Although the result is not always an AME state, an instance leading to 2-uniform states for odd local dimension $d$ reads,
\begin{equation}
O=\begin{pmatrix}
1 & 1 & 1\\
1 & d-1 & 1\\
1 & 1 & d-1
\end{pmatrix}.
\end{equation}

As a special example of 2-uniform six party states~\cite{Goyeneche_2018}, consider
\begin{equation}
|\psi'(6,d)\rangle=\frac{1}{\sqrt{d^3}}\sum_{i,j=0}^{d-1}\ket{i,j,i\oplus j,i\oplus 2j}\ket{\phi_{i,j}},
\end{equation}
defined by the generalized Bell basis
\begin{equation}
    \ket{\phi_{i,j}}= \sum_{l=0}^{d-1} \omega^{il} \ket{l\oplus j,l},
\end{equation}
where $\omega=e^{2\pi \mathrm{i} / d}$ and $\oplus$ denotes addition modulo $d$.

\subsection{3-uniform states}   

{\color{\redCom} 
A $3$-uniform state may  be derived from classical or quantum orthogonal {\sl Latin cubes}. Three  orthogonal Latin cubes
of size $d$ lead to a 
$3$-unitary matrix of order $d^3$
and generate state AME(6,$d$),
while four orthogonal cubes allows one to construct  the state AME(7,$d$).}

\begin{itemize}
    \item 
AME(6,2) state -- was investigated by
Borras et al.~\cite{Borras07}. 
Making use of the notation of quantum orthogonal arrays, introduced while discussing 2-uniform states --see Eq.~\eqref{eq:QOA}--, we can write the following equivalence
\begin{equation}
    \ket{\text{AME}(6,2)} = \ket{\text{QOA}(8,3_C+2_Q,2,3)} = \ket{\text{MOQLC}(2)},
\end{equation}
where we define mutually orthogonal quantum Latin cube (MOQLC) as in the array below~\cite{Goyeneche_2018}:

\begin{equation}\label{cube}
\begin{array}{l}
\begin{tikzpicture}
[
back line/.style={dashed},
cross line/.style={preaction={draw=white, -,
line width=6pt}}]
\node at (-4.7,0) {\text{MOQLC}(2)\quad =};
\matrix (m) [matrix of math nodes,
row sep=2.9em, column sep=0.1em,
text height=0.25ex,
text depth=0.25ex]
{
& \ket{\text{GHZ}_{100}} & & \ket{\text{GHZ}_{101}} \\
\ket{\text{GHZ}_{000}} & & \ket{\text{GHZ}_{001}} \\
& \ket{\text{GHZ}_{110}} & & \ket{\text{GHZ}_{111}} \\
\ket{\text{GHZ}_{010}} & & \ket{\text{GHZ}_{011}} \\
};
\path[-]
(m-1-2) edge [back line] (m-1-4)
edge [back line] (m-2-1)
edge [back line] (m-3-2)
(m-1-4) edge [back line] (m-3-4)
edge [back line] (m-2-3)
(m-2-1) edge [back line] (m-2-3)
edge [back line] (m-4-1)
(m-3-2) edge [back line] (m-3-4)
edge [back line] (m-4-1)
(m-4-1) edge [back line] (m-4-3)
(m-3-4) edge [back line] (m-4-3)
(m-2-3) edge [back line] (m-4-3);
\end{tikzpicture}
\end{array}.
\end{equation}
In the above cube construction, we have used a basis of three qubit GHZ states, defined as
\begin{equation}
    \ket{\text{GHZ}_{ijk}} = (-1)^{\alpha_{ijk}} \sigma_i \otimes \sigma_j \otimes \sigma_k \ket{\text{GHZ}},
\end{equation}
where $i,j,k\in \{0,1\}$ and $\alpha_{ijk}$ equals to 1 if $i=j=k$ and 0 otherwise. 
In the above notation, we have used the standard state $\ket{\text{GHZ}} = \frac{1}{\sqrt{2}} (\ket{000}+\ket{111})$, as well as the Pauli matrices $\sigma_0 = \sigma_x$ and $\sigma_1 = \sigma_z$~\cite{Goyeneche_2018}.

Alternatively, the same AME(6,2) state can be written using 3-unitary Hadamard matrix $O_8 \in \mathcal{O}(8)$~\cite{Goyeneche2015},
\begin{equation}
    \ket{\text{AME}(6,2)} =
     {\color{\redCom}\frac{1}{2\sqrt{2}}}
    \sum_{ijk} \ket{ijk}\otimes \ket{\text{GHZ}_{ijk}} = 
    {\color{\redCom}\frac{1}{2\sqrt{2}}}
    \sum_{ijk} \ket{ijk}\otimes O_8\ket{ijk}.
\end{equation}
The corresponding AME(6,2) graph state can be deduced from any of the graphs presented in Fig.~\ref{fig:graph_6_2}.

\begin{figure}[ht]
\centering
\begin{tikzpicture}

\node at (0,0) {\includegraphics[width=0.2\linewidth]{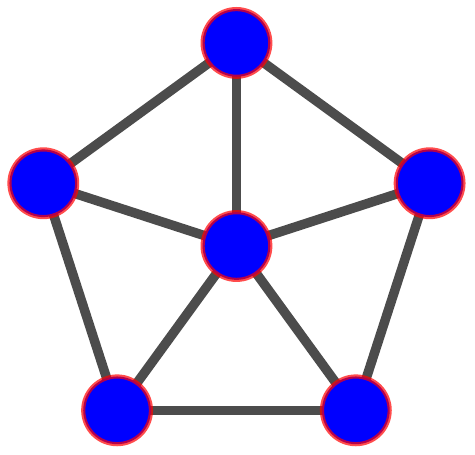}};

\node at (3.5,0) {\scalebox{2}{$\equiv$}};

\node at (7,0) {\includegraphics[width=0.24\linewidth]{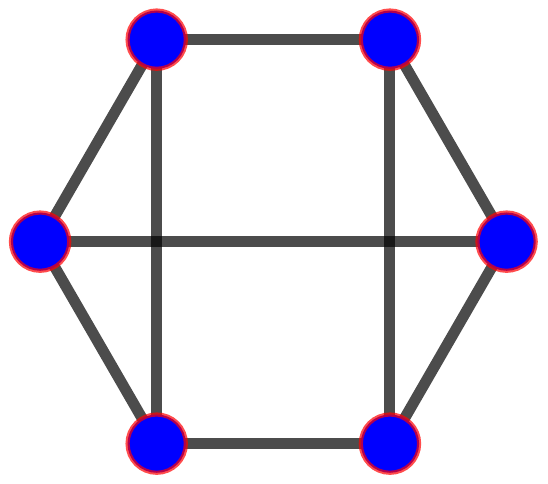}};

\end{tikzpicture}
\caption{
Six-vertex graphs that define two LU-equivalent AME(6,2) graph states
\cite{Helwig_2013,Ramadas_2024}.
Each 
edge represents a controlled Z gate between the corresponding qubits --
{\color{\redCom}
note that 
the left realization
requires ten gates
while the right 
one only nine.
The same state can be
 constructed
using  only seven
two-qubit gates~\cite{miller2024weightEnumExperiment},
provided some
additional single-qubit gates
are applied.
}}
\label{fig:graph_6_2}
\end{figure}

\item AME(6,4) state -- in this setup, absolutely maximally entangled states are equivalent to 3 mutually orthogonal Latin cubes~\cite{Goyeneche2015}, see Fig.~\ref{fig:OLS_6_4}.
One of them can be written as
\begin{equation}\label{eq:AME_6_4}
    \ket{\text{AME}(6,4)} = {\color{\redCom}\frac{1}{8}} \sum_{i,j,k=0}^3 \ket{i,j,k,i\oplus j \oplus k,i \oplus 2j \oplus 3k,i\oplus 3j\oplus 2k},
\end{equation}
with addition modulo six. The corresponding graphs are presented in Fig.~\ref{fig:graph_6_4} while the generating matrix reads
\begin{equation}\label{eq:generator_6_4}
	 G_{6,4}^\top = \left[
	\begin{array}{cccccc}
	1 & 0 & 0 & 1 & 1 & 1 \\
	0 & 1 & 0 & 1 & 2 & 3 \\
    0 & 0 & 1 & 1 & 3 &2
    \end{array}
	\right],
\end{equation}
see Appendix~\ref{app:graph_states} for more details on generating matrices.
\begin{figure}
\includegraphics[width=0.6\linewidth]{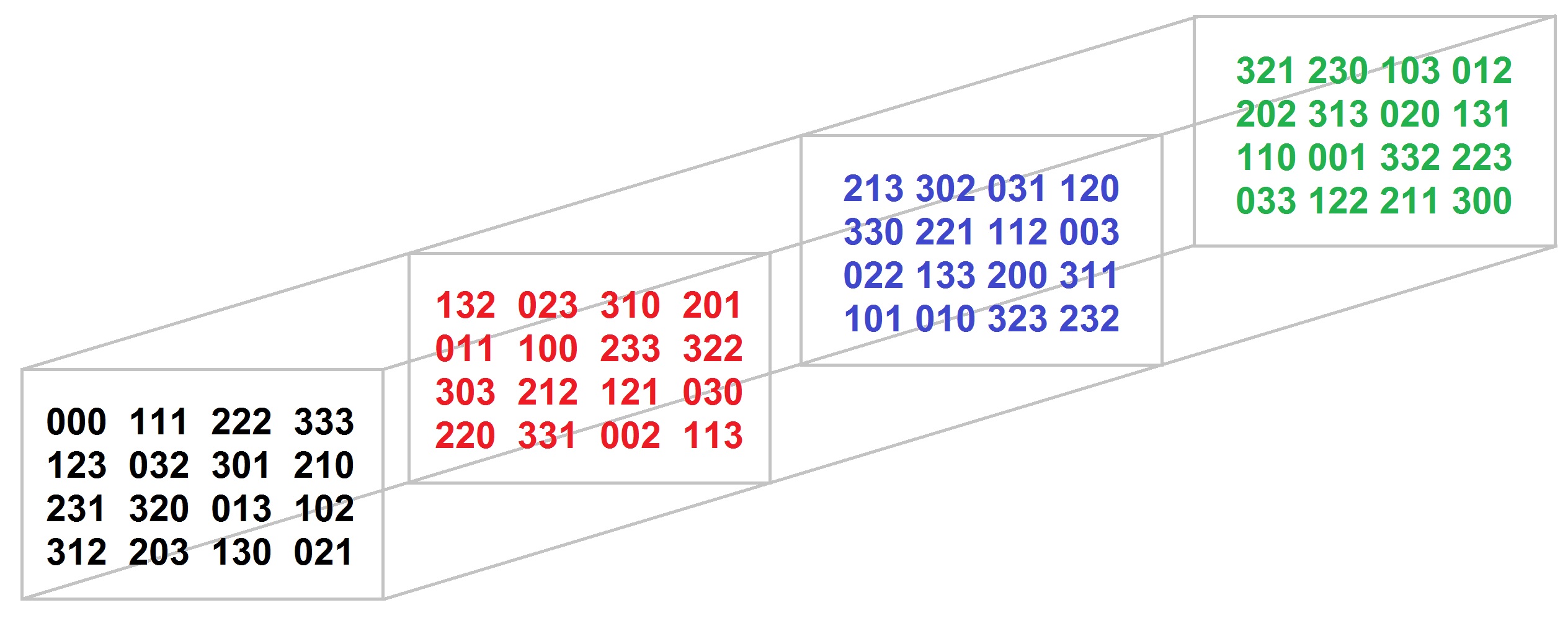}
\caption{
Three orthogonal Latin cubes of size four,
denoted as 4OLC(4), determine the structure of the AME(6,4) state as a superposition of 64 states
of six ququarts ~\cite{Goyeneche2015},
see Eq.~\eqref{eq:AME_6_4}.}
\label{fig:OLS_6_4}
\end{figure}

\begin{figure}
\centering
\begin{tikzpicture}

\node at (0,0) {\includegraphics[width=0.19\linewidth]{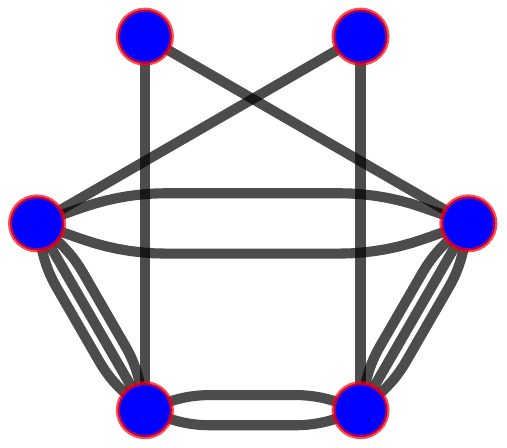}};

\node at (3.5,0) {\scalebox{2}{$\equiv$}};

\node at (7,0) {\includegraphics[width=0.19\linewidth]{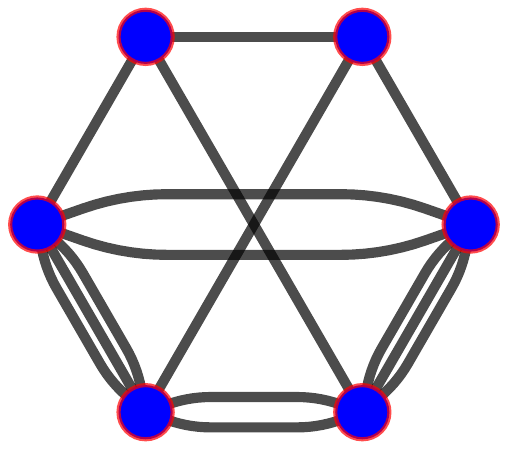}};

\end{tikzpicture}
\caption{Two equivalent ququart graph states, locally equivalent to minimal support AME$(6,4)$ state constructed form three mutually orthogonal Latin cubes of order four \cite{Goyeneche2015}. Note triple bounds characteristic to $d\ge 4$.}
\label{fig:graph_6_4}
\end{figure}

\item AME(6,$d$) state -- 
a method that is often used to construct AME states from smaller systems is by tensoring maximally entangled basis states to the terms superposed in a construction for an AME state \cite{Goyeneche2014}. For instance, two copies of AME(4,3) state
\eqref{eq:p9_state} allow one to
construct six-partite AME states shared among systems with odd local dimension $d$ given by
\begin{equation}
    \ket{\text{AME}(6,d)} = \frac{1}{d}\sum_{i,j=0}^{d-1}\ket{i,j,i\oplus j, i\oplus 2j}\ket{\phi_{i,j}}
\end{equation}
where $\ket{\phi_{i,j}}=\sum_{k=1}^{d}\omega_d^{ik}\ket{i\oplus k,k}$ with $\omega_d=e^{\frac{2\pi \text{i}}{d}}$.
For a special case of prime $p\coloneqq d$, AME(6,$p$) is given by 3 mutually orthogonal Latin cubes of size $p$, see Eq.~(39) in~\cite{Goyeneche_2018}.
At the same time, this arrangement is equal to a quantum orthogonal array
\begin{equation}
    3\text{MOQLS}(p) = \text{QOA}(p^3,3_c+3_q,p,3).
\end{equation}

{\color{\redCom}
\item AME(7,2) state does not exist \cite{Huber_2017} and this case serves
as an example of frustration effect \cite{FFMPP10}. Among all
seven-qubit pure states, several
strongly entangled cases were identified \cite{Borras07,TFCS09,ZYZ13,
Goyeneche2014,ZH22,YELL23}
}

\item AME(7,3) state -- in this case
the frustration is lifted due to a larger number of degrees of freedom. Such an AME state corresponds to a graph
found by Helwig~\cite{Helwig_2013} and
presented in Fig.~\ref{fig:graph_7_3}.

\begin{figure}[H]
\centering
\includegraphics[width=0.2\columnwidth]{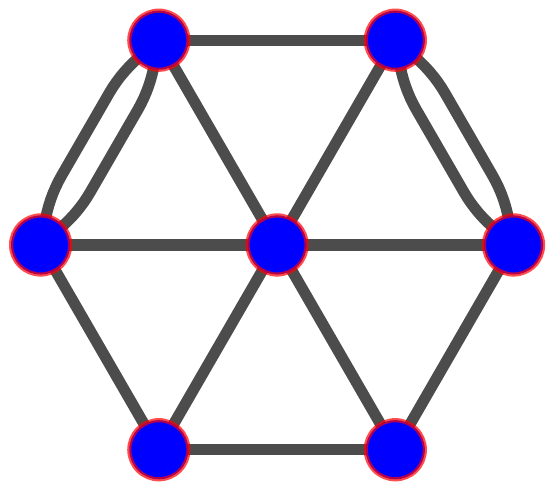}
\caption{Graph corresponding
to the AME$(7,3)$ state of seven qutrits 
contains two double bounds 
\cite{Helwig_2013,ZBE24}.}
\label{fig:graph_7_3}
\end{figure}

\end{itemize}

\end{widetext}


\bibliography{bibliography}

\end{document}